\documentclass[12pt]{article}
\usepackage[small,bf]{caption}
\usepackage[scaled=0.92]{helvet}
\usepackage[letterpaper, margin=1in]{geometry}
\usepackage{amsmath}
\usepackage{amssymb}
\usepackage{amsthm}
\usepackage{mathtools}
\usepackage{graphicx}
\usepackage{algorithm}
\usepackage{algpseudocode}
\usepackage{float}
\usepackage{subfig}
\usepackage{booktabs}
\usepackage{enumitem}
\usepackage{csquotes}
\usepackage{array}
\usepackage[T1]{fontenc}
\usepackage[swedish,english]{babel}
\usepackage{pdfpages}
\usepackage{colortbl}
\usepackage[backend=biber, style=alphabetic]{biblatex}

\usepackage{xcolor}
\usepackage{hyperref}
\hypersetup{
    colorlinks=true,
    citecolor=blue,
    urlcolor=blue,
    linkcolor=blue
}
\usepackage{bm}
\usepackage{siunitx}
\usepackage{titlesec}
\usepackage{listings}
\begin{document}

\begin{center}
{\fontsize{17}{23}\selectfont Simple Portfolios that Deliver Consistent Decent Returns}\\[1em]

{
\large
Oskar Edén Wallberg\textsuperscript{1,2}
\quad
Alexander Karolin\textsuperscript{1,2}
}\\[1em]

{\large February 5, 2026}
\end{center}

\begin{abstract}
\noindent
Portfolio construction has been studied extensively in both academia and industry. Its applications range from pension funds and hedge funds to private savings. The academic literature centers on methodology, commonly evaluated in settings that allow short selling, leverage, and frequent trading. Industry, on the other hand, operates under real constraints, but its data and methods are proprietary and largely unpublished. Neither is directly available to the private investor. This paper is an empirical study of established portfolio construction methods in the setting of the private investor. We consider a small investable universe of liquid US ETFs, where trading is monthly and carries a cost, and neither shorting nor leverage is permitted. Portfolio construction is posed as a convex optimization problem, its inputs are estimated from public data, and every parameter is interpretable. We denote such portfolios simple. We show that simple portfolios of as few as three ETFs deliver consistent and decent returns, without tuning to the backtest period. To find them, we have developed a framework, which we share publicly, able to backtest hundreds of portfolios on a standard laptop over a twenty year period or longer. When no return forecast is made, portfolios are constructed from risk alone, using methods such as risk parity and minimum volatility, built on a statistical factor model fitted with expectation maximization. When a forecast is made, we combine the Markowitz formulation with momentum signals and ridge regression. We report what works and what each component adds, weighing its gain against its complexity and the effort it demands of the investor. The findings are supported by numerical experiments over the past two decades, and the proposed portfolios outperform the standard 60/40 allocation.
\end{abstract}

\vspace{11em}
\noindent\textbf{Keywords:} Portfolio construction, exchange-traded funds, convex optimization, factor risk models, return forecasting, risk-based asset allocation, backtesting, retail investing

\begin{NoHyper}
\let\thefootnote\relax
\footnotetext{\textsuperscript{1}KTH Royal Institute of Technology, Stockholm, Sweden.}
\footnotetext{\textsuperscript{2}Visiting Student Researcher (VSR) at Stanford University under Professor Stephen P. Boyd.}
\footnotetext{Both authors contributed equally to this work.}
\footnotetext{This work was supported by the foundations of the Royal Swedish Academy of Sciences (Kungl. Vetenskapsakademien), through Stiftelsen Claes Adelskölds medalj- och stipendiefond, grant ES2025-0064.}
\end{NoHyper}

\thispagestyle{empty}

\clearpage
\tableofcontents
\thispagestyle{empty}
\newpage

\pagenumbering{arabic}
\section{Introduction}
Portfolio construction was first formalized by Harry Markowitz in 1952~\cite{Markowitz:1952}. The main idea behind portfolio construction is to balance risk and reward by diversifying investments. Markowitz was able to define diversification in mathematical terms, however the concept has been known for much longer. As far back as the third century, the Talmud advised splitting wealth three ways, one in land, one in business, and one held in reserve~\cite{markowitz1999early}. The core concepts still apply today, and investing has grown significantly in popularity. Professional investors such as Warren Buffett and Peter Lynch diversify across tens or hundreds of investments, while quantitative hedge funds such as Citadel hold positions in thousands of different financial instruments. For the private investor, tracking such quantities of assets requires time and effort beyond what is reasonable. In this paper, we present some modern portfolio construction techniques applied to portfolios with a more manageable number of assets, making them practical for private investors to replicate and maintain.

\subsection{Investing and diversification}\label{sec:investing}
A common problem all investors face is how to allocate their capital in a way that balances risk and return. There are many options to choose from in financial markets, including securities such as stocks, bonds, and mutual funds. A more recent addition, the exchange-traded fund (ETF), has become an increasingly popular alternative. An ETF is a fund traded like a stock, meaning that it can be bought and sold throughout the trading day, unlike a mutual fund. An ETF holds a basket of underlying assets such as stocks, bonds, or commodities, and typically tracks an index. The most traded ETF has the ticker (identifier) SPY~\cite{SPDR:SPY}, and tracks the Standard and Poor's 500 (S\&P~500). The S\&P~500 is a market index that tracks the performance of the 500 largest publicly traded companies in the US.

\begin{figure}[h]
    \centering
    \includegraphics[width=1.0\linewidth]{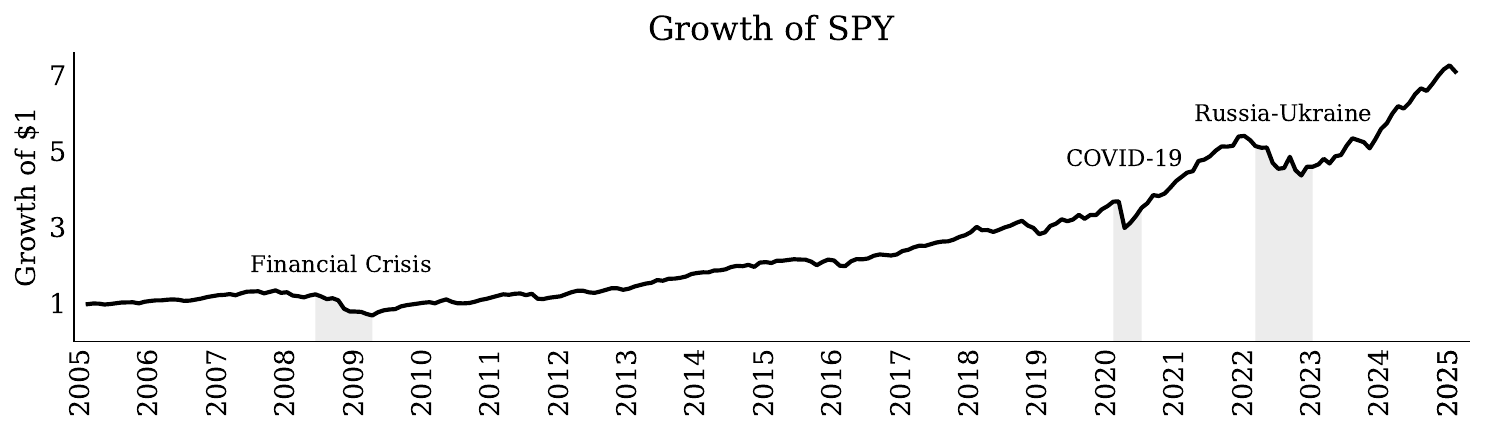}
    \caption{An investment in SPY grows roughly sevenfold over two decades.}
\label{fig:spy}
\end{figure}

A long-standing fact is that it is difficult to consistently outperform a broad market index such as the S\&P~500, even for professional managers~\cite{malkiel2003,spiva}. Warren Buffett famously bet a financial institute \$1 million that no funds they selected could outperform the S\&P~500, net of fees, over a 10-year period (2008--2017)~\cite{buffett2017}. Buffett won. A natural first investment is therefore to simply hold the S\&P~500, e.g., via SPY. Figure~\ref{fig:spy} shows the growth of SPY over the past two decades, with an initial investment growing roughly sevenfold.

\begin{figure}[h]
    \centering
    \includegraphics[width=1.0\linewidth]{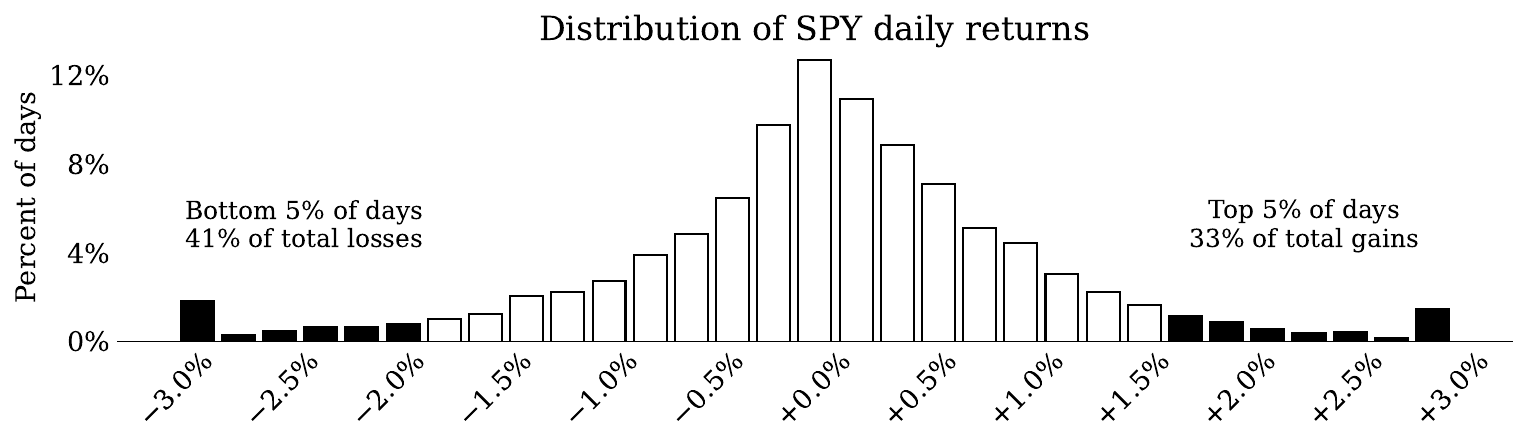}
    \caption{A small minority of days accounts for most of SPY's gains and losses.}
    \label{fig:spy_dist}
\end{figure}

\newpage
\noindent The path to a sevenfold total return is far from smooth. The average annual return of SPY over this period is $10\%$, meaning that if you were to invest \$100 at the beginning of a given year you would expect to have \$110 at the end of that year. This is however just an average. In reality the return of SPY will be higher or lower for some years. In fact, the majority of the losses and gains from SPY come from a small minority of days, as seen in Figure~\ref{fig:spy_dist}. 

\noindent We refer to the magnitude of these variations as the volatility, or the "riskiness" of investing in an asset. For SPY, the annual volatility is $19\%$, meaning that the annual returns typically fall within $-9\%$ and $29\%$ ($10\%\pm19\%$) in a given year, though not always. Some years have seen extreme drawdowns such as during the Financial Crisis in 2008, the COVID-19 pandemic in 2020, and the start of the Russia-Ukraine war in 2022. The largest drawdown was during the Financial Crisis when SPY fell by $55\%$ from its peak, wiping out an estimated $\$7$ trillion of market cap in the S\&P~500~\cite{brunnermeier2009deciphering}.

\begin{figure}[!h]
    \centering
    \includegraphics[width=1.0\linewidth]{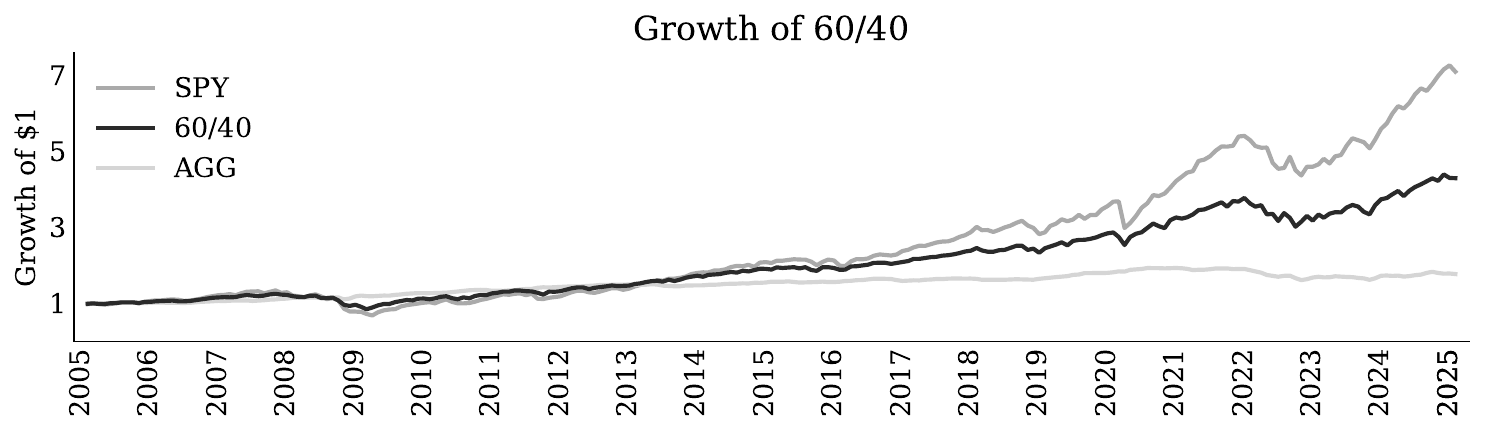}
    \caption{The 60/40 keeps most of SPY's growth at nearly half its volatility.}
    \label{fig:6040}
\end{figure}

One simple method for mitigating volatility and large drawdowns is to diversify investments into multiple assets that do not experience losses and gains at the same time. The most famous and very simple example of diversification is the 60/40 portfolio~\cite{cfa6040_2024}, which can be replicated with ETFs by combining SPY with AGG into a 60{:}40 ratio. AGG is an ETF that tracks an index of US aggregate bonds. Bonds have historically had a low or slightly negative correlation with equities such as SPY. Figure~\ref{fig:6040} shows the cumulative return of SPY, AGG, and the 60/40 portfolio over the same period.

\noindent Looking at the figure, SPY ends with the largest total return but is also the most volatile of the three. AGG, on the other hand, grows much more slowly with an average annual return of $3\%$, but is much more stable with a volatility of $5\%$. The 60/40 portfolio falls somewhere in between, with an average annual return of $8\%$ and a volatility of $11\%$. The same pattern shows up in the drawdowns. During the 2008 Financial Crisis AGG only fell by $13\%$ and the 60/40 by $36\%$, compared to SPY's $55\%$. By mixing in some bonds, you keep some of SPY's long-run growth while almost halving the volatility and taking a much smaller hit during bad years. This is the value of diversification, by combining assets whose returns are mostly independent you end up with a portfolio that gives you a better balance between risk and return than either asset on its own. The trade-off between return and risk is often expressed as a ratio, formally known as the Sharpe ratio~\cite{sharpe1966mutual}. The Sharpe ratio expresses the efficiency of the portfolio and can be interpreted as return per unit of risk taken. Between 2005 and 2025 the Sharpe ratios for AGG, SPY, and 60/40 were 0.23, 0.45, and 0.51, respectively.

\subsection{Quantitative trading}\label{sec:quant_trading}
So far we have only looked at risk and return with one famous example of a portfolio. In practice, investing involves many more considerations. Which assets should be held, in what proportions, and when should they be bought or sold? Most private investors have no systematic framework to answer these questions, constructing their investment philosophy instead from experience and word of mouth~\cite{Malmendier:2011, Hong:2004, Kaustia:2012}. Decisions made this way have historically underperformed the market~\cite{Dalbar:2024, Barber:2000, Frazzini:2008}. An alternative is to make decisions systematically. One example of this is quantitative trading, where every rule is mathematically defined and decisions are computed from historical data. The investor's job now reduces to choosing a trading strategy and following it.

\begin{figure}[H]
    \centering
    \includegraphics[width=1.0\linewidth]{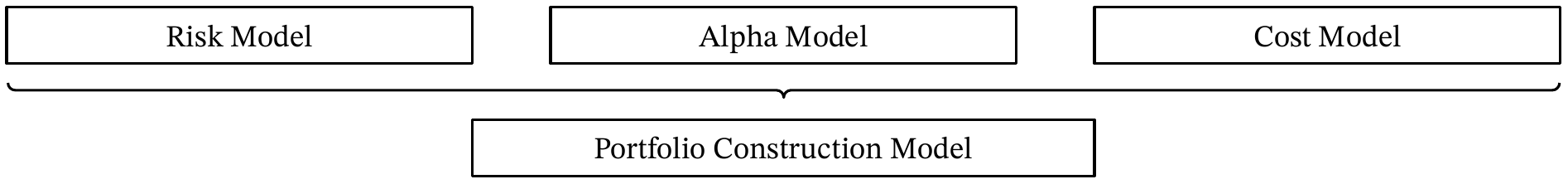}
    \caption{Basic structure of quantitative trading, adapted from~\cite[p.~17]{Narang:2024}.}
    \label{fig:quant_framework}
\end{figure}

To make these decisions systematically, a mathematical framework can be used. One widely used framework is the Markowitz formulation~\cite{Markowitz:1952} expressed as a convex problem. The original Markowitz formulation models the trade-off between return and risk, and adding cost gives us a complete model for quantitative trading. A simplification of this model can be expressed as a general convex problem,
\[
\begin{aligned}
    & \text{maximize}   && \text{Return} - \text{Risk} - \text{Cost} \\
    & \text{subject to} && \text{Trading limits},
\end{aligned}
\]
where the return is maximized while risk and cost are minimized, subject to limits on the trades. Each of the three terms can be modeled mathematically. The risk is captured by a risk model, the return by a forecast, and the cost by a cost model. Forecasted returns are commonly known as alpha, hence the forecast is described as an alpha model. Combining the three gives a complete portfolio construction model, as in Figure~\ref{fig:quant_framework}.

\begin{figure}[H]
    \centering
    \includegraphics[width=1.0\linewidth]{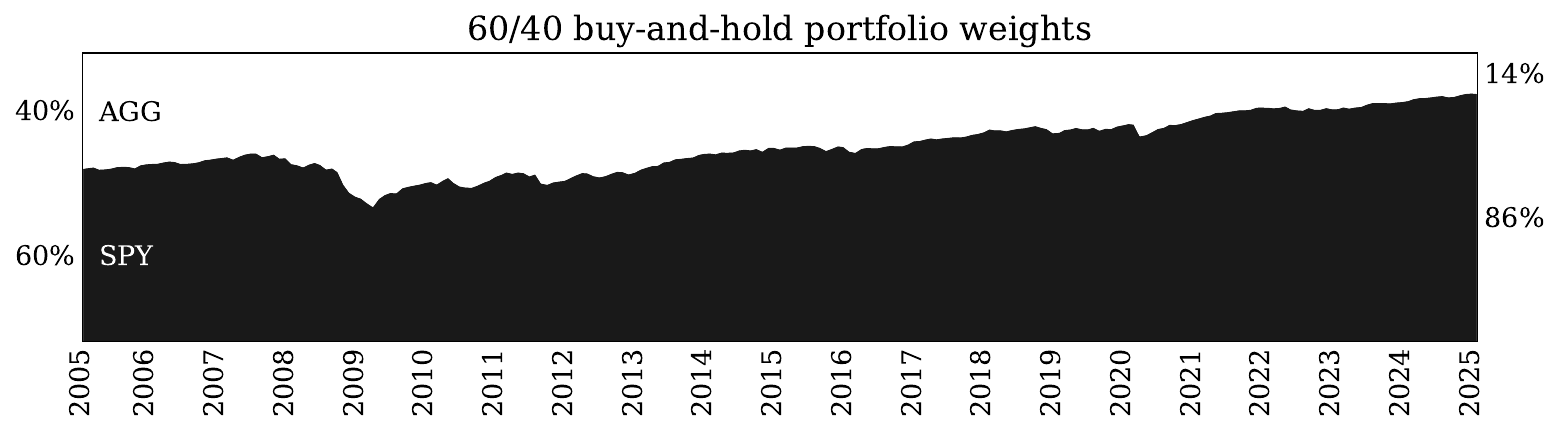}
    \caption{Without rebalancing, the 60/40 drifts to an 86:14 split}
    \label{fig:drift_6040}
\end{figure}

\paragraph{Trading.} The model of Figure~\ref{fig:quant_framework} decide what to buy at a given time, but how frequently a quantitative strategy trades varies. Some large firms utilize high frequency trading, executing trades multiple times per second, while others trade far less often. Regardless of strategy, you must trade at some frequency. Consider a simple strategy such as the 60/40 of Figure~\ref{fig:6040}. Simply buying SPY and AGG in a 60:40 ratio in 2005 and never touching the portfolio would have ended 2025 at an 86:14 ratio, as seen in Figure~\ref{fig:drift_6040}, because SPY outgrew AGG. The drifted portfolio is no longer the one chosen. Its volatility in 2025 was $11\%$ against $8\%$ for the maintained version, so without trading, the risk profile walks away from what the investor intended. For a fixed weight strategy such as the 60/40, restoring the target weights, selling what has grown and buying what has lagged, is known as rebalancing. Rebalancing also carries a second benefit, in that it systematically trims positions after they have risen and adds to them after they have fallen, without any attempt to time the market. The 60/40 is a deliberately simple example, where trading only ever restores the same fixed weights. A strategy can equally well decide to move capital from one asset to another based on another heuristic than weight drifting, and the trades then serve whatever the strategy dictates. 

\paragraph{Risk model.} In \S\ref{sec:investing} we saw that what matters for a portfolio is not just how much each asset moves, but also how the asset returns $r$ move together. The risk model collects both into a single object, the covariance matrix $\Sigma = \operatorname{Cov}(r)$. For a universe of $n$ assets the covariance matrix is an $n \times n$ grid,
\begin{equation*}
\Sigma =
\begin{bmatrix}
\sigma_1^2 & \sigma_{12} & \cdots & \sigma_{1n} \\
\sigma_{21} & \sigma_2^2 & \cdots & \sigma_{2n} \\
\vdots & \vdots & \ddots & \vdots \\
\sigma_{n1} & \sigma_{n2} & \cdots & \sigma_n^2
\end{bmatrix},
\end{equation*}
where each diagonal entry $\sigma_i^2$ is the variance of asset $i$ and each off-diagonal entry $\sigma_{ij}$ is the covariance between assets $i$ and $j$. Just like volatility, variance is also a measurement of risk and describes how much an asset moves on its own, while covariance describes how two assets move together. A large positive $\sigma_{ij}$ means two assets tend to rise and fall as one, while a negative value means one tends to rise when the other falls. This is the property diversification relies on, and from this single object the risk of any portfolio can be computed.

\paragraph{Alpha.} The risk model describes how asset returns move
together, but says nothing about where they are headed. That forecast is
the job of the alpha model. Within the financial literature alpha is the expected asset returns over the next period,
\begin{equation*}
    \alpha_{t} = \mathbf{E}[r_{t+1}].
\end{equation*}
Within the context of Markowitz, alpha does not have to be the exact future return of the assets. Rather it can be simplified and estimation errors can be reduced using robustification techniques. Some simplifications include reducing alpha to a yes or no question, whether the return of asset A will be positive or negative tomorrow. Another simplification is forecasting only the relative order of the investable assets. Simple forecasting methods use previous returns of each asset to model future returns. However, in industry alpha is more diffuse, and can be interpreted as some information about how the prices of assets will move in the future. Examples of data that have been used to achieve alpha are satellite imagery, weather forecasts, or social media sentiment. For the curious reader,~\cite{narang2020alpha} gives a more thorough breakdown, best enjoyed while taking lunch.

\begin{figure}[!h]
    \centering
    \includegraphics[width=1.0\linewidth]{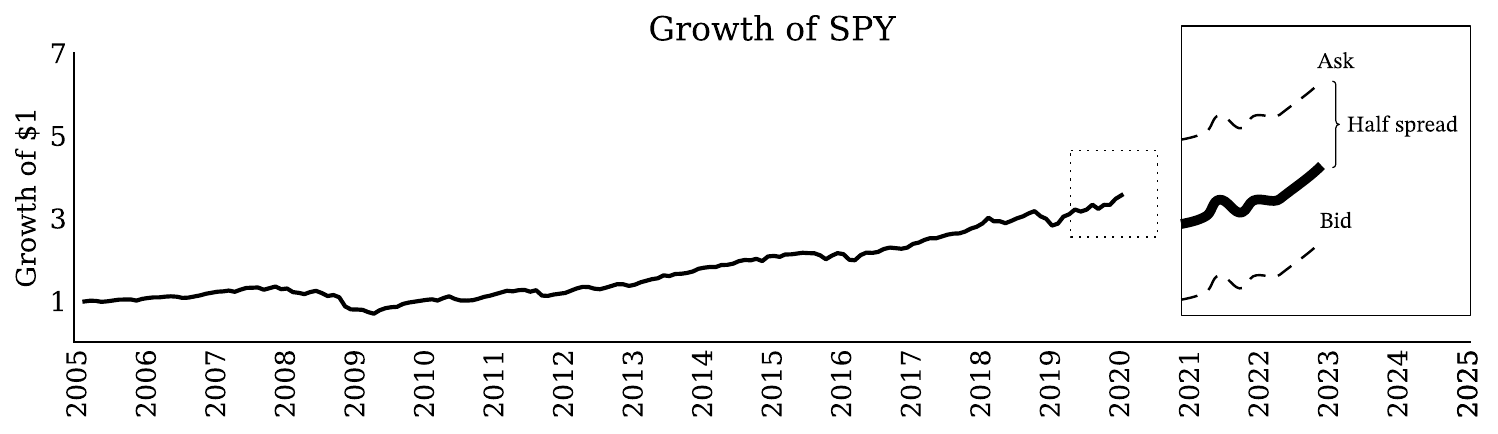}
    \caption{Each side of a trade pays the half-spread against the midpoint.}
    \label{fig:bid_ask}
\end{figure}

\paragraph{Cost model.} Trading is not free and each trade incurs two main costs. The first is the broker fee, or commission, charged to execute the trade. The second is the bid-ask spread. At any moment an asset has two quoted prices, the bid, the highest price a buyer will pay, and the ask, the lowest price a seller will accept. The spread is the gap between them, and it is a cost to both sides. Suppose you want to sell asset A. The ask price is \$101 and the bid price is \$99. The effective price sits in the middle at \$100, so selling at the \$99 bid costs you \$1 against that midpoint, half the \$2 spread. Spreads are usually quoted in basis points, where one hundred basis points equal 1\% of the price. A \$2 spread on a \$100 asset is a 200 basis point spread, and the half-spread paid by each side is 100 basis points. Figure~\ref{fig:bid_ask} illustrates this with the common example of SPY.

\noindent The figure shows the price of SPY, with the bid and ask quoted on either side as dashed lines. The true price lies in the middle, and the distance from the midpoint to either quote is the half-spread, what a buyer or seller actually pays. Cost models can be made considerably more detailed, accounting for fees, currency conversion, and the price impact of large orders, but they are often simplified to a fixed percentage charged on the value of every trade.

\newpage

\subsection{Previous work}\label{sec:previous_work}
This paper is an empirical study of portfolio construction applied to a universe of ETFs. Applying systematic portfolio methods to ETFs is itself an active area of research, with recent work studying multi-asset ETF portfolios built from prediction and optimization~\cite{linghu2025ipmo}, momentum and trend-following across asset classes~\cite{lu2025trendfolios}, and the effect of rebalancing on ETF-only portfolios~\cite{banyai2024rebalancing}. Our contribution is not a new method but a transparent application of well-established techniques to a simple, retail-accessible universe, with an emphasis on what each component contributes. Further work on the topic can be found in~\cite{devanathan:2026}, which was developed in parallel within the same group at Stanford.

\paragraph{Portfolio construction.} The original Markowitz formulation has been studied thoroughly, and many variations exist. A few examples include the Black-Litterman model~\cite{black1992global}, robust mean-variance optimization~\cite{goldfarb2003robust}, and the modern Markowitz at seventy formulation~\cite{Boyd:2024}. These variations were, among other reasons, developed to address the estimation sensitivity that return forecasting introduces in the original Markowitz formulation. There are however multiple methods that do not require an estimated future return. Common methods include minimum variance~\cite{Clarke:2006}, inverse volatility~\cite{Lee:2011}, risk parity~\cite{qian2005riskparity}, and maximum diversification~\cite{Choueifaty:2008}. These methods instead construct portfolios based on asset risk.

\paragraph{Risk model.} Many portfolio construction methods require a risk model which is a covariance matrix of asset returns estimated from a rolling window of historical data. Estimating this matrix well is difficult, and a large part of current financial research goes into producing robust estimates~\cite{tzikas2026enhancing, johansson2025convex, parshakova2025fitting, Cederberg:2025}. Using a naive sample covariance introduces multiple problems. One of the problems being that when the number of assets is large relative to the number of observations the estimate becomes unreliable. Another problem is that using large covariance matrices is computationally inefficient. One way to handle this problem is to use a factor model. Factor models have a long history. The concept of factor analysis was first introduced by Spearman in 1904~\cite{spearman1904}. It was later adopted in finance where asset returns are modeled as being driven by a few common factors plus idiosyncratic noise. While such models date back to the single-factor CAPM~\cite{sharpe1964capm}, the framework was popularized by Fama and French, who introduced empirically tested models with multiple factors~\cite{FamaFrench:1992, FamaFrench:1993}. Factor models are widely used today in industry~\cite{barra, axioma} and academia~\cite{fan2008highdim, fan2013poet, ledoitwolf2003} and results in interpretable, faster and more reliable covariance estimate. Most factor models use economically interpretable factors, such as the market. An alternative is to extract factors statistically from the returns themselves, for instance through principal component analysis or related latent-factor methods~\cite{connor1986performance}.

\paragraph{Alpha.} The return forecast, or alpha, is a central part of the Markowitz formulation, and receives a great deal of attention in both academia and industry~\cite{kozak2020shrinking, kelly2019characteristics, barunik2025forecasting, deprado2020machine}. Return prediction is difficult. The signal-to-noise ratio in financial data is extremely low, far lower than in most other domains where statistical learning is applied~\cite{deprado2018advances}. It is easy to overfit to noise rather than signal, and therefore it limits the complexity of the models that can be used. A common response to the low signal-to-noise ratio is regularization, which constrains the model to reduce overfitting. Examples include ridge regression~\cite{HoerlKennard:1970}, Laplacian-regularized models~\cite{tuck2021distributed}, and the broader use of shrinkage and sign constraints. Whether alpha prediction works is debated in academia. Goyal and Welch~\cite{GoyalWelch:2008} found that a long list of popular predictors fail to beat the historical mean out of sample, while Campbell and Thompson~\cite{campbell2008predicting} showed that the same predictors do beat it once weak restrictions are imposed. Within industry, alpha is achieved by having an edge against the market. Data and methods used to produce alpha are highly confidential and employees working with alpha prediction often have to sign long non-disclosure and non-compete agreements~\cite{zuckerman2019man}.

\paragraph{Software.} Software for solving optimization problems is almost as old as the methods themselves. The first nontrivial linear program ever solved took nine people around four months on hand calculators in the 1940s~\cite{bixby2012history}, a calculation that solvers now finish instantly.\footnote{Throughout this project we were challenged to solve at least one billion Markowitz problems, an ambition that would have made those nine people faint.} Today the field of convex optimization has matured and been adopted in many software libraries such as CVXPY 
for Python~\cite{diamond2016cvxpy}, Convex.jl for Julia~\cite{convexjl}, and CVXR for R~\cite{cvxr}. These libraries have produced a wave of accessible portfolio optimization tools. A few examples are Portfolio Visualizer~\cite{portfoliovisualizer}, PyPortfolioOpt~\cite{pyportfolioopt}, Riskfolio-Lib~\cite{riskfoliolib}, and CVXPortfolio~\cite{cvxportfolio}. Their coverage varies widely, from simple web-based backtesting and visualization to specific methods such as mean-variance optimization, risk parity~\cite{deprado2016hrp}, and multi-period construction. Commercial alternatives also exist and are widely used in industry~\cite{axioma, barra, aladdin}.

\subsection{This paper}\label{sec:this_paper}
\noindent This paper studies portfolios that are simple, and yields consistent, and decent returns. None of the methods we use are new. Techniques are gathered from across the portfolio construction literature and industry practitioners and we apply them to one universe under one set of rules, and report what works. The rules are the same for every portfolio. We use no leverage and no shorting, target an annual volatility of $8\%$, and earn the federal funds rate on cash. Strategies trade monthly with a $5$ basis points bid-ask spread and are backtested from 2005-01-01 to 2025-01-01. We do not tune parameters to this period. The results are presented for research and educational purposes and do not constitute investment advice.

By simple we mean three things. First, the strategies are implementable. They use only public data and a small set of well-known, liquid US ETFs. Second, the math is explainable. We use no deep learning and no construction methods that are hard to interpret. We take a clear set of ETFs, divide them into asset classes, and run well-understood methods on each. Third, the parameters are interpretable. We use quantities a reader can reason about, such as a hard target-volatility constraint and turnover limits stated as percentages. If you want a higher volatility target, you change one number.

By consistent we mean a portfolio should do well across the whole period, not just in a few good years. We measure this with the Sharpe ratio, maximum drawdown, and, as a stricter test, with the lower quantile of the twenty recorded annual Sharpe ratios. By decent we mean annual returns in the range of 6\% to 10\%. 

We study two families of portfolios. In the first family, we make no return forecast at all. The allocation methods are fixed weights, inverse volatility, minimum volatility, risk parity, and maximum diversification, either combined across different asset classes or applied to just a few select assets. In the second family, each asset produces its own alpha signal. We combine all of these signals and solve a single Markowitz problem for each trading day. In the final section we present reasonable portfolios for the different methods we have explored.

We benchmark against the reference $60/40$ portfolio. A backtest over two decades can look good simply because of how a portfolio handled one or two extreme events, so we also check how sensitive our proposed portfolios are to the choice of window. Each portfolio is rerun over two truncated periods, one with the first five years removed, which excludes the Financial Crisis, and one with the last five removed, which excludes the COVID-19 pandemic and the start of the Russia-Ukraine war. If a portfolio performs well in all three windows we can consider it to be less dependent on a specific event.

\paragraph{Software.} Along with this paper we release three resources. The first is a web-based tool that lets readers explore the historical performance of some of the simple portfolios we construct, with interactive controls for the relevant parameters. The second is our codebase, made open source so that the strategies described can be replicated, extended, or used as a starting point for new portfolio designs.  Both are available at
\begin{center}
  \url{https://github.com/AlexEmrik/consistent-decent-returns}.
\end{center}
The third is an open source package which lets anyone create a factor risk model through a simple python API. It is installed via
\begin{center}
  \href{https://pypi.org/project/covforge/}{\texttt{pip install covforge}}
\end{center}

\paragraph{Outline.} The rest of the paper follows the structure of the Markowitz formulation, from the assets it allocates across to the inputs it requires. In \S\ref{sec:assets} we define the asset universe, organized into categories by asset class and underlying structure. In \S\ref{sec:portfolio_construction} we set up the portfolio construction problem, present the simple trading strategies and the modified Markowitz formulation we use throughout, and describe how the portfolios are selected. In \S\ref{sec:risk_model} and \S\ref{sec:alpha_model} we cover the two inputs the strategies requires, the risk model, from simple covariance estimators to the factor model, and the return forecast, from simple alphas to ridge regression. Finally, in \S\ref{sec:numerical_experiments} we report the experiments, including the data, backtest, parameter choices, and results, and conclude in \S\ref{sec:conclusion}.

\paragraph{What we have found.} Simple strategies work surprisingly well, often performing as well as or better than more complicated approaches. Some quantitative methods for investing stand out as particularly useful. One example is volatility control, which lets the investor target a desired level of volatility. Among the simple strategies, fixed weights stand out as both effective and extremely simple given an educated selection of ETFs. Given that we are limited to open data, risk forecasting works surprisingly well while return prediction proves to be futile. However, with simple alpha signals and some robustification, Markowitz portfolio optimization is still able to deliver consistent decent returns.

\newpage
\section{Assets}\label{sec:assets}

Portfolio construction is the problem of deciding how to allocate
capital across a set of assets. This section defines the universe of
ETFs we invest in, organized into categories by asset class and underlying structure. From this universe the simple strategies of \S\ref{sec:portfolio_construction} later select portfolios of three to five ETFs, while the Markowitz formulation is given whole asset categories.

\subsection{Investable universe}\label{section:investable_universe}
Our universe consists of the 59 US-listed ETFs given in Table~\ref{tab:asset_categories}, spanning the three asset classes: equities, fixed income, and alternatives. We include only large, liquid ETFs from established issuers, and exclude leveraged and inverse products, which amplify an index's daily moves or bet against them. The ETFs themselves come in a few different kinds. Some track an index, such as SPY tracking the S\&P~500, and AGG tracking a broad index of US bonds. Others hold a single asset, so their value simply follows its price. GLD holds gold and IBIT holds bitcoin. A few hold more complicated financial instruments tied to the price of an asset rather than the asset itself. We return to these in \S\ref{section:asset_categories}. Together the ETFs cover most of what an investor can reach with an online brokerage account. This includes several kinds of equities and bonds, commodities, and the recent addition of cryptocurrencies. Table~\ref{tab:asset_universe} in the appendix gives a short description and summary statistics for each ETF.

\begin{table}[h]
    \footnotesize
    \centering
    \renewcommand{\arraystretch}{1.3}
    \setlength{\tabcolsep}{3pt}
    \caption{Asset universe by class.}
    \begin{tabular}{l*{12}{l}}
        \textbf{Asset class} & \textbf{ETFs} &      &      &     &      &      &      &      &      &      &      &      \\
        \toprule
        Equities             & SPY  & QQQ  & VTV  & VUG & MDY  & IWM  & SCHD & USMV & QUAL & XLK  & XLV  & XLF  \\
                             & XLY  & XLI  & XLP  & XLE & XLU  & XLB  & IBB  & IYR  & EWJ  & EWG  & EWU  & EWA  \\
                             & EWH  & EWS  & EWZ  & EWT & EWY  & EWP  & EWW  & EWI  & EWD  & EWL  & EWC  &      \\
        \midrule
        Fixed income         & AGG  & TLT  & IEF  & TIP & MUB  & LQD  & LQDH & HYG  & BNDX & EMB  & IAGG & VWOB \\
        \midrule
        Alternatives         & GLD  & SLV  & PPLT & CPER & USO & UGA  & CORN & WEAT & SOYB & CANE & IBIT & ETHA \\
        \bottomrule
    \end{tabular} \\
        \label{tab:asset_categories}
\end{table}

\paragraph{Survivorship bias.}
Our universe was selected from ETFs trading today, and none of the 59 were delisted during the 2005--2025 backtest window; it is therefore survivorship-biased by construction. Roughly one third of all ETFs ever launched in the US have been liquidated, with 189 closures in 2024 alone and approximately 1{,}550 over the preceding decade~\cite{Kiplinger:2024,Morningstar:2024}. However, closures concentrate disproportionately in segments excluded from our universe. The typical liquidated ETF has fewer than \$50M in assets under management, a sub-five-year track record, and underperformed broad benchmarks by several percentage points in the trailing year prior to liquidation~\cite{Morningstar:2023}. Leveraged and inverse products, which we exclude entirely, account for a closure rate of 52\% versus 31\% for regular ETFs~\cite{Schwab:2024}. We therefore expect the resulting upward bias on reported Sharpe ratios to be modest, though we do not estimate it formally.

\subsection{Asset categories}\label{section:asset_categories}
Within each asset class some ETFs are structurally very different. We therefore divide the three classes further into a total of nine categories as seen in Table~\ref{tab:asset_subclasses}, each grouping ETFs that are similar in what they hold and how they hold it. The categorization is our own rather than an industry standard, and reflects the nature of the ETFs in our universe.

\begin{table}[h]
    \footnotesize
    \centering
    \renewcommand{\arraystretch}{1.3}
    \setlength{\tabcolsep}{2pt}
    \caption{Asset universe by class and category.}
    \begin{tabular}{ll*{11}{l}}
        \textbf{Class} & \textbf{Category}  &      &      &     &      &      &      &      &      &     & &   \\
        \toprule
        Equities     & Market        & SPY  & QQQ  & VTV  & VUG & MDY  & IWM  & SCHD & USMV & QUAL &     &     \\
                     & Sector        & XLK  & XLV  & XLF  & XLY & XLI  & XLP  & XLE  & XLU  & XLB  & IBB & IYR \\
                     & Country       & EWJ  & EWG  & EWU  & EWA & EWH  & EWS  & EWZ  & EWT  & EWY  & EWP & EWW \\
                     &               & EWI  & EWD  & EWL  & EWC &      &      &      &      &      &     &     \\
        \midrule
        Fixed income & Government    & AGG  & TLT  & IEF  & TIP  & MUB  &      &      &     &      &     & \\
                     & Corporate     & LQD  & LQDH & HYG  &      &      &      &      &     &      &     & \\
                     & Global        & BNDX & EMB  & IAGG & VWOB &      &      &      &     &      &     & \\
        \midrule 
        Alternatives & Physical      & GLD  & SLV  & PPLT &      &      &      &      &     &      &     & \\
                     & Futures       & CPER & USO  & UGA  & CORN & WEAT & SOYB & CANE &     &      &     & \\
                     & Crypto        & IBIT & ETHA &      &      &      &      &      &     &      &     & \\
        \bottomrule
    \end{tabular}
    \label{tab:asset_subclasses}
\end{table}

\paragraph{Equities.} ETFs within the Equities asset class consist of broad baskets of stocks. Simply put, owning a stock equates to owning a share of a company. Thus, owning an ETF within the Equities asset class equates to owning many small shares in different companies. We split Equities into three categories: Market, Sector, and Country. 

The Market category consists of nine ETFs spanning US market-cap, style, and factor exposures~\cite{SSGA:ETFs, iShares:ETFs, Vanguard:ETFs, Schwab:ETFs, Invesco:QQQ}. ETFs in this category are generally considered good buy-and-hold investments for the investor who does not engage in active portfolio construction. They tend to yield the highest average annual returns out of the categories and perform consistently well, but suffer during market crashes. Correlation within the Market category is very high. What differs is the risk profile, with QQQ at one end having an annual volatility of 22\% while USMV at the other has only 14\%.

The Sector category consists of the nine original Select Sector SPDRs~\cite{SPDR:SelectSectors}, which together divide the companies of the S\&P~500 by sector~\cite{MSCI:GICS}. All of the SPDR ETF tickers begin with XL. We complement them with two ETFs covering biotechnology and real estate~\cite{iShares:ETFs}. The nine original SPDRs are often grouped into defensive and offensive ones. Defensive sectors, such as health care and consumer staples, are less volatile and hold up better in distressed markets. Offensive sectors, such as technology and financials, trend periodically, delivering strong returns in good years but deeper drawdowns in bad ones.

\begin{figure}[!h]
    \centering
    \includegraphics[width=1.0\linewidth]{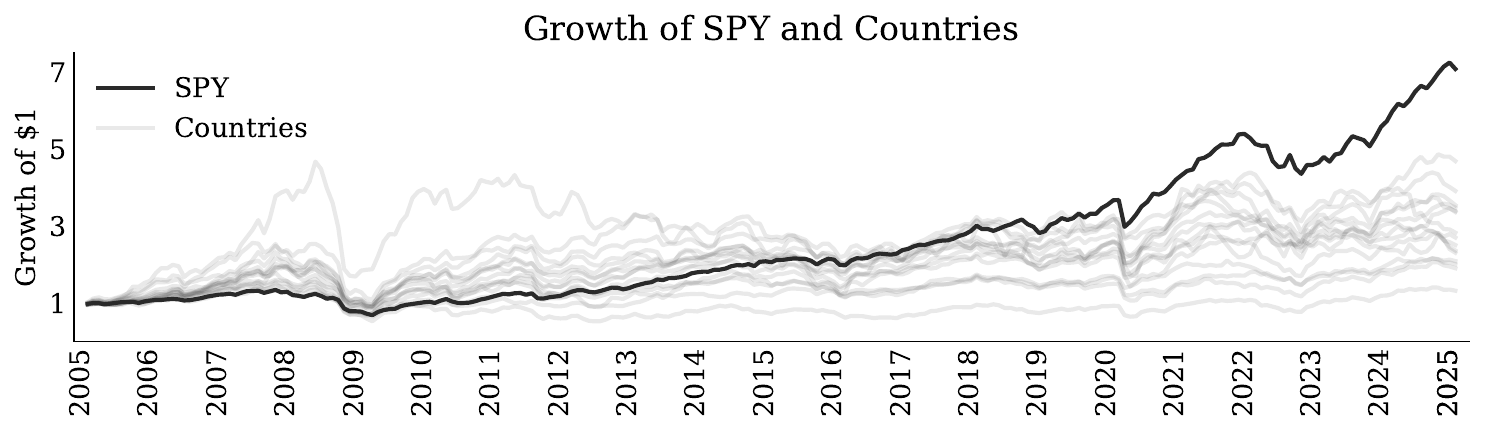}
    \caption{SPY outgrows all of the single country ETFs.}
    \label{fig:countries}
\end{figure}

The Country category consists of fifteen ETFs, with tickers beginning with EW, each tracking an MSCI index of a single country's equity market~\cite{iShares:CountryETFs}. They cover both developed markets, such as the UK and Sweden, and emerging markets, such as Brazil and South Korea. The ETFs trade in dollars while the stocks they hold trade in each country's own currency. They are unhedged, meaning this currency effect is not removed. The Sweden ETF EWD, for example, combines the performance of Swedish stocks with the move of the krona against the dollar, and loses value when the krona weakens even if the value of the underlying stocks is unchanged. The country ETFs are highly correlated with the US market and correlations with SPY range between 68\% and 84\%. Over shorter periods the country ETFs have outperformed SPY. However, over the past two decades every one of them has underperformed, with average annual returns ranging between 1\% and 8\% against SPY's 10\%. Figure~\ref{fig:countries} shows all fifteen ETFs together with SPY.

\paragraph{Fixed income.} ETFs within the Fixed income asset class consist of varying forms of bonds. A bond is a contract between the investor and an issuing institution, typically a government or a company. Investing in a bond equates to lending out money to that institution for an agreed period and receiving interest payments, known as coupons, in return. What makes bonds differ from normal loans is that the contract can be traded on the market, adding speculative value on top of the coupons and issue price. We split Fixed income into three categories depending on the type of issuer and region: Government, Corporate, and Global.

The Government category consists of five ETFs that hold bonds issued by the US government and related debt~\cite{iShares:ETFs}. Their returns are lower than those of equities, but so is their risk. Government bond correlations with SPY are among the lowest in the investable universe, some even being negative. The negative correlation can be explained by their valuation having an inverse relationship to the federal funds rate (FFR). The FFR tends to be cut during economic downturns when equities fall. A clear example of this was the 2008 Financial Crisis. As the FFR was cut, the fixed coupons of existing bonds became more attractive than the new lower rates, causing government bond prices to rise. Long-term Treasury bonds are the most sensitive to changes in the FFR. Between 2020 and 2023 TLT, which holds 20+ year Treasuries, lost close to half of its value as the FFR rose. 

The Corporate category consists of three ETFs that hold bonds issued by companies~\cite{iShares:ETFs}. A company is less certain to repay its loans than the US government, since companies can default and go bankrupt. Corporate bonds therefore pay higher interest to attract lenders. LQD lends to large, financially strong companies, while HYG lends to weaker companies that pay the highest interest. The third ETF, LQDH, holds LQD together with a hedge that removes its interest-rate risk. Corporate bonds are more correlated with the stock market than government bonds. LQD has a correlation of 19\% with SPY and HYG 68\%, the highest of any bond ETF, and HYG lost roughly a third of its value in the 2008 Financial Crisis. The average annual returns in the Corporate category range from 3\% to 5\%, slightly above the government ETFs.

The Global category consists of four ETFs that hold bonds issued outside the US~\cite{iShares:ETFs, Vanguard:ETFs}. BNDX and IAGG lend to developed-world governments and companies. Both are hedged, meaning the currency effect described for the Country category is removed. They have the lowest volatility of any bond ETF, 4\%, and near-zero correlation with SPY. EMB and VWOB instead hold bonds issued exclusively by governments in emerging markets, such as Mexico and Brazil. These governments issue some of their bonds in US dollars, and it is those that these ETFs hold, so the money is lent and repaid in dollars and the currency effect does not apply. The interest is higher because repayment is less certain. A government that has borrowed in dollars must find dollars to repay, which becomes harder when its own currency weakens. Because of this, the volatility of EMB and VWOB is higher than that of the US government bond ETFs, at 11\% and 9\% respectively, and their correlation with SPY is significantly higher.

\paragraph{Alternatives.} The Alternatives asset class consists of ETFs that hold commodities and cryptocurrencies. These are the ETFs where structure matters most. As noted in \S\ref{section:investable_universe}, some hold the asset itself while others hold financial instruments tied to its price. We therefore split Alternatives by structure into three categories: Physical, Futures, and Crypto.
 
The Physical category consists of three ETFs that each hold a precious metal in vaults: GLD holds gold, SLV silver, and PPLT platinum~\cite{SSGA:ETFs, iShares:ETFs, abrdn:PPLT}. A share corresponds to an amount of metal in the vault, so each ETF follows the spot price of its metal, less a small fee. All three are nearly uncorrelated with the rest of the universe. Among the precious metals, gold stands out. It has served as a store of value for millennia and remains a reserve asset for central banks today, and it is the only one of the three that has appreciated significantly over time. Put plainly, gold is the one physical commodity in our universe suited to buy and hold. 
 
The Futures category consists of seven ETFs covering energy, agriculture, and industrial metals~\cite{USCF:ETFs, Teucrium:ETFs}. 
\begin{figure}[!b]
    \centering
    \includegraphics[width=1.0\linewidth]{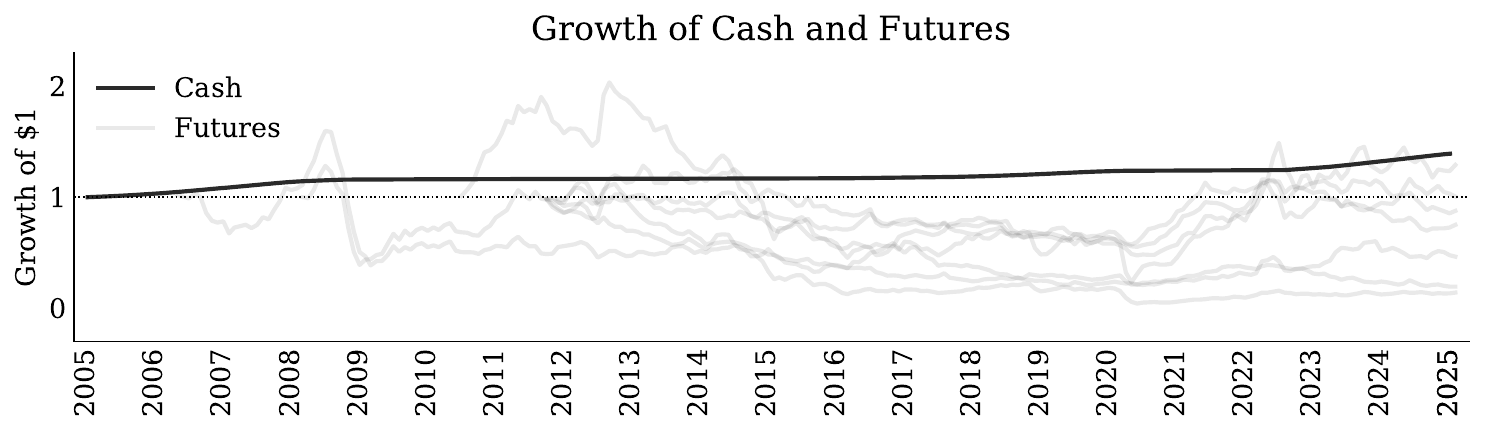}
    \caption{Holding cash outperforms all futures.}
    \label{fig:futures}
\end{figure}
As Figure~\ref{fig:futures} shows, every one of them has lost to holding plain cash earning the FFR over the backtest period, and the reason lies in how these ETFs are built. Unlike gold, these commodities are impractical to vault, so the ETFs in this category hold futures contracts. A futures contract is an agreement to buy a fixed amount of a commodity, at a fixed price, on a fixed date in the future. However, even though they could, managers of future-based ETFs never take delivery. As a contract approaches its delivery date, the fund sells it and buys a new contract with a later date, a procedure known as rolling. The later contract is usually the more expensive one, since whoever promises delivery further out must store, insure, and finance the commodity for longer and charges for it, a situation known as contango. When a futures-based ETF is in contango, the fund loses a little money on every roll, and the losses compound. Over days or weeks the ETF therefore tracks its respective commodity price closely, but over years it drifts steadily below it. These ETFs can therefore not be bought and held. They are instruments for taking commodity exposure over shorter horizons.

The Crypto category consists of two ETFs: IBIT holds bitcoin and ETHA holds ether, the two largest cryptocurrencies~\cite{iShares:ETFs}. The Crypto ETFs are the most recently listed ETFs in the investable universe. IBIT was listed in January 2024 and ETHA in July. Their annual volatilities and returns are many times those of SPY, and both have at one point lost more than 80\% of their value from an all time high. At the same time they are largely uncorrelated with the other assets in the universe. Crypto therefore requires special treatment in our experiments. Following~\cite{johansson2024crypto}, we record bitcoin and ether returns from 2017, extending the two ETFs backward with the prices of the coins they hold. This gives eight years of history rather than one. We nevertheless treat these observations with caution. Before 2024 crypto could not be reached through an ordinary brokerage account. Bitcoin has also returned over 50\% per year since 2017, meaning that large allocations can easily dominate historical portfolio performance. We therefore allow Crypto in selected experiments and portfolios under \S\ref{sec:simple_alphas}, but restrict it to modest allocations so that the results do not depend disproportionately on its exceptional historical returns.

\begin{figure}[!h]
    \centering
    \includegraphics[width=1.0\linewidth]{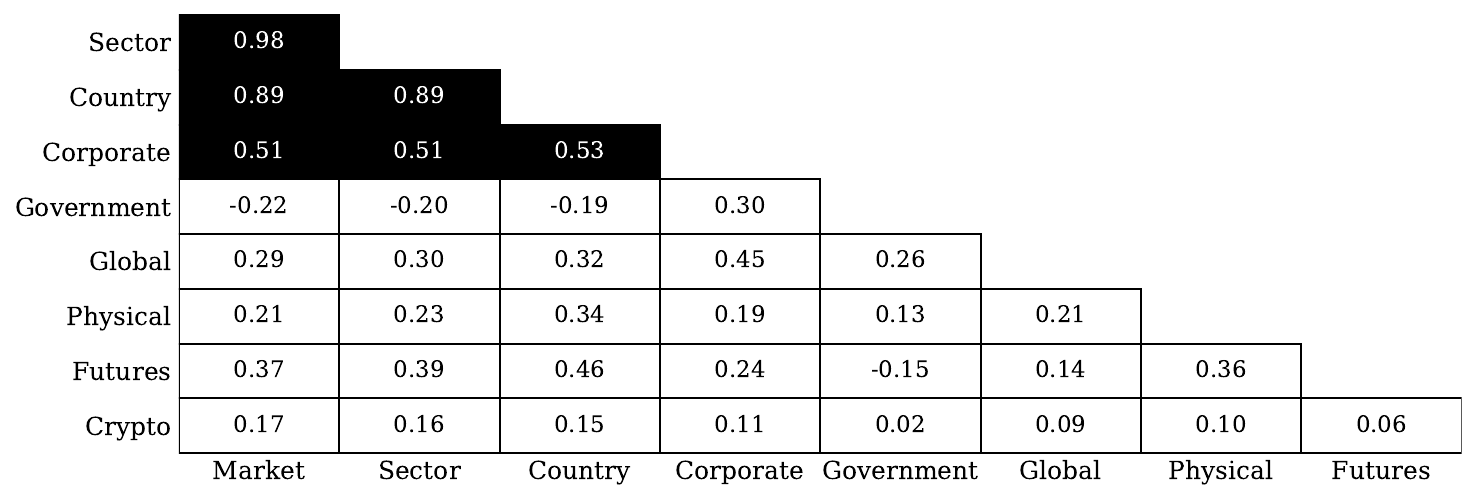}
    \caption{High correlation concentrates in the equity categories and corporate bonds.}
    \label{fig:corr_matrix}
\end{figure}

\paragraph{Correlation between categories.} Figure~\ref{fig:corr_matrix} shows the correlation between each of the nine categories. The three Equities categories together with corporate bonds show the highest correlation. Outside this block correlations are low. 
 
\section{Portfolio construction}\label{sec:portfolio_construction}
Portfolio construction concerns the problem of deciding how to allocate capital across a set of assets. This section begins by introducing the notation and terminology for portfolio weights, cash, returns, and risk, objects fundamental to portfolio construction regardless of the trading strategy. Our notation is similar to that of~\cite{Boyd:2024}. We then present a set of common trading strategies in increasing order of complexity: fixed weights, simple optimization, and our version of the Markowitz formulation.

\subsection{Mathematical model}\label{sec:mathematical_model}
\paragraph{Assets and cash weights.} We consider a universe of $n$ investable assets. A portfolio is described by a vector of asset weights $w \in \mathbf{R}^{n}$, where each weight $w_i$ denotes the fraction of capital invested in asset $i$, so $w_i = 0.10$ corresponds to allocating $10\%$ of the portfolio to asset $i$. The holdings sum to one, representing all available capital, and the fraction not invested in assets is held as cash $c \in \mathbf{R}$. Writing $\mathbf{1} \in \mathbf{R}^{n}$ for the vector with all entries one, and permitting neither short selling nor leverage, every portfolio in this paper satisfies the constraints
\[
\mathbf{1}^{\mathrm{T}}w + c = 1,
\qquad
w \ge 0,
\qquad
c \ge 0.
\]
\paragraph{Portfolio risk and return.} Each asset $i$ yields a return $r_i$ over the investment period, the fractional change in its price adjusted for dividends, splits, and other corporate actions. We collect these in the return vector $r = (r_1, \ldots, r_n) \in \mathbf{R}^{n}$, so that the return of a portfolio is the weighted sum $r^{\mathrm{T}}w$. Future returns are, however, uncertain, and the strategies of this section handle this uncertainty through two quantities, the portfolio risk and the forecasted portfolio return,
\[
w^{\mathrm{T}}\Sigma w,
\qquad
\alpha^{\mathrm{T}}w,
\]
where $\Sigma \in \mathbf{R}^{n \times n}$ is the covariance matrix of the returns and $\alpha \in \mathbf{R}^{n}$ is a forecast of the expected asset returns. The diagonal entries $\Sigma_{ii} = \operatorname{Var}(r_i)$ of the covariance matrix hold each asset's individual risk, how much the asset moves on its own, while the off-diagonal entries $\Sigma_{ij} = \operatorname{Cov}(r_i, r_j)$ hold the covariances, how each pair of assets moves together. The vector of asset volatilities $\sigma \in \mathbf{R}^{n}$, is defined as,
\[
\sigma_i = \sqrt{\Sigma_{ii}}.
\]The total portfolio risk is reduced either by holding lower risk assets or by holding assets with low or negative correlation, which is the diversification effect of \S\ref{sec:investing} in mathematical form. The risk enters nearly every strategy that follows, while the forecast $\alpha$ is used only by the Markowitz formulation of \S\ref{sec:markowitz}. The estimation of both is the subject of \S\ref{sec:risk_model} and \S\ref{sec:alpha_model}, and throughout this section they are treated as known.

\paragraph{Optimization problem formulation.} The strategies of this section differ in how they choose the weights. The simplest methods fix $w$ directly, through a fixed allocation, or decide $w$ as a function of $\sigma$. The rest obtain $w$ by solving an optimization problem, which in its most general form can be written as
\[
\begin{aligned}
    & \text{minimize}   && f_0(x) \\
    & \text{subject to} && f_i(x) \le 0, \quad i = 1,\ldots,m, \\
    &                   && g_i(x) = 0, \quad i = 1,\ldots,p,
\end{aligned}
\]
where $x$ denotes the optimization variables, $f_0$ is the objective function, and $f_i$ and $g_i$ are inequality and equality constraints. For us the variables $x$ are the portfolio weights $w$, the constraints are the portfolio constraints mentioned above among others, and the objective is built from the portfolio's risk and forecasted return. All of the problems we pose are convex, which means they can be solved reliably and in fractions of a second on a standard laptop~\cite{diamond2016cvxpy}.

\subsection{Simple trading strategies}\label{sec:simple_trading_strategies}
In this section we present five simple long only strategies, following the risk-based asset allocation framework of~\cite{Lee:2011}. Fixed weights (with and without volatility control), inverse volatility, minimum variance, risk parity, and maximum diversification. The first two do not use any optimization at all while the other three do. 

\paragraph{Fixed weights.} The simplest way of choosing your portfolio weights is to use a fixed weight strategy. The fixed weight trading strategy does not require a forecast of any kind. The weights are predetermined and trading only serves to keep the weights at their desired level. Our benchmark portfolio, the 60/40, is an example of this strategy, in which the weights of SPY and AGG are fixed to $w = (0.6, 0.4)$. A common general version of the fixed weight trading strategy is using equal weights. The equal weight strategy simply allocates capital equally across all $n$ assets as,
\begin{equation}
    w = \frac{\mathbf{1}}{n}.
\label{eq:equal_weights}
\end{equation}
It is well documented that the fixed weight strategy is hard to beat out of sample, because there is nothing in it to misestimate~\cite{DeMiguel:2009}. 

\paragraph{Volatility control.} Fixed weights control risk only implicitly, through the choice of assets but the actual risk of the portfolio is whatever the market delivers and changes over time. The 60/40 realized an average volatility of 8\% over the past two decades, but roughly 23\% during the 2008 Financial Crisis. Volatility control is a method that makes risk explicit, by treating cash as a risk-free asset and scaling down its positions in the portfolio so that the estimated volatility stays below a chosen target risk $\sigma_{\mathrm{tgt}}$. Scaling the weights $w$ by a factor $\theta$ and holding the rest as
cash scales the portfolio volatility by the same factor, so the target is met by holding $\theta w$ in assets and $c = 1 -\mathbf{1}^{T} \theta w$ in cash,
with
\begin{equation}
    \theta = \frac{\sigma_{\mathrm{tgt}}}{\sqrt{w^{\mathrm{T}}\Sigma w}},
\label{eq:vol_control}
\end{equation}
capped at $\theta = 1$ since we do not permit leverage. The portfolio is thus fully invested in calm markets and moves partly to cash in turbulent ones. This turns risk into an interpretable parameter. Investors who want a portfolio with a maximum risk of $8\%$ simply set $\sigma_{\mathrm{tgt}} = 8\%$. 

\paragraph{Inverse volatility.} The inverse-volatility portfolio sizes each position relative to its own risk, so that every asset contributes a similar amount of volatility to the portfolio~\cite{Lee:2011, LeoteDeCarvalho:2012}, 
\begin{equation}
    w = \frac{\sigma^{-1}}{\mathbf{1}^{\mathrm{T}}\sigma^{-1}}.
\label{eq:inverse_volatility}
\end{equation}
Lower risk assets therefore occupy larger holdings. Take the 60/40 as an example. SPY carries an annual volatility of $19\%$ against AGG's $5\%$, so despite the 60/40 split of capital, nearly all of the portfolio's movement comes from SPY. Sized by inverse volatility instead, the same two assets would roughly be split as $20\%$ SPY and $80\%$ AGG. The weights use only each asset's individual volatility and ignore correlations. When the assets are uncorrelated, or share the same pairwise correlation, inverse volatility coincides with the risk parity strategy later in this chapter~\cite{Maillard:2010}, but when correlations differ it misjudges the risk, since an asset that moves with the rest of the portfolio contributes more than its own volatility suggests. For this reason it is also known as naive risk parity~\cite{Shimizu:2020}. Its simplicity keeps it popular in practice, and the rule is applied across asset classes~\cite{Lee:2011, LeoteDeCarvalho:2012, Shimizu:2020}.

\paragraph{Minimum volatility.} The minimum volatility portfolio minimizes the total portfolio variance subject to being fully invested and our long-only constraints~\cite{Clarke:2006},
\begin{equation}
\begin{aligned}
    & \underset{w}{\text{minimize}} && w^{\mathrm{T}} \Sigma w \\
    & \text{subject to}             && \mathbf{1}^{\mathrm{T}} w = 1, \quad w \geq 0.
\end{aligned}
\label{eq:min_var}
\end{equation}
Minimum volatility is a cash free strategy, since cash is assumed to be risk free the strategy would simply collapse into cash only. It uses the full covariance matrix, so it can exploit assets that offset one another. However, with no notion of return, it concentrates capital in the assets with the lowest volatility. For example, in an investable universe that includes stable but low-yielding bonds, the minimum volatility portfolio invests almost exclusively in the bonds, accepting very low returns in exchange for low risk. Despite this, minimum volatility portfolios have historically delivered returns comparable to the broad market at considerably lower risk~\cite{Clarke:2006, Baker:2011}. The strategy is behind one of the ETFs in our own universe: USMV, the ETF that individually has the highest Sharpe ratio in the Market category, holding a minimum volatility portfolio of US large- and mid-cap stocks~\cite{iShares:ETFs}.

\paragraph{Risk parity.} The risk parity strategy penalizes each asset not only for its own volatility but for its covariance with the rest of the portfolio. As a result, two assets with the same volatility can receive different weights if their correlations with the rest of the portfolio differ. Unlike inverse volatility, weights are formed such that each asset holding contributes equally to the total portfolio risk~\cite{qian2005riskcontribution}. This equal risk contribution (ERC) condition can be expressed through the following convex formulation~\cite{Maillard:2010},
\begin{equation}
\begin{aligned}
    & \underset{w}{\text{minimize}} && \frac{1}{2} w^{\mathrm{T}}\Sigma w - \rho\, \mathbf{1}^{\mathrm{T}}\log(w) \\
    & \text{subject to} && \mathbf{1}^{\mathrm{T}}w + c = 1, \quad w > 0, \quad c \ge 0,
\end{aligned}
\label{eq:risk_parity}
\end{equation}
where the regularization parameter $\rho > 0$ balances the two objectives of minimizing portfolio risk and having a similar split between assets. When $\rho = 0$ it degenerates into  minimum volatility as in~(\ref{eq:min_var}), and for $\rho \gg 0$ the solution collapses into equal weights as in~(\ref{eq:equal_weights}). The logarithmic term forces every weight to stay strictly positive, causing all assets to be held to some degree throughout the entire investing period. When all asset pairs share the same covariance the ERC weights coincide with the inverse volatility portfolio of~(\ref{eq:inverse_volatility}). This is why inverse volatility has been called the naive risk parity~\cite{Maillard:2010}. Risk parity was popularized in conjunction with the All Weather portfolio of Bridgewater Associates~\cite{Bridgewater:AllWeather} and has since become a standard investment strategy among institutional investors~\cite{Roncalli:2013, Asness:2012, Chaves:2011}.

\paragraph{Maximum diversification.} The maximum diversification portfolio directly targets the reduction in portfolio risk that comes from combining assets. It does so by maximizing the diversification ratio, defined as the weighted average of the individual asset volatilities divided by the volatility of the portfolio they form. The ratio equals one for a single asset and increases as less correlated assets are combined, making it a natural measure of the diversification effect discussed in \S\ref{sec:investing}. The standard formulation of the Most Diversified Portfolio (MDP), introduced by \cite{Choueifaty:2008}, is
\begin{equation}
\begin{aligned}
& \underset{w}{\text{maximize}} && \frac{w^{\mathrm{T}} \sigma}{\sqrt{w^{\mathrm{T}} \Sigma w}} \\
& \text{subject to}             && \mathbf{1}^{\mathrm{T}} w = 1, \quad w \geq 0.
\end{aligned}
\label{eq:max_div}
\end{equation}
With the long-only constraint $w \ge 0$, the problem generally does not admit a closed-form solution. It can instead be reformulated as the convex quadratic program (QP)
\begin{equation}
\begin{aligned}
& \underset{y}{\text{minimize}} && y^{\mathrm{T}} C y \\
& \text{subject to}             && \mathbf{1}^{\mathrm{T}} y = 1, \quad y \geq 0,
\end{aligned}
\label{eq:max_div_reformulated}
\end{equation}
where $y$ are the portfolio weights expressed on a volatility-scaled basis and $C$ is the correlation matrix of returns. The full derivation is given in Appendix~\ref{app:mdp_qp_formulation}. The MDP is closely related to the minimum volatility portfolio of~(\ref{eq:min_var}) and the inverse volatility portfolio of~(\ref{eq:inverse_volatility}), and coincides with each under particular conditions~\cite{Choueifaty:2013}. When all assets have the same volatility, MDP coincides with minimum volatility. When all asset pairs have the same correlation, MDP instead coincides with inverse volatility. In the general case, it accounts for both differences in individual volatilities and differences in correlations, making it especially relevant when the correlation structure contains meaningful cross-sectional variation. Its theoretical appeal is that if all assets offer the same expected excess return per unit of standalone risk, the MDP also maximizes the portfolio Sharpe ratio~\cite{Choueifaty:2013}. It can therefore be viewed as extending the intuition of inverse volatility by incorporating the correlation structure explicitly.

\subsection{Markowitz}\label{sec:markowitz}
The simple strategies of \S\ref{sec:simple_trading_strategies} all assume no knowledge of future returns. Suppose instead you have a belief about how your assets will perform, e.g.\ that your equities will outperform your fixed income assets, or even an exact figure for each asset produced by a forecast. A natural extension of a method like minimum volatility is then to not only minimize risk but also maximize expected return. This exact trade-off was formalized by Markowitz~\cite{Markowitz:1952} in 1952. The work was recognized with the 1990 Nobel Memorial Prize in Economic Sciences~\cite{nobel1990}, and still underpins portfolio allocation today, both in academia~\cite{Kolm:2014, Kim:2021, Boyd:2024} and in industry~\cite{barra, axioma, aladdin}. At its core the idea is simple: collect the expected returns in a vector $\alpha$, and maximize the return it implies for your portfolio while weighing that against the portfolio risk. The problem can be expressed in several ways. We present two standard formulations, followed by the modified version we use throughout this paper.

\paragraph{Formulated with a soft risk constraint.} One way to express the trade-off is to subtract the portfolio risk from the expected return in the objective, scaled by a risk-aversion parameter $\gamma > 0$,
\begin{equation}
\begin{aligned}
    & \text{maximize}   && \alpha^{\mathrm{T}} w - \gamma\,w^{\mathrm{T}} \Sigma w \\
    & \text{subject to} && \mathbf{1}^{\mathrm{T}} w + c = 1, \quad c \ge 0, \quad w \ge 0.
\end{aligned}
\label{eq:markowitz_soft}
\end{equation}
The parameter $\gamma$ is a trade-off parameter between the two objective terms and is used to decide how much variance the investor is willing to accept per unit of expected return. A small $\gamma$ tolerates more risk, while a large $\gamma$ does the opposite. As $\gamma$ approaches zero the penalty disappears and the solver puts everything into the single asset with the highest expected return, and as $\gamma$ grows the penalty dominates until the portfolio only holds cash.

\paragraph{Formulated with a hard risk constraint.} The alternative is to take risk out of the objective and impose it as a constraint with a target volatility $\sigma_{\mathrm{tgt}}$,
\begin{equation}
\begin{aligned}
    & \text{maximize}   && \alpha^{\mathrm{T}} w \\
    & \text{subject to} && \mathbf{1}^{\mathrm{T}} w + c = 1, \quad c \ge 0, \quad w \ge 0, \\ 
    &                   && w^{\mathrm{T}}\Sigma w \le \sigma_{\mathrm{tgt}}^{2}.
\end{aligned}
\label{eq:markowitz_hard}
\end{equation}
Because cash can always be held, any target volatility is reachable, and the optimizer simply scales down its weights with cash until the constraint is met. The target volatility works as an upper limit, in the same way as the volatility control of~(\ref{eq:vol_control}) and realized volatility will sit close to $\sigma_{\mathrm{tgt}}$ except if the inherent risk of the portfolio is below that of $\sigma_{\mathrm{tgt}}$. One example of this would be using very large $\sigma_{\mathrm{tgt}}$ for a portfolio of low risk bond ETFs. Both formulations describe the same trade-off and can produce the same portfolios, so nothing is lost by choosing one over the other. What differs is the parameter you have to set. An investor who wants at most $8\%$ risk sets $\sigma_{\mathrm{tgt}} = 8\%$. However, the value of $\gamma$ that yields that same portfolio cannot be predetermined, and since it depends on the scale of $\alpha$ it has to be adjusted if the asset universe changes. For the sake of interpretability we therefore use the hard risk constraint formulation in our extended version below.

\paragraph{In this paper.} Throughout this paper we use a modified version of the Markowitz formulation, similar to that of~\cite{Boyd:2024}. In its original form the problem responds sharply to the return forecast, and small changes in $\alpha$ produce large changes in the weights~\cite{Michaud:1989, Best:1991, DeMiguel:2009}. This is particularly important since, as mentioned in \S\ref{sec:previous_work}, $\alpha$ is very difficult to estimate. We define our modified Markowitz formulation as,
\begin{equation}
\begin{aligned}
    & \text{maximize}   && \alpha^{\mathrm{T}}w - \kappa\,\|w - w_{\mathrm{pre}}\|_1 \\
    & \text{subject to} && \mathbf{1}^{\mathrm{T}} w + c = 1, \quad c \ge 0, \quad w \ge 0\\
    &                   && w^{\mathrm{T}} \Sigma w \le \sigma_{\mathrm{tgt}}^{2}, \quad \|w - w_{\mathrm{pre}}\|_1 \leq T_{\mathrm{tgt}}, \\
    &                   && |w - w_{\mathrm{pre}}| \le \delta_{w}, \quad 0 \leq w \le w_{\mathrm{max}}, \\
\end{aligned}
\label{eq:markowitz_full}
\end{equation}
where $w_{\mathrm{pre}}$ are the weights before trading and $\kappa$ is the trading cost, a small percentage charged on every dollar bought or sold. Since $\|w - w_{\mathrm{pre}}\|_1$ is the total amount traded, the penalty $\kappa\|w - w_{\mathrm{pre}}\|_1$ represents the total cost for a given trade. The caps $T_{\mathrm{tgt}}$ and $\delta_{w}$ limit how much of the portfolio, and how much of any single position, may change in one trade. Without them a negative entry in $\alpha$ would empty that entire position in one trade, which is a drastic response to a number we know to be noisy. Finally $w_{\mathrm{max}}$ limits the size of any single holding, which lets an investor cap exposure to an asset regardless of how attractive its forecast looks, for example holding crypto at no more than a few percent of the portfolio.

\paragraph{Regret bound.} As an upper bound on achievable performance, 
we also report a regret bound strategy that uses realized future returns 
as the alpha forecast in the Markowitz formulation~\cite{Boyd:2014}. 
This strategy is not implementable in practice but bounds the 
performance any return-forecasting model could achieve given the same 
risk model and constraints. Frequent trading while knowing future returns leads to unreasonably high Sharpe ratios. Therefore, the regret bound is only allowed to trade yearly to keep the comparison on a reasonable scale against implementable strategies.

\subsection{Portfolio selection}\label{sec:portfolio_selection}

\paragraph{Simple strategies.} The simple strategies can all be run on the full universe, but the result would not be a portfolio anyone would hold. Spreading capital equally across all 59 assets assumes every one of them is worth holding for two decades. It also lets the composition of the universe decide the allocation, since 35 of the 59 assets are equities the portfolio would hold roughly $60\%$ equities by construction, a split we never chose. We therefore run the simple strategies on small portfolios instead. Backtests are fast, so under the simplest strategies we search every combination of three, four, and five ETFs, less for a single winner than for what the top portfolios share, which asset classes they draw from, in what proportions, and which assets never appear. The simple optimization-based strategies are then examined further on portfolios found from these tests.

\paragraph{Markowitz.} Markowitz needs no such selection. It is built for large universes, and the optimizer allocates capital on its own, with the forecasted $\alpha$ deciding what it holds. The one selection left is at the category level. If the forecast cannot extract information from an entire category, it is better removed, which we evaluate through an ablation test in \S\ref{sec:forecast_quality}.

\newpage
\section{Risk model}\label{sec:risk_model}
Portfolio construction requires an estimate of future risk. While future market conditions cannot be observed directly, historical returns contain information about the volatility of individual assets and how they tend to move together, and a risk model uses this information to estimate the covariance structure that will prevail over the period for which the portfolio is constructed. This section explains what a risk model is and how the covariance is forecast from historical returns. We present two simple estimators, the rolling-window sample covariance and the exponentially weighted moving average (EWMA), followed by a more sophisticated approach based on a statistical factor model.

\subsection{Risk forecasting}\label{sec:risk_forecasting}
Risk forecasting is concerned with estimating the covariance matrix $\Sigma$, which describes both the variability of individual assets and their co-movement. Following the original formulation of \cite{Markowitz:1952}, diversification arises from the covariance structure between assets, making an accurate estimate of $\Sigma$ central to nearly every portfolio construction method. A useful way for the reader to think about the covariance matrix is simply as
\[
    \Sigma - \text{encodes future ``riskiness'' of assets.}
\]

\noindent The rolling-window and EWMA estimators are procedural approaches to covariance estimation. Both construct a risk forecast by aggregating observed historical returns according to a predetermined rule, require few assumptions, and are straightforward to implement.

The statistical factor model is particularly useful for large investment universes, where estimating every pairwise relationship directly becomes increasingly difficult. It introduces several important mathematical concepts, including statistical factors, regularization, and expectation maximization. These methods require a somewhat heavier mathematical treatment, and are deferred to Appendix~\ref{app:derivations}, where the reader can find details about the derivations. To keep the main discussion intuitive and readable, the focus here is on the structure of the factor model, why it is useful, and how its resulting covariance estimate should be interpreted.

\subsection{Simple covariance estimators}\label{sec:sample_covariance_predictors}

\begin{figure}[H]
    \centering
    \includegraphics[width=0.55\linewidth]{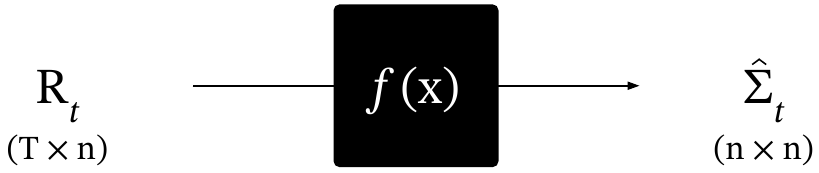}
    \caption{The covariance estimator represented as a black box operation on historical returns.}
    \label{fig:covariance_example}
\end{figure}

Although the future covariance matrix cannot be observed directly, historical asset returns provide valuable information about how assets tend to move together. A covariance estimator summarizes these historical relationships into a single matrix that serves as our best estimate of future risk. Figure~\ref{fig:covariance_example} illustrates the general covariance estimation process.
\noindent While no estimator can predict future market conditions perfectly, even simple statistical models often produce estimates that are sufficiently accurate for portfolio construction. Historical returns are collected in the return matrix $R_t \in \mathbf{R}^{T \times n}$, and a covariance estimator transforms these observations into an estimated covariance matrix $\hat{\Sigma}_t \in \mathbf{R}^{n \times n}$. We represent the estimator abstractly as a mathematical function,
\[
    \Sigma_{t+h} \approx \hat{\Sigma}_t = f(R_t),
\]
where the unknown future covariance matrix $\Sigma_{t+h}$ is approximated by the estimate $\hat{\Sigma}_t$ constructed from the historical return matrix $R_t$. Different covariance estimators correspond to different choices of the function $f(\cdot)$, each making different statistical assumptions about the observed returns. Throughout this section we will introduce several increasingly sophisticated choices of $f(\cdot)$, all of which map the same historical return matrix $R_t$ to an estimate of future risk. Here, $T$ denotes the number of historical return observations contained in $R_t$, while $n$ denotes the number of assets in the investment universe.

Regardless of the estimator, they all seek to approximate the same underlying quantity: the future covariance matrix $\Sigma_{t+h}$. Throughout this paper we assume zero-mean daily returns, $\mathbf{E}[r] = 0$, under which the covariance matrix equals the second moment,
\[
    \Sigma = \mathbf{E}[rr^{\mathrm{T}}],
\]
which must be estimated from a finite sample of historical returns. Different covariance estimators differ primarily in how they weight and summarize these observations. In this section we cover two simple but widely used estimators: the rolling-window sample covariance and the exponentially weighted moving average (EWMA)~\cite{Johansson:2023}. 

\begin{figure}[H]
    \centering
    \includegraphics[width=0.70\linewidth]{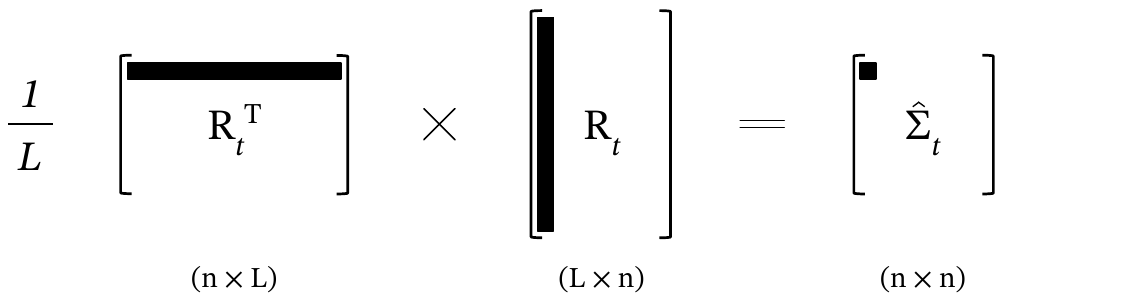}
    \caption{Illustration of rolling window covariance estimation, from (\ref{eq:covariance_example_rolling}).}
    \label{fig:covariance_example_rolling}
\end{figure}

 \paragraph{Rolling window estimate.}
The simplest choice of the covariance estimator $f(\cdot)$ is the uniformly weighted sample covariance. The underlying assumption is that the average covariance observed over the most recent L trading days provides a reasonable estimate of the covariance over the future forecasting horizon. In other words, the expected future risk is approximated by the sample average of the most recent observations, with each day receiving equal weight. Also that each observation within this lookback window is equally informative when forecasting future risk. The covariance forecast is therefore obtained by averaging the pairwise products of the historical asset returns,
\begin{equation}
    \hat{\Sigma}_t = \frac{1}{L}R_t^{\mathrm T}R_t,
    \label{eq:covariance_example_rolling}
\end{equation}
where $R_t=[r_{t-1},r_{t-2},\ldots,r_{t-L}]^{\mathrm T} \in \mathbf R^{L\times n}$ contains the trailing return observations over a lookback period of $L$ trading days. Figure~\ref{fig:covariance_example_rolling} illustrates this computation graphically. 

\noindent Each column of $R_t$ corresponds to the historical returns over $L$ days for one asset. The matrix multiplication $R_t^{\mathrm T}R_t$ computes the pairwise covariance between every asset in the universe, producing an $n\times n$ covariance matrix. The factor $1/L$ normalizes these accumulated products to adjust the covariance estimates to a daily scale.

The choice of the lookback window $L$ reflects a trade-off between estimation accuracy and responsiveness to changing market conditions. A short lookback adapts quickly to changes in volatility and correlation, but produces a noisier covariance estimate because it is based on fewer observations. Conversely, a longer lookback reduces estimation noise by averaging over more observations, but may incorporate outdated return relationships that are no longer relevant for forecasting future risk. The faster market conditions change, the shorter $L$ should generally be to avoid placing excessive weight on stale data. Consequently, there is no universally optimal choice of $L$, and its value depends on both the forecasting horizon $h$ and the persistence of the underlying covariance structure. Another limitation of the rolling-window estimator is that every observation within the lookback window receives exactly the same weight, so a return observed yesterday influences the estimate just as much as a return observed $L$ days ago, until it abruptly leaves the window altogether. 

The choice of $L$ is also constrained by the number of assets $n$. When $L \le n$, the estimated covariance matrix is guaranteed to be rank deficient, implying the existence of zero-risk portfolios that do not exist in the true market. Even when $L$ is only slightly larger than $n$, the covariance matrix may remain poorly conditioned and numerically unstable~, so in practice $L \gg n$ is generally preferred. For large investment universes, of say $n=1000$ assets, this would require multi-year lookback windows, which both increases computational cost and incorporates progressively more outdated market information.

\begin{figure}[H]
    \centering
    \includegraphics[width=0.90\linewidth]{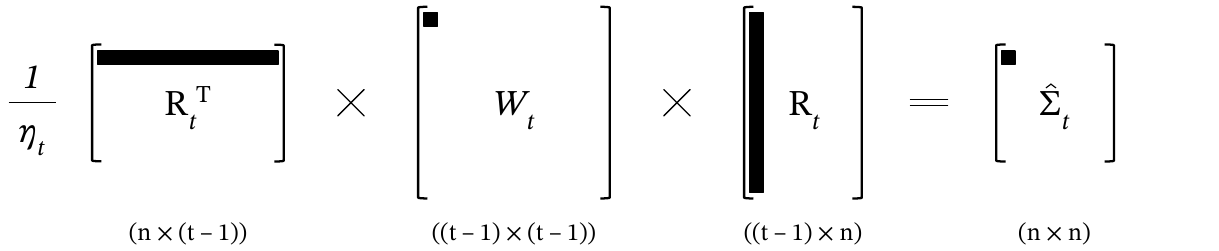}
    \caption{Illustration of EWMA covariance estimation, from (\ref{eq:ewma_covariance}).}
    \label{fig:covariance_example_ewma}
\end{figure}

\paragraph{EWMA covariance estimate.}
Another widely used choice of the covariance estimator $f(\cdot)$ is the exponentially weighted moving average (EWMA). Rather than assigning equal importance to every historical return within a fixed lookback window, the EWMA estimator assigns exponentially decreasing weights to older observations. Consequently, recent market movements have a greater influence on the covariance forecast, while older observations continue to contribute with gradually diminishing weight. The covariance forecast is computed as
\begin{equation}
    \hat{\Sigma}_t = \frac{1}{\eta_{t}}\,R_t^{\mathrm{T}} W_t R_t,
\label{eq:ewma_covariance}
\end{equation}
where $R_t=[r_{t-1},r_{t-2},\ldots,r_1]\in\mathbf{R}^{(t-1)\times n}$ contains all historical return observations available up to day $t-1$. Figure~\ref{fig:covariance_example_ewma} illustrates this computation.

\noindent Unlike the rolling-window estimator, which only considers the most recent $L$ observations, the EWMA estimator uses the entire return history while gradually reducing the influence of older observations through the diagonal weighting matrix
\[
    W_t=\operatorname{diag}(\lambda_{t-1},\lambda_{t-2},\ldots,\lambda_1).
\]
 Each diagonal entry $\lambda_i$ along the diagonal of $W$ specifies the weight assigned to one historical return observation $r_i$. Since $W_t$ is diagonal, every row of $R_t$ is weighted independently before the pairwise covariances are computed. The normalization constant $\eta_t$ is given by the sum of the EWMA weights, ensuring that the covariance estimate remains on the same scale regardless of the number of observations.

The observation weights are parameterized through a half-life $H$, measured in trading days. Rather than specifying every weight individually, the weights are generated according to the exponential decay $\lambda_{\tau}=0.5^{\tau/H}$, where $\tau$ denotes the age of the observation in days. Consequently, an observation that is $H$ days old receives half the weight of the most recent observation, while older observations receive progressively smaller weights. The half-life therefore provides an intuitive way of controlling how quickly historical information loses influence. Smaller values of $H$ place greater emphasis on recent observations, whereas larger values retain information from a longer history.

Unlike the rolling-window estimator, EWMA never discards observations abruptly. Instead, every historical return continues to contribute to the covariance estimate, with its influence gradually fading as it becomes older. Since the estimator uses the entire available return history, numerical stability is determined primarily by the total number of available observations rather than by a fixed lookback window. In practice, provided that sufficiently many observations are available, typically $t-1 \gg n$, the covariance estimate remains full rank while avoiding the abrupt discontinuities introduced by a rolling window. Reported half-life values range from a few days to several hundred days depending on the intended application~\cite{Longerstaey:1996,Pafka:2004,Bollen:2015}. Another practical advantage of EWMA is that the covariance estimate can be updated recursively as new return observations become available,
\[
    \hat{\Sigma}_t
    =
    \lambda_1\hat{\Sigma}_{t-1} + (1-\lambda_1)r_{t-1}r_{t-1}^{\mathrm T},
\]
rather than being recomputed from the entire return history each day. This makes the EWMA estimator computationally efficient while still incorporating all previously observed market information.

\subsection{Factor models}\label{sec:factor_models}

The covariance estimators introduced above estimate how every pair of assets moves together. This works well when the number of assets is small relative to the amount of historical data available. As the investment universe grows, however, the number of relationships that must be estimated grows quickly. For $n$ assets, the covariance matrix contains $n(n-1)/2$ distinct pairwise covariances. With 1,000 assets, this already amounts to almost half a million relationships, all estimated from the same finite history of returns. Many of these estimates will inevitably reflect noise rather than persistent relationships in the market.

Factor models address this problem by imposing structure on how assets are allowed to move together. Rather than estimating every relationship independently, a factor model assumes that much of the co-movement between assets can be explained by a small number of common sources of variation, called factors. For example, two equity ETFs may move together not because their relationship is unique, but because both respond to the same broad movement in the equity market. If many assets share such common movements, it can be more reliable to estimate these underlying factors than every pairwise relationship separately.

\begin{figure}[h]
    \centering
    \includegraphics[width=0.72\linewidth]{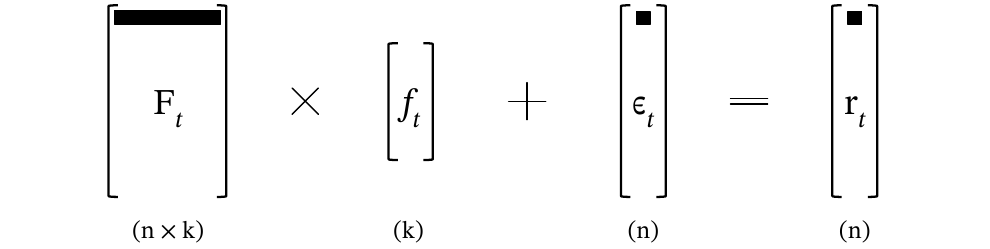}
    \caption{Illustration of the factor model in (\ref{eq:factor_model}).}
    \label{fig:factor_model}
\end{figure}

There are two common ways to define these factors~\cite{Cederberg:2025}. Economic factor models specify them explicitly, using interpretable quantities such as the market, size, and value factors of Fama and French~\cite{FamaFrench:1992,FamaFrench:1993}. Statistical factor models instead infer the common movements directly from observed asset returns. We use the latter approach, so the factors need not have an economic interpretation. However, the first factor tends to capture the market movement. The factor model we use follows the form of \cite{Cederberg:2025} and is given by,
\begin{equation}
    r_t = F_t f_t + \epsilon_t.
    \label{eq:factor_model}
\end{equation}
Figure~\ref{fig:factor_model} illustrates the factor model geometrically. The matrix $F_t\in\mathbf{R}^{n\times k}$ is called the factor loading matrix, but can simply be understood as a collection of weights: each row corresponds to one asset and describes how sensitive that asset is to each of the $k$ factor returns. Each entry of $f_t\in\mathbf{R}^{k}$ is the return of one common factor. The product $F_t f_t$ therefore combines the factor returns according to the sensitivities to produce the factor-driven return of each asset. 

\noindent This is essentially the familiar linear relation $y=mx+b$ in vectorized form, with $F_t$ playing the role of the slopes, $f_t$ the explanatory variables, and $\epsilon_t\in\mathbf{R}^{n}$ the remaining unexplained component. The number of factors $k$ is a hyperparameter, which we elaborate on in the following part.

\paragraph{Factor covariance estimate.}
Earlier we introduced a covariance estimator as a function $f(\cdot)$ that maps the historical return matrix $R_t$ to a covariance estimate $\hat{\Sigma}_t$. For the simple estimators above, this mapping could be written in a single equation, whereas the factor model requires several intermediate steps. The factor model $r=Ff+\epsilon$ in (\ref{eq:factor_model}) is an assumption about the structure of asset returns which lets us derive the decomposed form of the covariance matrix in (\ref{eq:factor_decomposed}), later used to construct the covariance estimate. The full estimation procedure involves volatility scaling, winsorization, EWMA correlation estimation, eigendecomposition, and expectation maximization (EM). We leave these mathematical details to Appendix~\ref{app:factor_estimation} and focus here on the resulting covariance estimate and its interpretation.

\begin{figure}[h]
    \centering
    \includegraphics[width=0.90\linewidth]{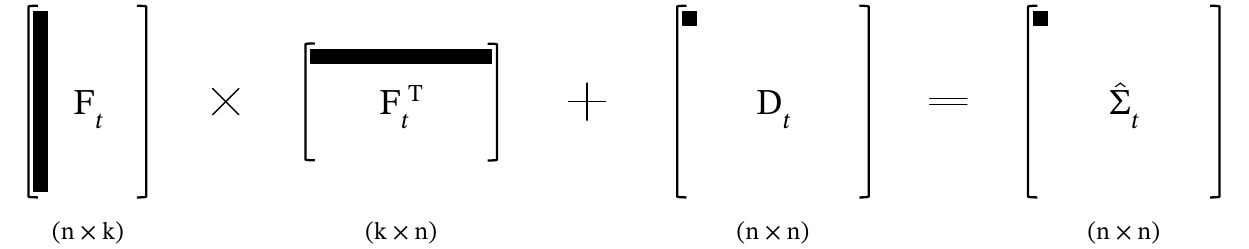}
    \caption{Illustration of the factor covariance estimate in (\ref{eq:factor_decomposed}).}
    \label{fig:covariance_example_factor}
\end{figure}

The assumption in (\ref{eq:factor_model}) has a useful consequence for the covariance matrix. The term $Ff$ captures the returns explained by the common factors, while $\epsilon$ captures the remaining asset-specific returns. If these two components are unrelated, their contributions to risk can also be separated. We therefore assume that the asset-specific returns are uncorrelated with the common factors and with each other, so that any co-movement between assets is only explained through the common factors. We also normalize the factors to have unit covariance. Under these assumptions, the covariance estimate takes the decomposed form
\begin{equation}
    \hat{\Sigma}_t = F_tF_t^{\mathrm{T}} + D_t,
    \label{eq:factor_decomposed}
\end{equation}
where $F_tF_t^{\mathrm{T}}$ captures the covariance explained by the common factors and $D_t\in\mathbf{R}^{n\times n}$ is a diagonal matrix containing the remaining asset-specific variances. Figure~\ref{fig:covariance_example_factor} illustrates how these two components combine to form the covariance estimate.

\noindent The important distinction from the simple covariance estimators is that the off-diagonal relationships are no longer estimated independently. Instead, they must be explained through the $k$ common factors. When $k\ll n$, this substantially reduces the flexibility of the model: rather than estimating $\mathcal{O}(n^2)$ covariance relationships, the common covariance structure is described by the $\mathcal{O}(nk)$ entries of $F$. This acts as a form of regularization. Relationships supported by common movements across many assets are retained, while isolated relationships caused primarily by sampling noise are less likely to enter the covariance estimate. The resulting estimate can therefore be more robust when the number of assets is large relative to the available data~\cite{ledoit2004}. For large asset universes, the resulting low-rank structure can also substantially reduce computational cost.

The factor model equation in (\ref{eq:factor_model}) and the covariance decomposition in (\ref{eq:factor_decomposed}) provide complementary views of the same underlying structure:
\[
    \underset{r}{\text{total asset returns}}
    \longleftarrow
    \underset{Ff}{\text{common factor-driven part}} + \underset{\epsilon}{\text{asset-specific part}},
\]
\[
    \underset{\hat{\Sigma}}{\text{total asset risk}}
    \longleftarrow
    \underset{FF^{\mathrm{T}}}{\text{common factor-driven part}} + \underset{D}{\text{asset-specific part}}.
\]
The factor model equation separates asset returns into movements shared across the investment universe and movements specific to individual assets. The covariance decomposition carries this same structure over to risk: $FF^{\mathrm{T}}$ captures the risk arising from shared movements, while $D$ captures the remaining risk specific to each asset. Consequently, co-movement between different assets is explained through the common factors. The factor model therefore reduces a large covariance estimation problem to estimating a small set of shared movements, how strongly each asset responds to them, and whatever risk remains is attributed as  specific to each asset.

\newpage
\section{Alpha model}\label{sec:alpha_model}

Markowitz portfolio optimization requires, outside of the risk estimate, also some estimate of the future attractiveness of investments. An alpha model attempts to summarize the information contained in historical data into such a forecast. This section explains what an alpha model is and how a return forecast is constructed from historical prices. We present two simple signals, momentum and a moving average, followed by a fitted model that uses ridge regression.

\subsection{Return forecasting}\label{sec:return_forecasting}
\begin{figure}[h]
    \centering
    \includegraphics[width=1\linewidth]{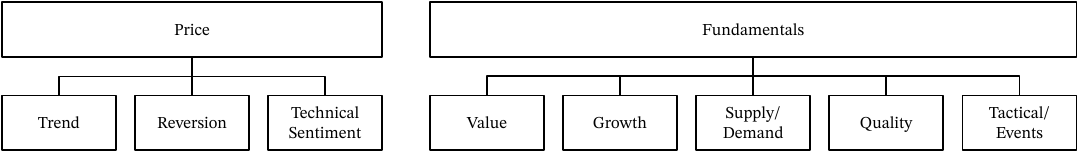}
    \caption{Taxonomy of alpha models, adapted from~\cite[p.~29]{Narang:2024}.}
    \label{fig:rishi_alpha}
\end{figure}

Return forecasting is concerned with estimating $\alpha$, meant to capture the performance of investments over the coming investing period. Unlike risk, returns generally exhibit much weaker statistical structure, making useful alpha estimation considerably more difficult~\cite{Merton:1980,GoyalWelch:2008}. Following the terminology of Grinold and Kahn~\cite{grinold2000}, alpha is often introduced as a forecast of expected excess return. Throughout this paper, however, it is helpful to adopt a slightly broader interpretation as,
\[
    \alpha - \text{encodes future ``goodness'' of assets.}
\]
The attractiveness of an investment can be captured through many different signals, as shown in Figure~\ref{fig:rishi_alpha}, of which we focus on trend and reversion in this paper.

\noindent Momentum and the moving-average signal are procedural alphas constructed directly from historical prices. Both apply predetermined rules to recently observed price movements and carry the resulting information forward as an indication of future attractiveness. They require no fitted statistical model, are straightforward to implement, and provide an intuitive starting point.

Statistical return forecasting instead estimates the relationship between information available today and subsequently realized returns from historical data. This provides a considerably more flexible framework: rather than specifying beforehand how a particular signal should translate into alpha, a statistical model attempts to learn that relationship from the data. However, any forecasting model is ultimately constrained by the data it has access to. Institutional investors may combine sophisticated statistical models with proprietary datasets, alternative information, or other signals unavailable to the average market participant. Our objective is different. We restrict the forecasting problem to public, historically observable market data so that the resulting models remain reproducible and accessible. The regression models introduced later should therefore only be viewed as increasingly flexible methods for extracting whatever predictive information is present in the available features.

\subsection{Simple alphas}\label{sec:simple_alphas}
There are countless ways to construct an alpha signal from historical market data. Prices and returns can be aggregated and transformed in many ways, producing a jungle of possible signals. The purpose of this section is not to exhaustively search for the best-performing alpha. Instead, we consider two simple and interpretable textbook alphas to investigate whether useful information can be extracted from recent prices: momentum, which retains the magnitude of recent returns, and a moving-average signal, which reduces the forecast to the direction of the recent price trend.

\paragraph{Momentum.} The simplest alpha we consider is momentum, where the return over a recent period is carried forward directly as the forecast. An asset that has recently appreciated receives a positive alpha, while an asset that has declined receives a negative one. The minimal underlying assumption is that price movements exhibit some persistence, such that assets performing well relative to their recent history may continue to do so over the near future. We use one-month momentum, computed from the log return over the preceding $L$ trading days,
\begin{equation}
\mathrm{mom}_t = \log\!\left(\frac{p_{t-1}}{p_{t-L}} \right) = \sum_{\tau=t-L}^{t-1} \log(r_{\tau}),
\end{equation}
where $p_{t-1}$ is the most recently observed price and $p_{t-L}$ is the price $L$ trading days earlier. Their log ratio is the continuously compounded return over the lookback period, so the signal preserves both direction and magnitude. Momentum is inherently backward-looking and reacts slowly to changes, progressively tilting the portfolio toward or away from an asset rather than attempting to anticipate the turning point itself. The lookback $L$ controls the trade-off between reactivity and stability. Short windows react quicker to changing trends but are more exposed to short-term noise, while longer windows produce smoother signals but respond more slowly when conditions change.

\paragraph{Moving average.} The moving-average signal instead reduces the alpha to a directional forecast. Rather than extrapolating the magnitude of recent returns, the signal asks only whether the current price lies sufficiently above or below its recent trend. Letting $L$ denote the lookback window, we define
\begin{equation}
\mathrm{mov}_t = 2\times\mathbf{1}\!\left( p_{t} > \frac{\rho}{L}\sum_{\tau=t-L}^{t-1}p_{\tau} \right) - 1,
\end{equation}
where the trailing average provides a slowly moving reference level for the current price and $\rho$ controls the threshold required to classify the trend as positive. For example, with $L=50$ and $\rho=1.2$ the alpha only gives a buy signal when the price is 20\% above the 50-day moving average. The signal takes only two values, $+1$ when the price is sufficiently above its trailing average and $-1$ otherwise, so unlike momentum, the distance from the average carries no additional information. The signal therefore acts as a simple gate, participating while the price remains in a favorable trend and moving away as soon as that trend disappears. Since the current price is compared directly with the slower-moving reference, the signal reacts more quickly than momentum.

\subsection{Regression models}\label{sec:regression_models}
Returning to the textbook definition of alpha as an expected future return, we can construct a set of features available at day $t$ and estimate how these features have historically related to returns over the subsequent horizon of $h$ days. The fitted relationship can then be applied to the information available today to produce a return forecast $\hat{r}_{t+h}$, which is used directly, or after some transformation, as the alpha supplied to Markowitz. This turns alpha forecasting into a conventional supervised learning problem, but an unusually difficult one. 

\begin{figure}[h]
    \centering
    \includegraphics[width=1.0\linewidth]{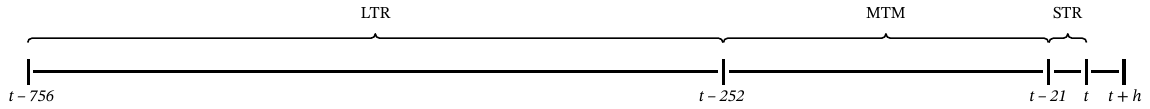}
    \caption{The three features cover non-overlapping windows of past returns.}
    \label{fig:regression_features}
\end{figure}

\paragraph{Features and dataset.} Any return forecasting model is ultimately limited by the information made available to it. We restrict this information to public and historically observable market data, keeping the forecasts reproducible and consistent with the simple investment framework considered throughout this paper. Letting $p_{t}$ denote the price of asset $i$ at day $t$,
\begin{equation}
    x^{(S, L)}_{t} = \log \!\left(\frac{p_{t-S}}{p_{t-L}} \right),
\end{equation}
is a feature analogous to the simple momentum alpha mentioned in \S\ref{sec:simple_alphas} with the two lookback parameters $S$ and $L$. Here, $x^{(S, L)}_{i,t}$ is a data feature which covers the period between $t-L$ to $t-S$, thus a window of $L-S$ days. With the addition of the parameter $S$ we are able to create a dataset of features covering non-overlapping periods of historical returns. The three very simple and classical textbook features we use in our experiments are,
\begin{equation}
    \mathrm{STR}_{t} = -\log \!\left(\frac{p_{t-1}}{p_{t-21}} \right),
    \qquad
    \mathrm{MTM}_{t} = \log \!\left(\frac{p_{t-21}}{p_{t-252}} \right),
    \qquad
    \mathrm{LTR}_{t} = -\log \!\left(\frac{p_{t-252}}{p_{t-756}} \right),
\end{equation}
where STR is the short-term reversal feature, MTM is the medium-term momentum feature, and LTR is the long-term reversal features, also shown in Figure~\ref{fig:regression_features}. 

\noindent The reversal effect is captured by the minus signs, and is based on the assumption that the future price will move in the opposite direction of the short- and long-term trends. For each asset $i$ in the asset universe we form a dataset $X \in \mathbf{R}^{T \times 3}$ consisting of the three aforementioned features, specific to that asset. That dataset is then used in a rolling window fashion to backtest the regression model while producing an alpha over the period 2005--2025.

\begin{figure}[h]
    \centering
    \includegraphics[width=1.0\linewidth]{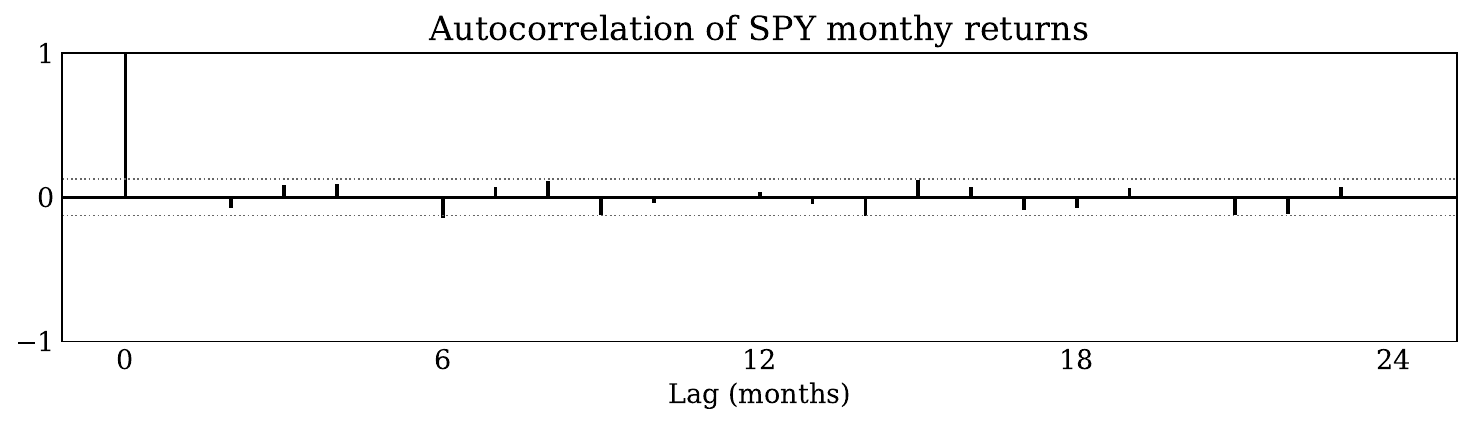}
    \caption{SPY shows little to no autocorrelation due to the stochastic nature of returns.}
    \label{fig:spy_acf}
\end{figure}

\paragraph{Return predictability.} Predicting returns is difficult not because statistical models cannot fit historical financial data, but because there is very little persistent signal to fit. A useful way to conceptualize a future return is as the sum of a component that can, in principle, be inferred from information available today and a component that cannot,
\begin{equation}
r_{t+h} = f(X_t) + \varepsilon_{t+h},
\end{equation}
where $f(X_t)$ represents the predictable component contained in the feature vector $X_t$, while $\varepsilon_{t+h}$ collects the remaining unpredictable variation. For financial returns, the central difficulty is that the latter is generally much larger than the former,
\begin{equation}
\operatorname{Var}\left[f(X_t)\right] \ll \operatorname{Var}\left[\varepsilon_{t+h}\right].
\end{equation}
The signal-to-noise ratio is consequently very low. Even a model that estimates $f(X_t)$ reasonably well can therefore produce forecasts that appear weak when compared with individual realized returns, since each realization remains dominated by $\varepsilon_{t+h}$. This weak dependence is already visible in the return series themselves. If past returns contained strong information about future returns, substantial dependence would remain across nearby lags. For broad and liquid assets such as SPY, however, return autocorrelations tend to be small, leaving comparatively little structure for a forecasting model to exploit. Figure~\ref{fig:spy_acf} illustrates this through the autocorrelation function (ACF) of SPY returns across successive lags.

\noindent Low signal-to-noise also changes how model complexity should be viewed. A flexible model has more freedom to discover relationships in the training data, but when most historical variation is noise, it also has more freedom to explain patterns that occurred by chance. A good in-sample fit can therefore be obtained without learning a relationship that persists out of sample. This is the familiar bias--variance trade-off, but it becomes particularly important for return forecasting. Reducing model bias by adding flexibility can quickly increase estimation variance as the model adapts to noise. Model complexity is therefore not automatically an advantage. The objective is instead to use enough flexibility to capture the weak systematic component of returns while restricting the model sufficiently that random historical variation is not mistaken for persistent structure.

\paragraph{Ridge regression.}
We use a linear model to estimate the relationship between the features and subsequent returns. For a feature vector $x_t \in \mathbf{R}^{F}$, holding the $F$ features observed at day $t$, the return over the following $h$ days is modeled as
\begin{equation}
    \hat{r}_{t+h} = x_t^{\mathrm{T}}\theta,
\label{eq:ridge_model}
\end{equation}
where $\theta \in \mathbf{R}^{F}$ contains one coefficient for each feature. Each coefficient describes the direction and magnitude of the estimated relationship between its feature and the return that follows. This makes the model transparent, since the contribution of every feature to a forecast can be read directly from its coefficient.

Every asset is given its own model, fitted on its own dataset. A row of the dataset pairs the features observed at day $t$ with the return realized over the subsequent $h$ days,
\begin{equation}
    x_t = \left(\mathrm{STR}_t,\, \mathrm{MTM}_t,\, \mathrm{LTR}_t\right),
    \qquad
    y_t = r_{t+h},
\label{eq:ridge_dataset}
\end{equation}
so in our case $F = 3$. Stacking the rows over time gives the arrays $X \in \mathbf{R}^{T \times F}$ and $y \in \mathbf{R}^{T}$, one pair per asset.

At each forecasting date $t$ the coefficients are re-estimated from the $H$ most recent pairs whose outcome is known. A pair with feature date $s$ is complete only once its target $r_{s+h}$ has been realized before the forecast is made, so the newest usable feature date is $t-h-1$, and the split becomes
\begin{equation}
    X_{\mathrm{trn}} = X_{[t-h-H \,:\, t-h-1]},
    \qquad
    y_{\mathrm{trn}} = y_{[t-h-H \,:\, t-h-1]},
    \qquad
    x_{\mathrm{tst}} = x_t,
    \qquad
    y_{\mathrm{tst}} = y_t,
\label{eq:ridge_split}
\end{equation}
with $X_{\mathrm{trn}} \in \mathbf{R}^{H \times F}$ and $y_{\mathrm{trn}} \in \mathbf{R}^{H}$. The gap of $h$ days between the last training pair and today is what keeps future information out of the fit. The features observed today, $x_{\mathrm{tst}}$, are held out for the forecast, and their outcome $y_{\mathrm{tst}}$ is realized only $h$ days later, entering the evaluation of \S\ref{sec:forecast_quality} but never the training sample.

The simplest estimate of $\theta$ is ordinary least squares (OLS), which minimizes the squared prediction error over the training set. The simplicity is attractive but does not prevent overfitting. OLS is free to assign large coefficients to relationships that arose largely from noise, the problem grows as the number of parameters rises relative to the amount of training data, and correlated features make the estimates sensitive to small changes in the sample. We therefore use ridge regression, which adds an $L_2$ penalty on the magnitude of the coefficients,
\begin{equation}
    \hat{\theta}
    =
    \arg\min_{\theta}\, \|y_{\mathrm{trn}} - X_{\mathrm{trn}}\theta\|_2^2 + \lambda\,\|\theta\|_2^2
    =
    \left(X_{\mathrm{trn}}^{\mathrm{T}}X_{\mathrm{trn}} + \lambda I_F\right)^{-1} X_{\mathrm{trn}}^{\mathrm{T}}\,y_{\mathrm{trn}},
\label{eq:ridge_closed_form}
\end{equation}
where $\lambda \geq 0$ sets the strength of the regularization and $I_F$ is the identity matrix. The first equality defines the estimator and the second is its closed form solution. At $\lambda = 0$ the penalty disappears and the solution reduces to OLS. As $\lambda$ grows, large coefficients become progressively more costly, and a feature must show stronger evidence in the training data before it earns influence over the forecast. The shrinkage deliberately trades bias for lower estimation variance. A relationship that appears predictive only by chance would otherwise receive a large coefficient and swing as the training window moves forward, and ridge presses such coefficients toward zero. In the low signal-to-noise setting described above, a forecast near zero is not necessarily a failure. It can simply reflect that the data offers too little evidence for a stronger prediction.

The fitted coefficients applied to the held out features give the forecast, $\hat{r}_{t+h} = x_{\mathrm{tst}}^{\mathrm{T}}\hat{\theta}$, which directly or after a transformation is the alpha supplied to Markowitz. The closed form matters in practice. The model is re-estimated at every trading date, separately for every asset and specification, and avoiding iterative optimization keeps the cost of these repeated fits low. More flexible nonlinear models could capture relationships a linear model cannot, but they bring additional computational cost, more modeling choices, and greater capacity to fit the noise that already dominates the data. For our purposes ridge regression is the compromise between regularization, interpretability, and computational efficiency.

\newpage
\section{Numerical experiments}\label{sec:numerical_experiments}
This section presents the numerical experiments that we have run to test the different trading strategies discussed above. We start by defining how our backtest is setup and runs. In \S\ref{sec:simple_strategies_experiments} we give intuition behind the choice of ETFs for the simple trading strategies defined in \S\ref{sec:simple_trading_strategies}. In \S\ref{sec:forecast_quality} we evaluate the risk models presented in \S\ref{sec:risk_model} and selection of alpha in \S\ref{sec:alpha_model}. We end by showing the performance of a smaller set of reasonable portfolios and strategies derived from the conclusions of the previous sections. 

\subsection{Data and backtest}\label{sec:data_and_backtest}

\paragraph{Data.} Daily adjusted close prices for all assets are sourced from Yahoo Finance~\cite{yahoofinance} via the yfinance Python package~\cite{yfinance}. Macroeconomic time series, such as the federal funds rate, are retrieved from the Federal Reserve Economic Data (FRED) database~\cite{fred} accessed via the fredapi package~\cite{fredapi}.

\paragraph{Backtest.} We evaluate our portfolio construction methods through a daily backtest covering 2005-01-01 to 2025-01-01, a total of $T = 5{,}032$ trading days $t = 1, 2, \ldots, T$. Every trade in it is decided by one of the strategies of \S\ref{sec:portfolio_construction}, and we consider the universe of $n$ assets of \S\ref{section:investable_universe} plus cash. Let $h_t \in \mathbf{R}^n$ denote the dollar holdings in each asset and $c_t \in \mathbf{R}$ the cash balance at the start of day $t$. We allow no shorting and no leverage, so $h_t \geq 0$ and $c_t \geq 0$. The net asset value (NAV) and the portfolio weights are
\[
    v_t = \mathbf{1}^{\mathrm{T}} h_t + c_t,
    \qquad
    w_t = h_t / v_t,
\]
with the remaining fraction $1 - \mathbf{1}^{\mathrm{T}} w_t$ held as cash. Each day the holdings drift with the realized return vector $r_t \in \mathbf{R}^n$ and cash earns the federal funds rate $r_{f,t}$,
\begin{equation}
    h_{t+1} = \mathrm{diag}(1 + r_t)\, h_t,
    \qquad
    c_{t+1} = (1 + r_{f,t})\, c_t.
    \label{eq:backtest_drift}
\end{equation}
On a day without trading this is the entire update. On a trading date, trades are assumed to execute at the start of day $t$ at the
previous day's closing price, after which holdings drift. Writing $w_{pre}$ for
the weights carried into the trade, the turnover on day $t$ is
\begin{equation}
    \tau_t = \tfrac{1}{2} \| w_t - w_{pre} \|_1 ,
\end{equation}
so that moving capital out of one asset and into another counts once rather than
twice. Every dollar bought or sold pays half of a bid-ask spread of
$\kappa = 5$ basis points, and the cost $\kappa \, \tau_t \, v_t$ is deducted
from cash. The Python code for the backtest loop and each of trading strategies it uses to update the portfolio weights can be found in Appendix~\ref{code:backtest_loop} and~\ref{code:strategies}.

Not every ETF in the universe of \S\ref{section:investable_universe} was listed on the first day of the backtest, and the exact listing dates are given in Table~\ref{tab:asset_universe}. The backtest is simply not allowed to buy an asset before it was listed, so the number of assets a strategy can choose from grows over the period. We also require assets to have a minimum amount of 126 trading days before it becomes available, since the risk model and the return forecasts both need a trailing window of observations.

\paragraph{Evaluation.} We evaluate strategies using six performance metrics. These are the annualized geometric excess return $r_e$, the annualized realized volatility $\sigma$, the maximum drawdown (MDD), the Sharpe ratio (SR), the average annual turnover (TO), and the lower quartile (LQ) of the per-year Sharpe ratio. Let $r_{p,t} \in \mathbf{R}$ denote the realized portfolio return on day $t$. We define $r_e$ and $\sigma$ as,
\begin{equation}
    r_e = \left( \prod_{t=1}^{T} \frac{1 + r_{p,t}}{1 + r_{f,t}} \right)^{\!252/T} - 1,
    \qquad
    \sigma = \left( \frac{252}{T} \sum_{t=1}^{T} (r_{p,t} - \bar{r}_p)^2 \right)^{1/2},
    \label{eq:annualized_return_and_vol}
\end{equation}
where $\bar{r}_p = (1/T)\sum_{t=1}^{T} r_{p,t}$ is the sample mean and the factor $252$ annualizes the per-period estimate under the assumption that daily returns are uncorrelated across time. The return is measured net of the risk-free rate while the volatility is measured on the portfolio return itself. We use the geometric mean because it equals the realized compound return, whereas the arithmetic mean overstates compound performance whenever returns vary~\cite{Fernholz:2002}. We write MDD, SR, and LQ as,
\begin{equation}
    \mathrm{MDD} = \max_{1 \leq t \leq T} D_t,
    \qquad \mathrm{SR} = \frac{r_e}{\sigma},
    \qquad \mathrm{LQ} = Q_{0.25}\big(\{\mathrm{SR}_y\}_y\big),
\end{equation}
where $D_t = \max_{1 \leq \tau \leq t} (V_\tau - V_t)/V_\tau$ is the running drawdown at time $t$, $V_t$ is the portfolio NAV at time $t$, MDD is the worst such drawdown over the backtest, and $\mathrm{SR}_y$ is the annualized Sharpe ratio within calendar year $y$. LQ is the empirical $25$th percentile of the per-year Sharpe ratios and is our measure of weak-year consistency. Turnover is reported as the average over calendar years of $\sum_{t \in y} \tau_t$, so a TO of $0.20$ means that a fifth of the portfolio changed in a typical year.

\paragraph{Parameters.} The portfolio construction methods of \S\ref{sec:portfolio_construction} and the risk and alpha models of \S\ref{sec:risk_model} and \S\ref{sec:alpha_model} introduce a number of free parameters. We fix these once and use the same values throughout, unless stated otherwise. None are finely tuned. In each case we take a round number within the range reported in the literature, rather than optimizing for any single metric. Table~\ref{tab:parameters} lists the parameters central to the backtest and the trading strategies, while the remaining model parameters are given in~\S\ref{sec:forecast_quality}.

\begin{table}[H]
    \footnotesize
    \centering
    \caption{Parameters central to the backtest and the trading strategies.}
    \begin{tabular}{ll|l|l}
        \toprule
        \textbf{Model} & \textbf{Parameter} & \textbf{Description} & \textbf{Value} \\
        \midrule
        Backtest   & $r_{f}$                 & Return on cash                      & FFR                      \\
                   & $T$                     & Trading days                        & 5{,}032                  \\
                   & $W_{\mathrm{backtest}}$ & Backtest window                     & 2005-01-01 to 2025-01-01 \\
                   & $H_{\mathrm{min}}$      & Minimum history before an asset is investable & 126 days       \\
        \midrule
        Strategies & $\sigma_{\mathrm{tgt}}$ & Target annual volatility            & 8\%                      \\
                   & $\rho$                  & Regularization in risk parity       & $1 \times 10^{-5}$       \\
                   & $t_{\mathrm{freq}}$     & Trading frequency                   & Monthly                  \\
        \midrule
        Markowitz  & $T_{\mathrm{tgt}}$      & Turnover limit per rebalance        & 10\%                     \\
                   & $\delta_{w}$            & Position change limit per rebalance & 5\%                      \\
                   & $w_{\mathrm{max}}$      & Maximum weight in a single asset    & 10\%                     \\
                   & $\kappa$                & Trading cost                        & $5 \times 10^{-4}$       \\
        \bottomrule
    \end{tabular}
    \label{tab:parameters}
\end{table}

\newpage

\subsection{Simple strategies}\label{sec:simple_strategies_experiments}
The simple strategies of \S\ref{sec:simple_trading_strategies} each need a smaller portfolio to run on. This section examines the candidate portfolios for each strategy. The strategies are ordered by complexity, and each addresses a specific shortcoming of the strategy before it. For each strategy we report what it addresses, our findings on portfolio composition, and which assets are reasonable to include.
 
\paragraph{Fixed weights.} 
We begin with equal weights. We backtest every combination of three, four, and five ETFs, rebalancing monthly, and rank the resulting portfolios twice: first by SR alone and then by SR followed by LQ. We exclude two groups of assets from the search. First, the Country and Futures categories, since they are not suitable buy-and-hold assets, as established in \S\ref{section:asset_categories}, and the fixed-weight strategy can do little beyond buying, holding, and rebalancing. Second, we exclude any ETF that was not listed at the start of the backtest. An investor in 2005 could not have selected an asset that did not yet exist, and introducing new ETFs as they became available would change the portfolio composition during the experiment. These restrictions leave 23 ETFs. Rather than testing every possible allocation for every combination, we initially restrict the search to equal weights. This separates two questions that would otherwise be mixed together: which assets should a fixed-weight portfolio contain, and how should capital be divided among them? Here we focus on the first question and return to the choice of proportions afterwards. Table~\ref{tab:exhaustive_search_equal_weights} in the appendix reports the top 60 portfolios under both ranking criteria for each portfolio size, while Table~\ref{tab:top_by_criterion} presents the best portfolios under each criterion.

\begin{table}[h]
    \centering
    \footnotesize
    \caption{Top portfolios of 3, 4, and 5 ETFs, ranked by SR and by SR$+$LQ.}
    \begin{tabular}{ll|SSSSSS}
        \toprule
        \textbf{Criteria} & \textbf{ETFs} & {$\boldsymbol{r}$} & {$\boldsymbol{\sigma}$} & \textbf{MDD} & \textbf{TO} & \textbf{SR} & \textbf{LQ} \\
        \midrule
        SR    & QQQ/IEF/GLD         & 0.10 & 0.09 & 0.22 & 0.18 & 0.81 & 0.44 \\
              & QQQ/XLP/TLT/GLD     & 0.10 & 0.09 & 0.22 & 0.19 & 0.84 & 0.62 \\
              & QQQ/XLP/IEF/TIP/GLD & 0.08 & 0.08 & 0.19 & 0.16 & 0.84 & 0.50 \\
        \midrule
        SR+LQ & XLK/XLV/GLD         & 0.12 & 0.13 & 0.33 & 0.18 & 0.74 & 0.73 \\
              & QQQ/XLU/AGG/GLD     & 0.09 & 0.10 & 0.27 & 0.18 & 0.74 & 0.80 \\
              & SPY/QQQ/XLP/TLT/GLD & 0.10 & 0.10 & 0.26 & 0.18 & 0.77 & 0.81 \\
        \bottomrule
    \end{tabular}
    \label{tab:top_by_criterion}
\end{table}
\noindent The two ranking criteria favor somewhat different portfolios. When ranked by SR, the strongest portfolios combine an aggressive equity position, typically QQQ, with intermediate- or long-term government bonds such as IEF or TLT and an allocation to gold through GLD. These portfolios reach Sharpe ratios around $0.80$--$0.84$ at roughly $8\%$--$9\%$ volatility, with maximum drawdowns around $19\%$--$22\%$. Ranking by SR+LQ instead shifts the portfolios toward equities. For example, the highest-ranked three-ETF portfolio replaces the bond allocation with XLV, combining the growth-oriented XLK with a less volatile, defensive equity sector and gold. This shift comes at the cost of higher overall volatility and larger maximum drawdowns, but substantially improves LQ: compared with the top three-ETF portfolio ranked by SR, volatility rises from $9\%$ to $13\%$ and MDD from $22\%$ to $33\%$, while LQ increases from $0.44$ to $0.73$. The difference reflects what the two consistency measures capture. LQ is based on the lower quartile of annual Sharpe ratios and therefore rewards portfolios that perform consistently across years, while a single extreme loss has limited influence on the measure. MDD behaves in the opposite way: it is determined entirely by the most severe peak-to-trough loss. Since the backtest contains three major market crashes, the two measures can therefore favor different portfolio compositions. Government bonds such as IEF and TLT provide protection during severe equity downturns and improve MDD, whereas replacing some of this protection with additional equity exposure, such as XLV or XLU, can improve performance across a larger share of individual years and therefore increase LQ.

\paragraph{Common characteristics.}
Despite these differences, the strongest portfolios share a clear structure and generally draw from all three asset classes. Gold appears particularly consistently: GLD, the only Alternatives ETF with sufficient history to enter the search, is present in every one of the top fifteen portfolios at each portfolio size and in most of the remaining top 60. Fixed-income allocations are dominated by government bonds. Although the corporate bond ETF LQD occasionally appears, government bond alternatives rank higher for the same equity-and-gold combinations. The equity allocation, meanwhile, is typically concentrated in growth-oriented assets, with larger portfolios often adding either a defensive equity sector or a second government bond position. Adding more ETFs therefore provides only limited improvement. Moving from three to four or five assets increases the best Sharpe ratio only slightly, from $0.81$ for three ETFs to $0.84$ for both four and five ETFs, and the additional holdings are usually variations of exposures already present in the portfolio. The main result is consequently not a particular winning combination of ETFs, but a recurring allocation across distinct sources of risk: growth-oriented equities provide return, government bonds provide protection during equity downturns, and gold adds a third source of diversification.
 
\paragraph{Choosing the split.} 
The equal-weight search identifies which assets work well together, but leaves open whether their exact proportions matter. This is particularly relevant for fixed-weight portfolios such as the 60/40, where the allocation itself is part of the strategy. To isolate this effect, we select SPY, AGG, and GLD as representative ETFs for equities, fixed income, and alternatives, and backtest every allocation between them in ten-percentage-point increments. Table~\ref{tab:optimal_split_sbg} in the appendix reports the full sweep, while Figure~\ref{fig:splits_sbg} shows the average allocation of the ten best-performing splits under each ranking criterion.

\begin{figure}[h]
    \centering
    \includegraphics[width=1.0\linewidth]{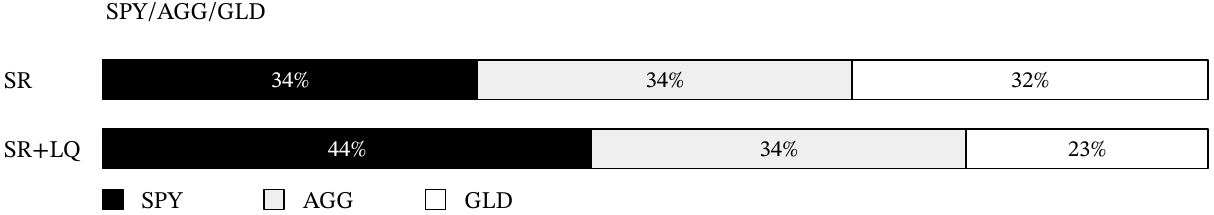}
    \caption{The best fixed splits hold all three assets in comparable proportions.}
    \label{fig:splits_sbg}
\end{figure}
\noindent The results suggest that the exact allocation matters less than holding all three asset classes. When ranked by SR, the best-performing splits are centered close to an equal allocation between SPY, AGG, and GLD. Ranking by SR+LQ shifts some weight from gold toward equities, but the same three-way structure remains. The sweep also allows us to compare these portfolios directly with the traditional 60/40 portfolio. The 60/40 portfolio has a Sharpe ratio of $0.51$ and a maximum drawdown of $0.36$. By adding gold and adjusting the allocation across all three asset classes, the best portfolio with a 30/40/30 split increases the Sharpe ratio to $0.67$ and reduces the maximum drawdown to $0.22$.

\begin{figure}[h]
    \centering
    \includegraphics[width=1.0\linewidth]{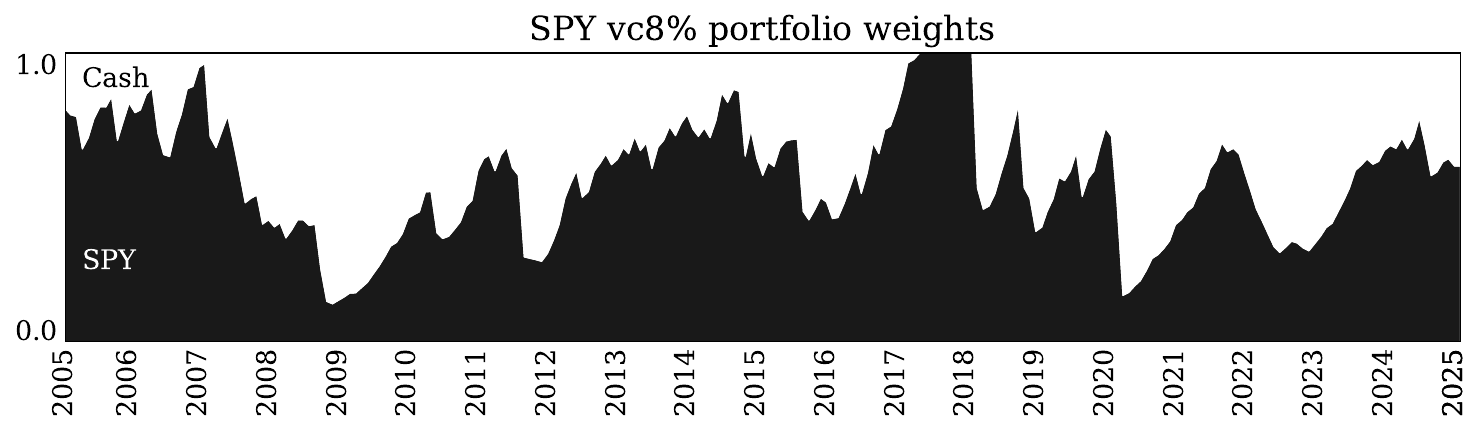}
    \caption{Volatility control moves the portfolio into cash when volatility rises.}
    \label{fig:spy_vc8_weights}
\end{figure}

\paragraph{Volatility control.}
We next use the risk model to control the overall volatility of the portfolio. The simplest example is a portfolio containing only SPY. Figure~\ref{fig:spy_vc8_weights} shows how the allocation between SPY and cash changes when an $8\%$ volatility target is applied.
 
\begin{figure}[h]
    \centering
    \includegraphics[width=1.0\linewidth]{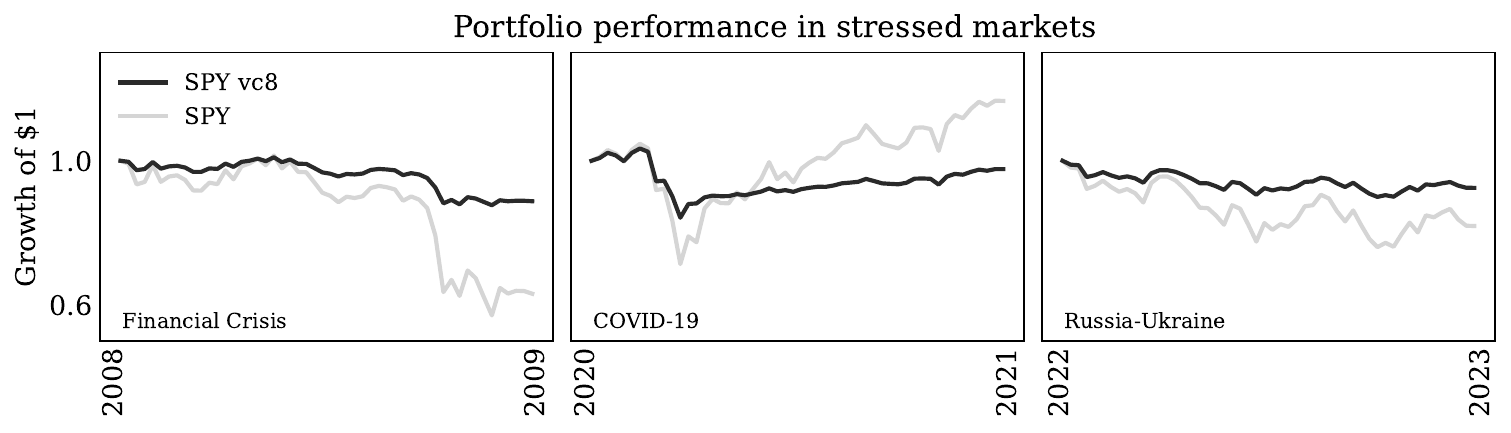}
    \caption{Volatility control softens all three major drawdowns}
    \label{fig:vc_crises}
\end{figure}

\noindent The volatility control in~(\ref{eq:vol_control}) does not change the relative allocation between the assets in a portfolio. Instead, it scales the entire portfolio against cash until its estimated volatility reaches the target $\sigma_{\mathrm{tgt}}$. For SPY, this means investing more in the market when volatility is low and moving more of the portfolio into cash when volatility rises. This behavior is particularly visible during the turbulent periods of 2008--2009 and 2020--2021. Figure~\ref{fig:vc_crises} compares SPY with its volatility-controlled counterpart during the three major market crashes in the backtest.
 
\noindent In all three crashes, volatility control reduces the drawdown by lowering market exposure as volatility increases. The trade-off is that the strategy can be slow to increase its exposure again once the market begins to recover, causing it to miss part of the rebound. This is particularly clear during the rapid recovery following the COVID-19 crash in the second panel. Table~\ref{tab:optimal_split_sbg} in the appendix repeats the SPY/AGG/GLD allocation sweep with volatility control and shows three main effects. First, the strategy brings the realized risk of the different allocations much closer together. Without volatility control, realized volatility ranges from approximately $5\%$ to $17\%$. With an $8\%$ target, the portfolios are instead limited to approximately this level, with deviations of around $\pm1\%$ due to errors in the risk forecast. Second, maximum drawdown decreases across the allocations, with the largest improvements occurring in the most aggressive portfolios. For example, the near-all-equity 90/0/10 allocation sees its maximum drawdown fall from $50\%$ to $20\%$. Third, this reduction in risk does not necessarily come at the expense of the Sharpe ratio. The best allocation retains an SR of $0.67$, while the SR of the 90/0/10 allocation increases from $0.50$ to $0.57$ as the volatility control reduces its exposure during the worst periods. Volatility control therefore provides a simple way to manage how much risk the portfolio takes, but it does not determine where that risk should come from. Since every holding is scaled by the same factor, the investor must still decide which assets to hold and how to allocate between them. The next strategy takes a step toward solving this problem by sizing each position according to its individual risk.

\paragraph {Inverse volatility.}
Inverse volatility replaces the static allocation of fixed weights with an allocation that adapts to changes in risk. Under~(\ref{eq:inverse_volatility}), each asset is weighted inversely to its estimated volatility, giving larger positions to less volatile assets and smaller positions to more volatile ones. These weights are recomputed at every trade, allowing the portfolio to adjust as market conditions change.

We first apply inverse volatility to the best equal-weight portfolios from the previous section. The results are shown in Table~\ref{tab:iv_on_winners}.

\begin{table}[h]
    \centering
    \footnotesize
    \caption{Inverse volatility on the top equal weight portfolios, ranked by SR and SR+LQ.}
    \begin{tabular}{ll|SSSSSS}
        \toprule
        \textbf{Criteria} & \textbf{ETFs} & {$\boldsymbol{r}$} & {$\boldsymbol{\sigma}$} & \textbf{MDD} & \textbf{TO} &  \textbf{SR} & \textbf{LQ} \\
        \midrule
        SR    & QQQ/IEF/GLD         & 0.07 & 0.07 & 0.19 & 0.28 & 0.84 & 0.21 \\
              & QQQ/XLP/TLT/GLD     & 0.10 & 0.09 & 0.21 & 0.33 & 0.91 & 0.70 \\
              & QQQ/XLP/IEF/TIP/GLD & 0.06 & 0.06 & 0.16 & 0.29 & 0.80 & 0.17 \\
        \midrule
        SR+LQ & XLK/XLV/GLD         & 0.11 & 0.13 & 0.32 & 0.32 & 0.75 & 0.73 \\
              & QQQ/XLU/AGG/GLD     & 0.07 & 0.07 & 0.19 & 0.30 & 0.76 & 0.64 \\
              & SPY/QQQ/XLP/TLT/GLD & 0.10 & 0.10 & 0.21 & 0.32 & 0.86 & 0.83 \\
        \bottomrule
    \end{tabular}
    \label{tab:iv_on_winners}
\end{table}
\noindent The main result is that inverse volatility works best when the assets have reasonably similar levels of volatility. In this case, no single asset dominates the allocation, but the weights can still adjust as their relative risks change. For example, TLT has a volatility of around $15\%$, which is comparable to the equity and gold ETFs it is combined with. In portfolios containing TLT, inverse-volatility weighting improves the allocation: for QQQ/XLP/TLT/GLD, the Sharpe ratio increases from $0.84$ under equal weights to $0.91$, while LQ increases from $0.62$ to $0.70$. When all assets have very similar volatility, the inverse-volatility allocation instead remains close to equal weights and performance changes little. The weakness of the strategy appears when one asset is much less volatile than the others. Since a low volatility directly translates into a large weight, a sufficiently calm asset can dominate the portfolio. IEF, for example, has a volatility of around $7\%$ and receives roughly $60\%$ of QQQ/IEF/GLD. Although the Sharpe ratio increases slightly, the portfolio becomes much more dependent on bonds: annual return falls from $10\%$ to $7\%$, while LQ falls from $0.44$ to $0.21$. The same problem is even more pronounced for low-volatility assets such as AGG. Inverse volatility therefore does not necessarily favor the asset that provides the most useful diversification; it simply favors the asset with the lowest individual volatility.

To see whether this pattern holds more generally, we repeat the exhaustive search using inverse-volatility weights. The top 60 portfolios for each portfolio size are reported in Table~\ref{tab:exhaustive_search_inverse_volatility}. The same broad portfolio structure found under equal weights remains: the strongest portfolios combine growth-oriented equities, defensive equities, government bonds, and gold. However, the best Sharpe ratios increase from $0.81$, $0.84$, and $0.84$ under equal weights to $0.84$, $0.91$, and $0.90$ under inverse volatility for portfolios of three, four, and five ETFs, respectively. TLT also performs better in these portfolios than lower-volatility aggregate bonds, supporting the earlier result that inverse volatility is most effective when the assets provide different sources of risk without having vastly different volatility levels. This leads to the main limitation of inverse volatility: it considers each asset separately. The strategy knows how volatile an asset is, but not whether it moves together with the other assets in the portfolio. Adding several similar equity ETFs therefore gives each of them its own allocation and can unintentionally increase the portfolio's total equity exposure. Likewise, including a very low-volatility bond ETF can cause fixed income to dominate the portfolio. Inverse volatility can determine the size of each position, but it cannot determine which assets should be held together. The remaining strategies address this limitation by using the full covariance structure, allowing both individual risk and the relationships between assets to influence the allocation.

\paragraph{Minimum volatility.}
Minimum volatility of~(\ref{eq:min_var}) is the first strategy to use the full covariance matrix, which addresses exactly what inverse volatility lacked. It can distinguish a diversifying asset from a duplicate. What it cannot see is return, and this section shows both sides of that trade-off. We run it through three tests that the two remaining strategies will repeat, a duplicate asset, a set of complementary assets, and an extreme asset. Table~\ref{tab:min_vol_duplicates} starts with SPY and AGG, the assets of the 60/40, and then adds VTV and VUG.

\begin{table}[H]
    \centering
    \footnotesize
    \caption{Adding equities similar to SPY does not change the split between equities and AGG.}
    \begin{tabular}{l|SSSSSS}
        \toprule
        \textbf{ETFs} & {$\boldsymbol{r}$} & {$\boldsymbol{\sigma}$} & \textbf{MDD} & \textbf{TO} & \textbf{SR} & \textbf{LQ} \\
        \midrule
        SPY/AGG         & 0.04 & 0.05 & 0.18 & 0.18 & 0.38 & -0.03 \\
        SPY/VTV/VUG/AGG & 0.04 & 0.05 & 0.16 & 0.28 & 0.39 & -0.04 \\
        \bottomrule
    \end{tabular}
    \label{tab:min_vol_duplicates}
\end{table}

\begin{figure}[H]
    \centering
    \includegraphics[width=1.0\linewidth]{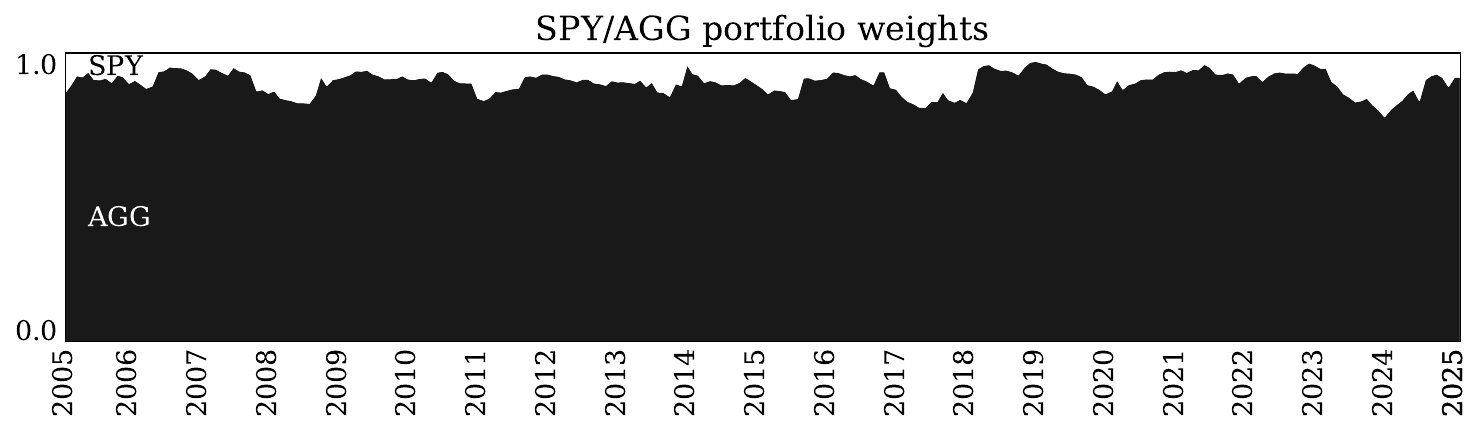}
    \caption{AGG dominating the minimum volatility portfolio due to its low volatility.}
    \label{fig:min_vol_equity_agg}
\end{figure}

\noindent AGG has a volatility of $5\%$ against SPY's $19\%$, and with no notion of return, minimum volatility holds the pair in a near 10/90 split, as seen in Figure~\ref{fig:min_vol_equity_agg}.

\noindent The portfolio earns $4\%$ per year at a Sharpe ratio of $0.38$, well below the $0.51$ of the fixed 60/40 split of the same two assets. This is the concentration problem described in \S\ref{sec:simple_trading_strategies}. Adding VTV and VUG changes almost nothing since minimum volatility recognizes that the three equities are highly correlated and splits the same small allocation between them.

The concentration in AGG comes from the asset selection as much as from the method, since AGG is the only low volatility asset available. Table~\ref{tab:min_vol_complements} instead replaces AGG with TLT and GLD, complementary assets whose volatilities are closer to SPY's.

\begin{table}[H]
    \centering
    \footnotesize
    \caption{Complementary assets of similar risk give minimum volatility something to balance.}
    \begin{tabular}{l|SSSSSS}
        \toprule
        \textbf{ETFs} & {$\boldsymbol{r}$} & {$\boldsymbol{\sigma}$} & \textbf{MDD} & \textbf{TO} & \textbf{SR} & \textbf{LQ} \\
        \midrule
        SPY/AGG         & 0.04 & 0.05 & 0.18 & 0.18 & 0.38 & -0.03 \\
        SPY/TLT         & 0.08 & 0.09 & 0.30 & 0.37 & 0.65 &  0.38 \\
        SPY/GLD         & 0.11 & 0.13 & 0.32 & 0.56 & 0.70 &  0.61 \\
        \midrule
        SPY/TLT/GLD     & 0.09 & 0.09 & 0.23 & 0.59 & 0.87 &  0.56 \\
        \bottomrule
    \end{tabular}
    \label{tab:min_vol_complements}
\end{table}

\noindent With similar volatilities their covariance decides the allocation. TLT with SPY raises the Sharpe ratio from $0.38$ to $0.65$ and GLD with SPY reaches $0.70$, and combining all three reaches $0.87$ with an LQ of $0.56$, on par with the best equal weight portfolios. Minimum volatility thus finds good proportions within complementary assets of similar risk, and can be trusted with more, as USMV of \S\ref{sec:simple_trading_strategies} shows on US stocks.

In the last experiment we add IBIT, the bitcoin ETF extended back to 2017 as described in \S\ref{section:asset_categories}. Table~\ref{tab:min_vol_ibit} shows the result.

\begin{table}[H]
    \centering
    \footnotesize
    \caption{Minimum volatility does not hold IBIT.}
    \begin{tabular}{l|SSSSSS}
        \toprule
        \textbf{ETFs} & {$\boldsymbol{r}$} & {$\boldsymbol{\sigma}$} & \textbf{MDD} & \textbf{TO} & \textbf{SR} & \textbf{LQ} \\
        \midrule
        SPY/TLT/GLD      & 0.09 & 0.09 & 0.23 & 0.59 & 0.87 & 0.56 \\
        SPY/TLT/GLD/IBIT & 0.09 & 0.09 & 0.23 & 0.59 & 0.86 & 0.56 \\
        \bottomrule
    \end{tabular}
    \label{tab:min_vol_ibit}
\end{table}

\noindent Nothing changes. To an objective that only minimizes risk, bitcoin is only risk, and the optimizer sets the position to zero. The next strategy instead enforces a more equal split of the risk.

\paragraph{Risk parity.}
Risk parity uses the same covariance matrix but with a different objective. The formulation in~(\ref{eq:risk_parity}) forces every asset to contribute the same share of the total portfolio risk, so no asset can be dropped and no asset can dominate. We run the same three tests, starting with the duplicates in Table~\ref{tab:risk_parity_duplicates}.

\begin{table}[H]
    \centering
    \footnotesize
    \caption{Each added equity claims its own share of the total risk.}
    \begin{tabular}{l|SSSSSS}
        \toprule
        \textbf{ETFs} & {$\boldsymbol{r}$} & {$\boldsymbol{\sigma}$} & \textbf{MDD} & \textbf{TO} & \textbf{SR} & \textbf{LQ} \\
        \midrule
        SPY/AGG	        & 0.05 & 0.06 & 0.18 & 0.20 & 0.58 & 0.19 \\
        SPY/VTV/VUG/AGG	& 0.07 & 0.08 & 0.22 & 0.30 & 0.61 & 0.22 \\
        \bottomrule
    \end{tabular}
    \label{tab:risk_parity_duplicates}
\end{table}

\begin{figure}[H]
    \centering
    \includegraphics[width=1.0\linewidth]{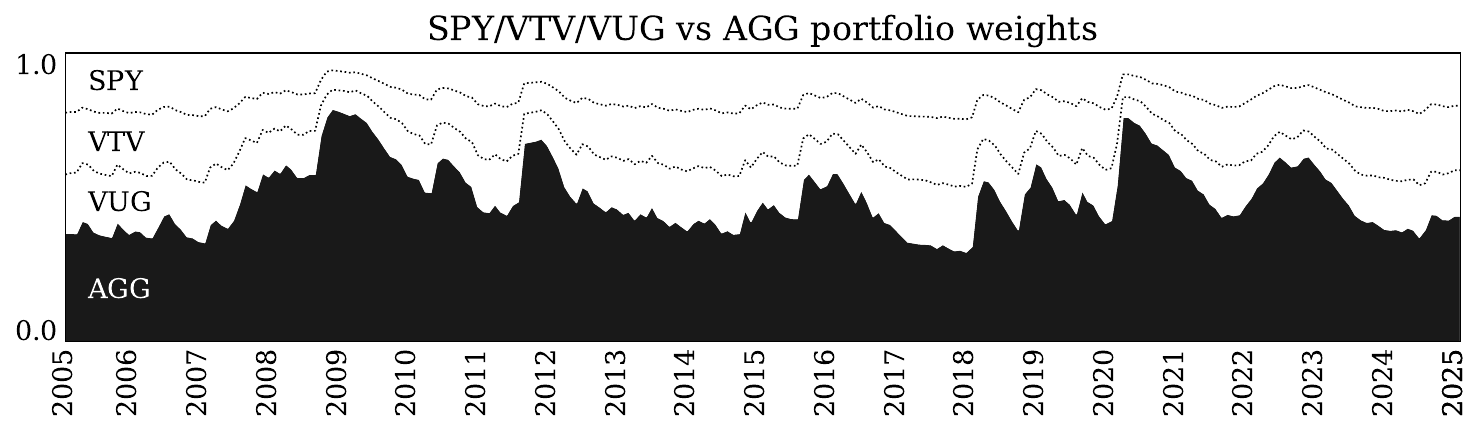}
    \caption{Added equities account for more of the total portfolio weight and are split evenly.}
    \label{fig:risk_parity_equity_agg}
\end{figure}

\noindent The equal risk condition changes both rows. SPY and AGG now split the risk evenly instead of the capital, which gives SPY a meaningful weight and raises the Sharpe ratio from minimum volatility's $0.38$ to $0.58$. Adding VTV and VUG shows how risk parity handles more assets of similar risk profile. Each equity claims its own equal share, so three quarters of the portfolio risk now comes from equities, as seen in Figure~\ref{fig:risk_parity_equity_agg}.

\noindent Volatility, however, rises from $6\%$ to $8\%$. The problem that was present in inverse volatility returns, now in risk rather than in capital. Table~\ref{tab:risk_parity_complements} repeats the complementary asset test.

\begin{table}[H]
    \centering
    \footnotesize
    \caption{Risk parity is the first strategy that handles AGG well.}
    \begin{tabular}{l|SSSSSS}
        \toprule
        \textbf{ETFs} & {$\boldsymbol{r}$} & {$\boldsymbol{\sigma}$} & \textbf{MDD} & \textbf{TO} & \textbf{SR} & \textbf{LQ} \\
        \midrule
        SPY/AGG	         & 0.05 & 0.06 & 0.18 & 0.20 & 0.58 & 0.19 \\
        SPY/TLT	         & 0.08 & 0.09 & 0.30 & 0.27 & 0.63 & 0.43 \\
        SPY/GLD	         & 0.11 & 0.13 & 0.32 & 0.40 & 0.69 & 0.50 \\
        \midrule
        SPY/AGG/GLD	     & 0.07 & 0.07 & 0.18 & 0.28 & 0.74 & 0.64 \\
        SPY/TLT/GLD	     & 0.09 & 0.09 & 0.23 & 0.32 & 0.79 & 0.25 \\
        \bottomrule
    \end{tabular}
    \label{tab:risk_parity_complements}
\end{table}

\noindent The notable rows are the ones holding AGG. Since no asset can contribute more than its equal share of the risk, the low volatility bond can no longer dominate, and risk parity is the first strategy where AGG improves the portfolio. SPY/AGG/GLD reaches a Sharpe ratio of $0.74$ with an LQ of $0.64$, the most consistent of the SPY portfolios in these tests. Replacing AGG with TLT raises the Sharpe ratio to $0.79$ but LQ falls to $0.25$, so we keep the AGG version for the final test in Table~\ref{tab:risk_parity_ibit}.

\begin{table}[H]
    \centering
    \footnotesize
    \caption{Risk parity must hold IBIT and benefits from it.}
    \begin{tabular}{l|SSSSSS}
        \toprule
        \textbf{ETFs} & {$\boldsymbol{r}$} & {$\boldsymbol{\sigma}$} & \textbf{MDD} & \textbf{TO} & \textbf{SR} & \textbf{LQ} \\
        \midrule
        SPY/AGG/GLD	     & 0.07 & 0.07 & 0.18 & 0.28 & 0.74 & 0.64 \\
        SPY/AGG/GLD/IBIT & 0.08 & 0.08 & 0.21 & 0.31 & 0.84	& 0.72 \\
        \bottomrule
    \end{tabular}
    \label{tab:risk_parity_ibit}
\end{table}

\noindent Where minimum volatility excluded bitcoin, risk parity must hold it, and the equal risk condition decides how much. IBIT enters at a weight of a few percent, and since bitcoin is nearly uncorrelated with the rest, the Sharpe ratio rises from $0.74$ to $0.84$ and LQ from $0.64$ to $0.72$. Being forced to hold every asset, works in the investor's favor here, but the same rule means risk parity would, if allowed, also hold the futures ETFs of \S\ref{section:asset_categories}. The final strategy can again exclude assets, this time for being redundant rather than volatile.

\paragraph{Maximum diversification.}
Maximum diversification also uses the full covariance matrix as minimum volatility and risk parity but instead seeks the portfolio where combining assets results in the largest diversification ratio of~(\ref{eq:max_div}). The reformulation in~(\ref{eq:max_div_reformulated}) is solved on the correlation matrix, with every asset scaled by its own volatility beforehand, so the volatility of an asset no longer decides how much of it is held. Table~\ref{tab:max_div_duplicates} starts with the duplicate test.

\begin{table}[H]
    \centering
    \footnotesize
    \caption{Maximum diversification drops SPY as redundant.}
    \begin{tabular}{l|SSSSSS}
        \toprule
        \textbf{ETFs} & {$\boldsymbol{r}$} & {$\boldsymbol{\sigma}$} & \textbf{MDD} & \textbf{TO} & \textbf{SR} & \textbf{LQ} \\
        \midrule
        SPY/AGG	        & 0.05 & 0.06 & 0.18 & 0.19 & 0.52 & 0.21 \\
        SPY/VUG/VTV/AGG	& 0.05 & 0.06 & 0.17 & 0.35 & 0.50 & 0.15 \\
        \bottomrule
    \end{tabular}
    \label{tab:max_div_duplicates}
\end{table}

\begin{figure}[H]
    \centering
    \includegraphics[width=1.0\linewidth]{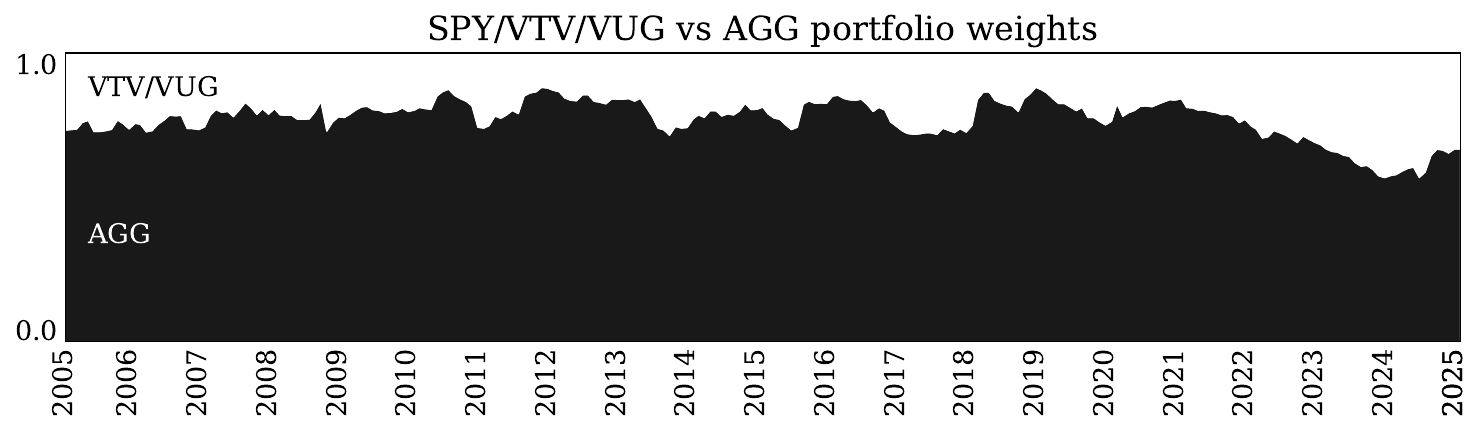}
    \caption{SPY is replaced by VTV and VUG as the more diversifying constituents.}
    \label{fig:max_div_equity_agg}
\end{figure}

\noindent The strategy allocates its capital in a unique way. VTV and VUG together essentially form SPY, and to the diversification ratio the two parts are the marginally more diversified way of holding the same exposure, so SPY is dropped as redundant, as seen in Figure~\ref{fig:max_div_equity_agg}.

\noindent Performance is close to unchanged, which is expected. The strategy holds the same exposure through different tickers. The SPY/AGG pair is also worth noting. It reaches a Sharpe ratio of $0.52$ against minimum volatility's $0.38$, because the objective works on correlations rather than volatilities, so AGG no longer dominates the portfolio even though it is still the only low volatility asset available. Table~\ref{tab:max_div_complements} repeats the complementary asset test.

\begin{table}[H]
    \centering
    \footnotesize
    \caption{The choice of bond matters less under maximum diversification.}
    \begin{tabular}{l|SSSSSS}
        \toprule
        \textbf{ETFs} & {$\boldsymbol{r}$} & {$\boldsymbol{\sigma}$} & \textbf{MDD} & \textbf{TO} & \textbf{SR} & \textbf{LQ} \\
        \midrule
        SPY/AGG     & 0.05 & 0.06 & 0.18 & 0.19 & 0.52 & 0.21 \\
        SPY/TLT     & 0.08 & 0.09 & 0.29 & 0.25 & 0.64 & 0.41 \\
        SPY/GLD	    & 0.11 & 0.13 & 0.32 & 0.32 & 0.68 & 0.37 \\
        \midrule
        SPY/TLT/GLD & 0.09 & 0.09 & 0.24 & 0.38 & 0.82 & 0.54 \\
        \bottomrule
    \end{tabular}
    \label{tab:max_div_complements}
\end{table}

\noindent Which bond sits next to SPY matters much less than before. Under minimum volatility, SPY/AGG achives a Sharpe ratio of $0.38$ and SPY/TLT $0.65$, a gap driven almost entirely by their volatility levels. Here the same pair of portfolios achieve $0.52$ and $0.64$, and the remaining difference comes from correlation. TLT has a correlation of $-0.32$ with SPY against AGG's $-0.01$ in Table~\ref{tab:asset_universe}, and a negative correlation is exactly what the diversification ratio rewards, so the strategy still prefers TLT, but now for a better reason. Adding GLD completes the trio, and SPY/TLT/GLD reaches a Sharpe ratio of $0.82$. Table~\ref{tab:max_div_ibit} adds IBIT.

\begin{table}[H]
    \centering
    \footnotesize
    \caption{IBIT is nearly uncorrelated with the other assets, which the diversification ratio rewards.}
    \begin{tabular}{l|SSSSSS}
        \toprule
        \textbf{ETFs} & {$\boldsymbol{r}$} & {$\boldsymbol{\sigma}$} & \textbf{MDD} & \textbf{TO} & \textbf{SR} & \textbf{LQ} \\
        \midrule
        SPY/TLT/GLD      & 0.09 & 0.09 & 0.24 & 0.38 & 0.82 & 0.54 \\
        SPY/TLT/GLD/IBIT & 0.10 & 0.09 & 0.28 & 0.42 & 0.90 & 0.54 \\
        \bottomrule
    \end{tabular}
    \label{tab:max_div_ibit}
\end{table}

\noindent This is the setting the strategy was built for. IBIT is close to uncorrelated with everything else in the universe, which is exactly what the objective rewards, and because every asset is scaled by its own volatility, the position stays small on its own. Nothing forces the strategy to hold bitcoin, as risk parity did, and nothing caps the position, as minimum volatility effectively did by excluding it. The optimizer simply holds a small amount because it improves the portfolio, and the Sharpe ratio rises from $0.82$ to $0.90$ with LQ unchanged.

Every portfolio so far has used a smaller subset of assets. The last question is whether the strategy can be trusted with a larger selection, which minimum volatility and risk parity could not. As a test, we hand maximum diversification all 30 ETFs in the Market, Sector, Government, Physical, and Crypto categories, the same categories we previously included in the exhaustive search. No three to five ETFs are chosen, instead the strategy gets everything in each category and must find the allocation on its own. Table~\ref{tab:max_div_categories} reports the result.

\begin{table}[H]
    \centering
    \footnotesize
    \caption{Maximum diversification run on five whole asset categories.}
    \begin{tabular}{l|SSSSSS}
        \toprule
        \textbf{Categories} & {$\boldsymbol{r}$} & {$\boldsymbol{\sigma}$} & \textbf{MDD} & \textbf{TO} & \textbf{SR} & \textbf{LQ} \\
        \midrule
        Market/Sector/Government/Physical/Crypto & 0.07 & 0.07 & 0.21 & 1.50 & 0.73 & 0.34 \\
        \bottomrule
    \end{tabular}
    \label{tab:max_div_categories}
\end{table}

\begin{figure}[H]
    \centering
    \includegraphics[width=1.0\linewidth]{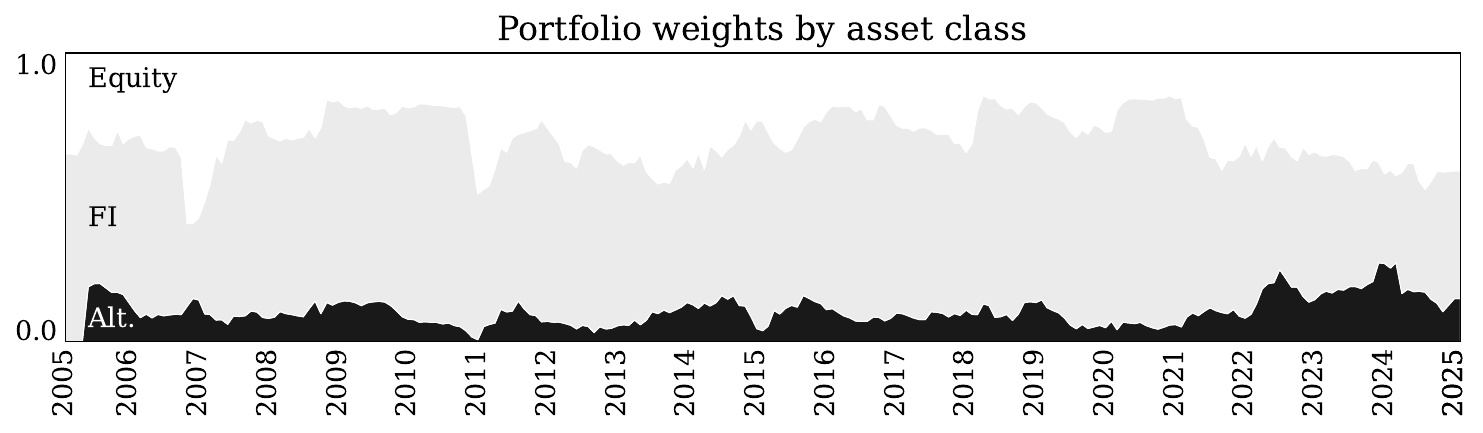}
    \caption{Maximum diversification successfully holding a broad ETF portfolio.}
    \label{fig:categories_max_div}
\end{figure}

\noindent It works. The strategy holds a broad mix across the asset classes throughout, as seen in Figure~\ref{fig:categories_max_div}, and delivers a Sharpe ratio of $0.73$ on a selection where minimum volatility would have concentrated in the lowest volatility bonds. The cost is turnover. Rebalancing thirty positions every month turns the portfolio over one and a half times per year, and the smaller selected portfolios above reach higher Sharpe ratios at a fraction of the effort. Selecting at the category level is possible, but for an investor willing to pick a handful of ETFs, it is not worth the trading.

\noindent Finally, for the curious reader, we show the portfolio of five assets that achieved the highest Sharpe ratio for any simple strategy we found. This portfolio used maximum diversification, and the assets included is not far from what we consider to be reasonable. The portfolio and related metrics can be seen in Table~\ref{tab:max_div_best}.

\begin{table}[H]
    \centering
    \footnotesize
    \caption{The best simple strategy portfolio found.}
    \begin{tabular}{l|SSSSSS}
        \toprule
        \textbf{ETFs} & {$\boldsymbol{r}$} & {$\boldsymbol{\sigma}$} & \textbf{MDD} & \textbf{TO} & \textbf{SR} & \textbf{LQ} \\
        \midrule
        QQQ/XLP/TLT/GLD/IBIT\textcolor{red}{*} & 0.11 & 0.09 & 0.25 & 0.54 & 1.02 & 0.60 \\
        \bottomrule
    \end{tabular}
    \label{tab:max_div_best}
\end{table}

\noindent QQQ and XLP as offensive and defensive equities, TLT, GLD, and a small position in IBIT, weighted by maximum diversification, reach a Sharpe ratio of $1.02$ with an LQ of $0.60$. We report this combination as a reference rather than propose it, since it was found by searching over the backtest period itself, and its performance is therefore a product of hindsight. It nevertheless summarizes the section well. The best simple portfolios keep returning to the same core concept. The concept being, two equities that complement each other, a long term bond, gold, and a small position in an uncorrelated extreme asset such as bitcoin, held together by a rule that uses the full covariance matrix. Simple and implementable. However, there is more to portfolio construction. Every strategy in this section is blind to return, which is why the assets had to be picked by hand, with no futures and no single country funds among them. The Markowitz formulation of \S\ref{sec:markowitz} removes that restriction by letting a forecast decide which assets are worth holding, and the next two sections cover the two forecasts it needs, including the covariance matrix that has been trated as given in this section above.

\newpage

\subsection{Risk and alpha model}\label{sec:forecast_quality}
Before the portfolio backtests of \S\ref{sec:performance} we evaluate the two forecasts on their own. Following~\cite{Johansson:2023}, a forecast can be judged directly, by statistical measures of how well it describes the quantity it predicts, or indirectly, by the performance of the portfolios built on it. This section does the former, and states the parameter values used for each model.

\paragraph{Risk model.} We first fix the number of factors for the factor model of \S\ref{sec:risk_model} by examining the cumulative share of variance explained by the leading principal components. The first component alone explains half of all variance in the universe, which can be interpreted as the market factor, and the first ten explain $89\%$, after which each additional component adds around one percentage point. We use $k = 10$ throughout. Table~\ref{tab:parameters_risk} lists all of the relevant risk model parameters.

\begin{table}[H]
    \footnotesize
    \centering
    \caption{Risk model parameters.}
    \begin{tabular}{l|l|l}
        \toprule
        \textbf{Parameter} & \textbf{Description} & \textbf{Value} \\
        \midrule
        $L$                   & Lookback, rolling window covariance & 126 days           \\
        $H_{\mathrm{var}}$    & EWMA half-life, variances           & 42 days            \\
        $H_{\mathrm{cov}}$    & EWMA half-life, correlations        & 126 days           \\
        $k$                   & Number of statistical factors       & 10                 \\
        \bottomrule
    \end{tabular}
    \label{tab:parameters_risk}
\end{table}

\noindent We evaluate our covariance forecast against the true realized volatility. At the end of each month every model produces a forecast $\hat{\Sigma}_t$, and the target is the realized covariance $\Sigma_t$ of the following 21 trading days, since we trade monthly. The distance between the two is the Frobenius norm, relative to the magnitude of the target,
\[
    \|A\|_{\mathrm{F}} = \Big( \sum_{i}\sum_{j} A_{ij}^{2} \Big)^{1/2},
    \qquad
    d_t = \frac{\|\hat{\Sigma}_t - \Sigma_t\|_{\mathrm{F}}}{\|\Sigma_t\|_{\mathrm{F}}},
\]
so the empty forecast $\hat{\Sigma}_t = 0$ scores exactly one, a perfect forecast zero, and the reported value is the fraction of the realized covariance the model fails to capture. Table~\ref{tab:cov_frobenius} reports the average of $d_t$ over the backtest for each model.

\begin{table}[H]
    \footnotesize
    \centering
    \caption{Normalized Frobenius distance between forecast and realized covariance.}
    \begin{tabular}{lrrr}
        \toprule        
        \textbf{Class} & \textbf{Sample} & \textbf{EWMA} & \textbf{K10} \\
        \midrule
        Equities      & 0.191 & 0.188 & \textbf{0.181} \\
        Fixed income  & 0.208 & 0.213 & \textbf{0.205} \\
        Alternatives  & 0.225 & 0.221 & \textbf{0.218} \\
        \midrule
        Full universe & 0.228 & 0.227 & \textbf{0.221} \\
        \bottomrule
    \end{tabular}
    \label{tab:cov_frobenius}
\end{table}

\noindent The factor model performs better for every asset class and over the full universe, but the gain is marginal, and all three models leave roughly a fifth of the realized covariance unexplained. For a small universe of, say, three assets, either simple estimator is a fully reasonable substitute for the factor model. We, nevertheless, use the factor model proposed in all experiments above in \S\ref{sec:simple_strategies_experiments} and that follow.

\begin{figure}[H]
    \centering
    \includegraphics[width=1.0\linewidth]{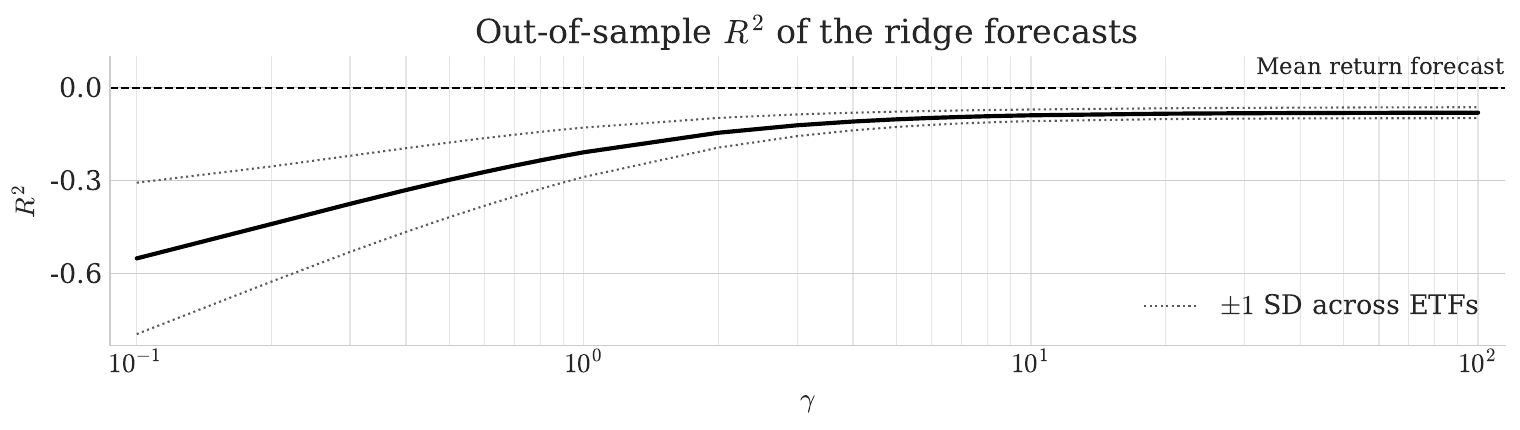}
    \caption{Regularization improves $R^2$ by shrinking the forecast toward zero mean guess.}
    \label{fig:r2_vs_gamma}
\end{figure}

\paragraph{Regression model.} The regression forecasts of \S\ref{sec:regression_models} are evaluated out of sample. At each trading date the model is fitted on the trailing window, using the parameters of Table~\ref{tab:parameters_ridge}, which is then used our prediction of returns for the following period.
\begin{table}[H]
    \footnotesize
    \centering
    \caption{Regression model parameters.}
    \begin{tabular}{l|l|l}
        \toprule
        \textbf{Parameter} & \textbf{Description} & \textbf{Value} \\
        \midrule
        $L_{\mathrm{ridge}}$ & Lookback, regression training window & 252 days \\
        $H_{\mathrm{ridge}}$ & Half-life, training sample weights   & 252 days \\
        $h$                  & Forecast horizon                     & 21 days  \\
        \bottomrule
    \end{tabular}
    \label{tab:parameters_ridge}
\end{table}
\noindent The predictions $\hat{r}_t$ are then compared with the actual realized returns $r_t$ through the out-of-sample $R^2$,
\[
    R^2 = 1 - \frac{\sum_t (r_t - \hat{r}_t)^2}{\sum_t r_t^2}.
\]
Since returns are treated as zero mean, the denominator is the error of the forecast that always predicts zero. This gives us that a $R^2 = 1$ represents a perfect forecast, while a $R^2 = 0$ is exactly as good as always saying zero, and $R^2 < 0$ is worse than saying nothing at all. To evaluate the importance of the regulization parameter$\lambda$, we sweep it across several orders of magnitude and evaluate the model at each level. Figure~\ref{fig:r2_vs_gamma} shows the result, reporting the mean $R^2$ across all ETFs with one standard deviation plotted with dashed lines.
\noindent Every entry in the figure is negative, so at no level of regularization does the model beat the zero forecast for a single asset. The $R^2$ nevertheless improves as the regularization grows, since increasing $\lambda$ shrinks the coefficients and thus the predictions toward zero, whose $R^2$ is zero by construction. This matters for how the forecast should be judged. A portfolio built on the model may well backtest decently, but a model whose statistical score improves only as it regularizes away its parameters offers no evidence that such performance is more than noise. We therefore conclude that our attempts at fitting a model bear no statistical support, though this does not mean return prediction is impossible. Our code is open for the reader to search for statistical support themselves, but at this point we defer to Markowitz's own words when asked about estimating expected returns, ``That's your job, not mine''~\cite{Boyd:2024}.

\paragraph{Simple signals.} The simple signals of \S\ref{sec:simple_alphas}, with parameters given in Table~\ref{tab:parameters_signals}, cannot be scored with the out-of-sample $R^2$ above. The moving average only takes the values $\pm1$ and momentum only assumes that recent returns persist, so neither makes a claim about the size of future returns.

\begin{table}[H]
    \footnotesize
    \centering
    \caption{Simple signal parameters.}
    \begin{tabular}{l|l|l}
        \toprule
        \textbf{Parameter} & \textbf{Description} & \textbf{Value} \\
        \midrule
        $L_{\mathrm{mom}}$ & Lookback, momentum        & 21 days \\
        $L_{\mathrm{mov}}$ & Lookback, moving average  & 50 days \\
        $\rho$             & Threshold, moving average & 1.0     \\
        \bottomrule
    \end{tabular}
    \label{tab:parameters_signals}
\end{table}

\noindent We therefore evaluate them indirectly, through the performance of the portfolios they build. Each signal is run through the Markowitz problem of \S\ref{sec:markowitz} on the full universe of \S\ref{section:investable_universe}, first with every asset receiving the one-month momentum signal, then with every asset receiving the moving average signal. Table~\ref{tab:markowitz_full_performance} reports both.

\begin{table}[H]
    \footnotesize
    \centering
    \caption{Each signal run through Markowitz on the full universe.}
    \begin{tabular}{l|SSSSSS}
        \toprule
        \textbf{Portfolio on full universe} & {$\boldsymbol{r}$} & {$\boldsymbol{\sigma}$} & \textbf{MDD} & \textbf{TO} & \textbf{SR} & \textbf{LQ} \\
        \midrule
        1-month momentum signal      & 0.06 & 0.09 & 0.20 & 1.08 & 0.43 & -0.03 \\
        50-day moving average signal & 0.05 & 0.08 & 0.23 & 1.06 & 0.45 &  0.48 \\
        \bottomrule
    \end{tabular}
    \label{tab:markowitz_full_performance}
\end{table}

\noindent Both portfolios perform considerably worse than the simple strategies of \S\ref{sec:simple_strategies_experiments}. The Sharpe ratios of $0.43$ and $0.45$ are roughly those of holding SPY alone, and both portfolios trade more than their full value in a typical year. The momentum is always invested somewhere and its LQ of $-0.03$ means its weak years do not beat cash, while the moving average keeps an LQ of $0.48$, since it moves the portfolio into cash when few assets are trending. A likely explanation is the universe itself, which still contains the categories that \S\ref{section:asset_categories} argued against holding, and a monthly signal under the turnover limits of~(\ref{eq:markowitz_full}) cannot always keep them out. We therefore run an ablation test, removing one category at a time and rerunning the entire backtest, nine runs per signal. Tables~\ref{tab:markowitz_one_month_momentum_ablation} and~\ref{tab:markowitz_moving_average_ablation} report the results, with the categories we later remove highlighted in grey.

\begin{table}[H]
    \footnotesize
    \centering
    \caption{Momentum ablation. Each row removes one category from the universe. Grey rows mark the categories we later remove.}
    \begin{tabular}{l|SSSSSS}
        \toprule
        \textbf{Removed category} & {$\boldsymbol{r}$} & {$\boldsymbol{\sigma}$} & \textbf{MDD} & \textbf{TO} & \textbf{SR} & \textbf{LQ} \\
        \midrule
        Market     & 0.06 & 0.09 & 0.21 & 1.08 & 0.44 & -0.01 \\
        \rowcolor{gray!10}
        Sectors    & 0.06 & 0.09 & 0.20 & 1.04 & 0.47 &  0.00 \\
        \rowcolor{gray!10}
        Countries  & 0.06 & 0.09 & 0.19 & 1.05 & 0.51 & -0.22 \\
        Government & 0.06 & 0.09 & 0.25 & 1.04 & 0.44 &  0.15 \\
        Corporate  & 0.06 & 0.09 & 0.19 & 1.08 & 0.43 & -0.01 \\
        Global     & 0.06 & 0.09 & 0.20 & 1.08 & 0.42 & -0.10 \\
        Physical   & 0.05 & 0.09 & 0.19 & 1.07 & 0.41 &  0.10 \\
        \rowcolor{gray!10}
        Futures    & 0.06 & 0.09 & 0.20 & 1.02 & 0.53 & -0.12 \\
        Crypto     & 0.04 & 0.09 & 0.24 & 1.07 & 0.29 & -0.14 \\
        \bottomrule
    \end{tabular}
    \label{tab:markowitz_one_month_momentum_ablation}
\end{table}

\noindent Each row is the full universe with one category removed, so a Sharpe ratio above the $0.45$ and $0.43$ baselines means the category hurt the overall performance. For momentum, removing Futures gives the largest improvement, to $0.53$, and removing Countries the second largest, to $0.51$. Countries is the only category whose removal improves both signals, in line with Figure~\ref{fig:countries}. The improvement from removing Sectors, to $0.47$, is the one result we did not expect. A possible explanation is that sectors that trend have larger drawdowns that a one-month window cannot anticipate. At the other end, removing Crypto lowers the Sharpe ratio to $0.29$, the largest change in either test. Momentum works best on assets with strong trends, and bitcoin has clearly had strong trends. 

\begin{table}[H]
    \footnotesize
    \centering
    \caption{Moving average ablation. Each row removes one category from the universe. Grey rows mark the categories we later remove.}
    \begin{tabular}{l|SSSSSS}
        \toprule
        \textbf{Removed category} & {$\boldsymbol{r}$} & {$\boldsymbol{\sigma}$} & \textbf{MDD} & \textbf{TO} & \textbf{SR} & \textbf{LQ} \\
        \midrule
        Market     & 0.05 & 0.07 & 0.20 & 1.06 & 0.42 & 0.32 \\
        Sectors    & 0.05 & 0.07 & 0.20 & 1.03 & 0.40 & 0.43 \\
        \rowcolor{gray!10}
        Countries  & 0.06 & 0.08 & 0.23 & 1.05 & 0.48 & 0.23 \\
        \rowcolor{gray!10}
        Government & 0.06 & 0.09 & 0.27 & 0.98 & 0.50 & 0.44 \\
        Corporate  & 0.05 & 0.08 & 0.23 & 1.08 & 0.44 & 0.20 \\
        \rowcolor{gray!10}
        Global     & 0.05 & 0.08 & 0.23 & 1.05 & 0.46 & 0.51 \\
        Physical   & 0.05 & 0.08 & 0.23 & 1.05 & 0.45 & 0.42 \\
        Futures    & 0.05 & 0.08 & 0.23 & 1.02 & 0.45 & 0.39 \\
        Crypto     & 0.05 & 0.08 & 0.22 & 1.06 & 0.43 & 0.34 \\
        \bottomrule
    \end{tabular}
    \label{tab:markowitz_moving_average_ablation}
\end{table}

\noindent For the moving average, removing Countries, Government, or Global improves the portfolio, while removing Market or Sectors makes it worse. The signal handles bonds poorly since a signal that only asks whether the price is above its trend gives a bond ETF returning $3\%$ little to no room for inclusion. We remove the highlighted categories and rerun both portfolios. Table~\ref{tab:markowitz_after_ablation_performance} shows both portfolios after the removals.

\begin{table}[H]
    \footnotesize
    \centering
    \caption{Both signals after removing the categories flagged by their ablations.}
    \begin{tabular}{l|SSSSSS}
        \toprule
        \textbf{Portfolio after ablation} & {$\boldsymbol{r}$} & {$\boldsymbol{\sigma}$} & \textbf{MDD} & \textbf{TO} & \textbf{SR} & \textbf{LQ} \\
        \midrule
        1-month momentum signal      & 0.08 & 0.08 & 0.16 & 0.81 & 0.78 & 0.40 \\
        50-day moving average signal & 0.06 & 0.08 & 0.23 & 1.04 & 0.51 & 0.30 \\
        \bottomrule
    \end{tabular}
    \label{tab:markowitz_after_ablation_performance}
\end{table}

\noindent Momentum improves substantially. The Sharpe ratio rises from $0.43$ to $0.78$, the drawdown falls from $0.20$ to $0.16$, LQ rises from $-0.03$ to $0.40$, and turnover falls to $0.81$, still several times that of the simple strategies. The moving average improves to $0.51$ while its LQ falls from $0.48$ to $0.30$. The removals were chosen using the backtest results themselves, so some of the improvement is still due to selection. However, Futures and Countries were argued against in \S\ref{section:asset_categories} on structural grounds, before any backtest was run. We therefore consider the selection bias introduced by the ablation to be modest, though not zero. Momentum is also a standard signal in industry practice~\cite{Narang:2024}, and its resulting portfolio holds 26 ETFs across six categories, with no individual asset selected by hand. We call this the Momentum Universe and carry it to \S\ref{sec:performance}.

\newpage

\subsection{Proposed portfolios}\label{sec:performance}
\begin{figure}[h]
    \centering
    \includegraphics[width=1.0\linewidth]{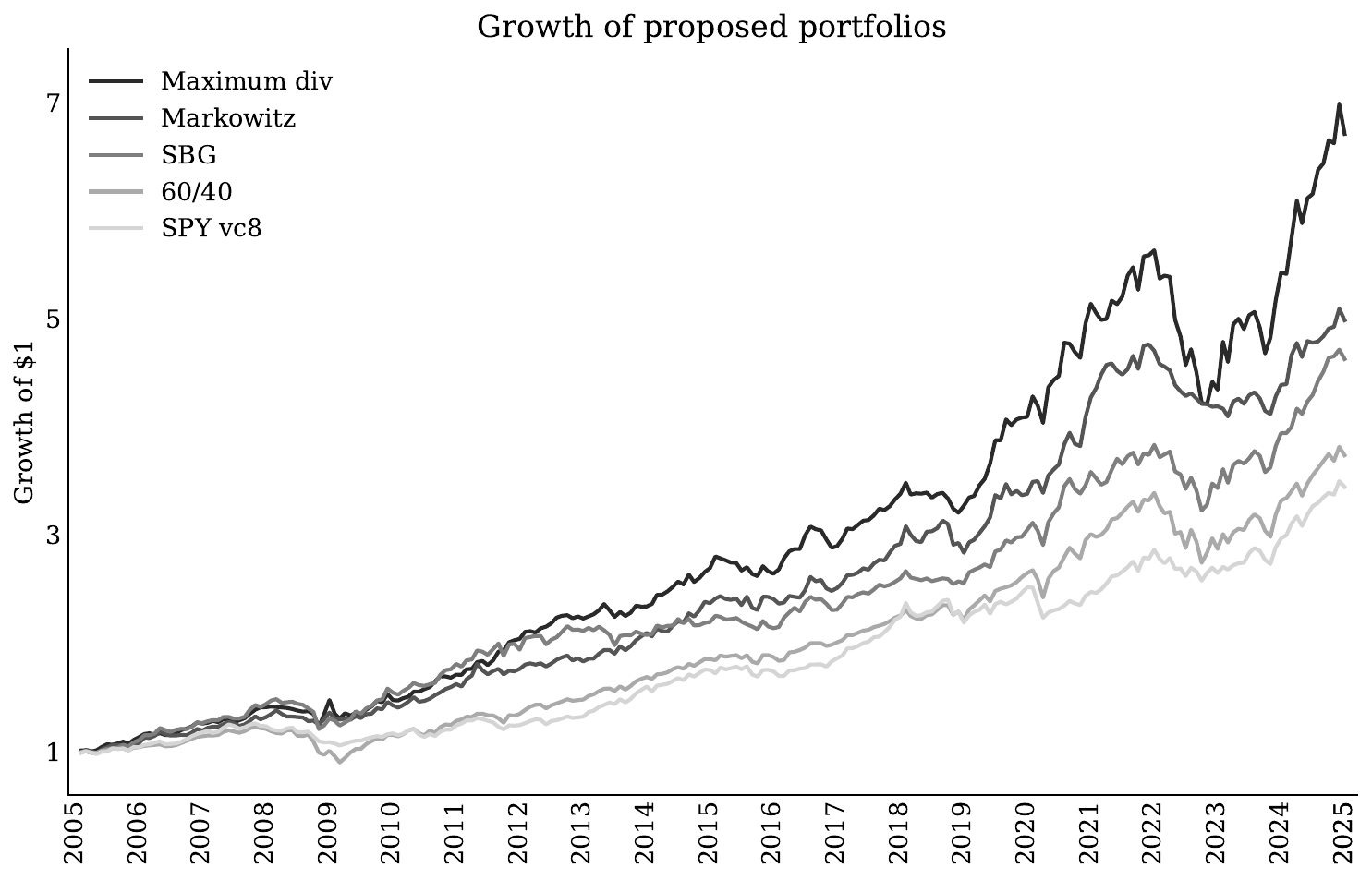}
    \caption{Growth of the proposed portfolios over the full backtest.}
    \label{fig:proposed_portfolios}
\end{figure}

From the experiments above we propose four portfolios, listed in increasing order of the involvement they demand of the investor. SPY under 8\% volatility control, SPY/AGG/GLD in an equal weight split, SPY/TLT/GLD/IBIT under maximum diversification, and the Momentum Universe of \S\ref{sec:forecast_quality} under Markowitz. We benchmark against the 60/40, and the regret bound of \S\ref{sec:markowitz} marks an upper bound. Figure~\ref{fig:proposed_portfolios} shows the growth of each portfolio and Table~\ref{tab:final_portfolios_performance} reports the metrics. Table~\ref{tab:final_portfolios_sensitivity_check} in the appendix repeats them over two truncated windows, and the ordering between the portfolios holds.

\begin{table}[h]
    \footnotesize
    \centering
    \caption{Performance of the proposed portfolios over the full backtest.}
    \begin{tabular}{ll|SSSSSSS}
        \toprule
        \textbf{Portfolio} & \textbf{Strategy} & {$\boldsymbol{r}$} & {$\boldsymbol{\sigma}$} & \textbf{MDD} & \textbf{TO} & \textbf{SR} & \textbf{LQ} \\
        \midrule
        SPY/AGG (60/40)   & Fixed weights                  & 0.07 & 0.10 & 0.30 & 0.12 & 0.52 &  0.13 \\
        SPY               & Buy-and-hold vc8\%             & 0.06 & 0.09 & 0.20 & 0.29 & 0.53 & -0.03 \\
        SPY/AGG/GLD       & Equal weights                  & 0.08 & 0.09 & 0.24 & 0.17 & 0.68 &  0.49 \\
        SPY/TLT/GLD/IBIT  & Maximum diversification        & 0.10 & 0.09 & 0.28 & 0.42 & 0.90 &  0.54 \\
        Momentum Universe & Markowitz                      & 0.08 & 0.08 & 0.16 & 0.81 & 0.78 &  0.40 \\
        \midrule
        Full universe     & Regret bound\textcolor{red}{*} & 0.43 & 0.26 & 0.36 & 0.83 & 1.57 &  1.38 \\
        \bottomrule
    \end{tabular}
    \label{tab:final_portfolios_performance}
\end{table}

\newpage
\section{Conclusions}\label{sec:conclusion}
This paper studied portfolio construction from the perspective of the private investor. All methods were evaluated under the same setting: 59 liquid US ETFs, no shorting or leverage, monthly trading with a 5 basis point spread, an 8\% volatility target, and data from 2005--2025, with no parameters tuned to the evaluation period. Under these conditions, simple portfolios containing as few as three ETFs achieve consistent and decent returns, and the portfolios we propose outperform the 60/40 benchmark.

The clearest result concerns asset selection. The strongest simple portfolios combine growth-oriented equities, government bonds, and gold, while the exact allocation appears less important than including all three. An equally weighted portfolio of SPY, AGG, and GLD raises the Sharpe ratio from 0.52 for the 60/40 to 0.68 and reduces maximum drawdown from 30\% to 24\%, while requiring only one additional asset.

The simple strategies also perform well relative to more sophisticated alternatives. Fixed weights is difficult to improve upon for a well-chosen set of ETFs, while volatility control provides a direct and interpretable way to manage risk. The optimization-based strategies offer different advantages depending on the assets included. Minimum volatility is most suitable for complementary assets with comparable risk, while risk parity handles large differences in volatility more naturally. Maximum diversification is better suited to broader universes with more varied correlation structures. The strongest portfolio we propose uses maximum diversification across SPY, TLT, GLD, and IBIT, achieving a Sharpe ratio of 0.90.

The risk and alpha models lead to different conclusions. Risk forecasting proved useful, with the factor model producing the strongest covariance estimates, closely followed by simpler estimators. Return forecasting was demonstrated within the same simple and transparent framework, using only publicly available price data rather than proprietary information. In this setting, ridge regression never outperforms the zero forecast out of sample and improves only as regularization shrinks its predictions toward zero. This does not rule out richer return models or data sources, but highlights the difficulty of extracting predictive information from price history alone. The simpler momentum signal, when applied to a restricted universe, allows the Markowitz portfolio to achieve a Sharpe ratio of 0.78 and the smallest maximum drawdown among the proposed portfolios, although with substantially higher turnover. Markowitz also reduces the need to select individual ETFs in advance beyond broad asset categories.

These results should be interpreted with some considerations in mind. All performance is historical, the ETF universe is survivorship biased, and some use of prior knowledge when defining the universe is unavoidable. A strategy with perfect knowledge of future returns and yearly rebalancing achieves a Sharpe ratio of 1.57, as shown by the regret bound. The sensitivity analysis shows that the proposed portfolios do not depend on a single market crisis, although the relative performance of the 60/40 improves when the Financial Crisis is excluded. Finally, the framework relies entirely on public data, open-source code, and interpretable parameters, allowing the results to be reproduced and extended.

\clearpage
\section{Acknowledgments}
\label{sec:acknowledgements}

We would like to thank Professor Stephen P. Boyd for hosting us in his lab group at Stanford University, for supervising this work, and for teaching us the ways of simplicity. His guidance helped shape this project and his overall mentorship will remain invaluable to us for years to come. We are especially grateful for the many discussions we had with Stephen throughout the project, as well as with members of the group, whose suggestions, feedback, and ideas directed this work as it developed. We would especially like to acknowledge the members of the lab with whom we had the opportunity to work most closely. Thank you, Daniel Cederberg for your cheerful spirit and for carefully breaking down problems in an intuitive way, Nikhil Devanathan for your patience with our many questions and insightful and structured answers, Alexandros Tzikas for your ability to identify promising new directions and always checking in on us, and Maximilian Schaller for your profound insights and thorough illustrative explanations of optimization problems. We would also like to direct a thank you to the entirety of the CVX group and the people we met. Thank you Logan Bell, Jaewook Lee, Jimmy Tran, Clara Baynham, and Kasper Johansson. 

This work was supported by the foundations of the Royal Swedish Academy of Sciences (Kungl. Vetenskapsakademien), through Stiftelsen Claes Adelskölds medalj- och stipendiefond, grant ES2025-0064. We thank them for their generous contribution to our research and stay at Stanford University.

\vspace{0.5em}

\begin{figure}[h]
    \centering
    \includegraphics[width=0.9\linewidth]{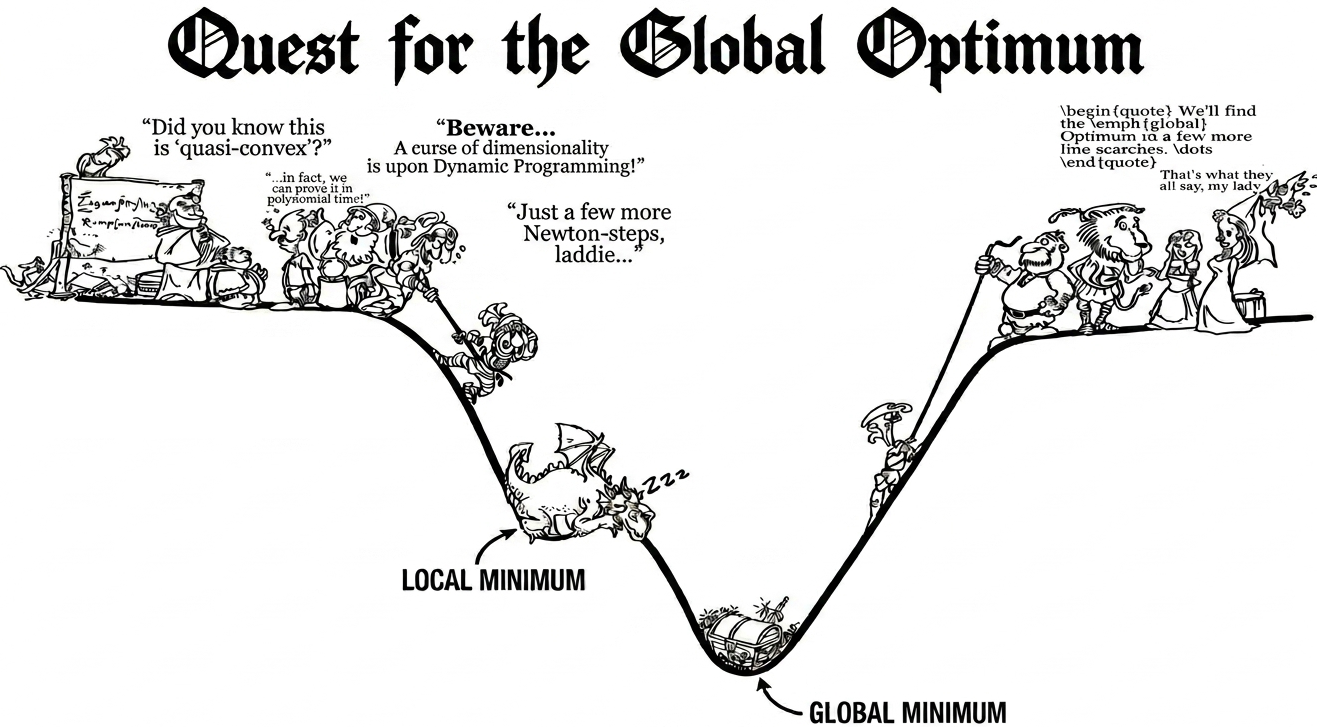}
    \caption{Our quest for the global optimum.}
    \label{fig:placeholder}
\end{figure}

\noindent It is difficult to capture an experience like this in just a few words. Instead, we turn to the hand-drawn illustration above, which hangs framed in the lab and, in many ways, captures the spirit of our time there. This past semester has truly been a journey, and we wish everyone the very best as they continue their own quest for the global optimum.

\newpage
\printbibliography

\newpage
\appendix
\section{Mathematical derivations}\label{app:derivations}

\subsection*{Derivation of the MDP QP formulation}
\label{app:mdp_qp_formulation}
From the original formulation of the MDP in (\ref{eq:max_div}), we rewrite $y = D_\sigma w $, where $\sigma^{2} = \operatorname{diag}\,(\Sigma)$ and $D_{\sigma} = \operatorname{diag}\,(\sigma)$. Substituting $w = D^{-1}_{\sigma}y$ gives, 

\begin{equation*}
    \frac{w^{\mathrm{T}} \sigma}{\sqrt{w^{\mathrm{T}} \Sigma w}}
    = \frac{(D_{\sigma}^{-1}y)^{\mathrm{T}} \sigma}{\sqrt{(D_{\sigma}^{-1}y)^{\mathrm{T}} \Sigma (D_{\sigma}^{-1}y)}}
    = \frac{\mathbf{1}^\mathrm{T} y}{\sqrt{y^{\mathrm{T}}(D_{\sigma}^{-1} \Sigma D_{\sigma}^{-1})y}} 
    = \frac{\mathbf{1}^\mathrm{T} y}{\sqrt{y^\mathrm{T} C y}},
\end{equation*}
and the constraint $\mathbf{1}^{\mathrm{T}}w = \mathbf{1}^{\mathrm{T}}(D_{\sigma}^{-1}y)$. The objective is scale invariant with respect to $w$, letting us scale the constraint $\mathbf{1}^{\mathrm{T}}w$ to become $\mathbf{1}^{\mathrm{T}}(D_{\sigma}w) = \mathbf{1}^{\mathrm{T}}(D_{\sigma}D_{\sigma}^{-1}y) = \mathbf{1}^{\mathrm{T}}y$ before solving for $y$ and then rescaling the solution back to $w$. We reformulate the optimization problem with the new constraints as, 
\begin{equation*}
\begin{aligned}
    & \underset{y}{\text{maximize}} && \frac{\mathbf{1}^{\mathrm{T}}y}{\sqrt{y^{\mathrm{T}} C y}} \\
    & \text{subject to}             && \mathbf{1}^{\mathrm{T}} y = 1, \quad y \geq 0.
\end{aligned}
\end{equation*}
The new constraint fixes the numerator, making this maximization problem equivalent to minimizing the denominator. This gives the final formulation,
\begin{equation*}
    \begin{aligned}
    & \underset{y}{\text{minimize}} && y^{\mathrm{T}} C y \\
    & \text{subject to}             && \mathbf{1}^{\mathrm{T}} y = 1, \quad y \geq 0.
\end{aligned}
\end{equation*}
After solving for $y^*$, undo the change of variables and normalize, 

\begin{equation*}
    w^* =  \frac{D^{-1}_\sigma y^*}{\mathbf{1}^{\mathrm{T}}D^{-1}_\sigma y^*}.
\end{equation*}

\newpage
\subsection*{Derivation of the factor model decomposition}
\label{app:factor_decomposition}

Consider the factor model
\[
    r = Ff + \epsilon,
\]
with $\mathbf{E}[f]=0$, $\mathbf{E}[\epsilon]=0$,
$\operatorname{Cov}(f,\epsilon)=0$, and
$\operatorname{Cov}(f)=I_k$. Taking the covariance gives
\begin{align*}
    \Sigma
    &= \operatorname{Cov}(Ff+\epsilon) \\
    &= F\operatorname{Cov}(f)F^{\mathrm{T}}
       + \operatorname{Cov}(\epsilon) \\
    &= FF^{\mathrm{T}} + D,
\end{align*}
where the cross-covariance terms vanish because
$\operatorname{Cov}(f,\epsilon)=0$. Assuming that the idiosyncratic returns
are mutually uncorrelated gives
\[
    D = \operatorname{Cov}(\epsilon)
      = \operatorname{diag}(\sigma_1^2,\ldots,\sigma_n^2).
\]
Hence,
\[
    \Sigma = FF^{\mathrm{T}} + D.
\]

\subsection*{Estimation of the statistical factor model}
\label{app:factor_estimation}

At each time $t$, the factor model is estimated from the historical return matrix $R_t$. Let
\[
    D_{\sigma,t}
    =
    \operatorname{diag}(\hat{\sigma}_{1,t},\ldots,\hat{\sigma}_{n,t})
\]
contain the EWMA volatility estimates with half-life $H_{\mathrm{var}}$. After winsorization, the returns are standardized as
\[
    \tilde{R}_t = R_tD_{\sigma,t}^{-1}.
\]
Using the EWMA weighting matrix $W_t$ with half-life $H_{\mathrm{cov}}$, the weighted second-moment matrix is
\[
    S_t = \tilde{R}_t^{\mathrm{T}}W_t\tilde{R}_t.
\]
Because the volatility and correlation estimates use separate EWMA half-lives, the diagonal of $S_t$ is not necessarily equal to one. We therefore normalize it to obtain the correlation matrix
\[
    C_t = D_{S,t}^{-1}S_tD_{S,t}^{-1},
    \qquad
    D_{S,t}=\operatorname{diag}(S_t)^{1/2}.
\]

\noindent The EM algorithm is initialized from the leading $k$ eigenpairs of
\[
    C_t = Q_t\Lambda_tQ_t^{\mathrm{T}}.
\]
Retaining the leading $k$ components gives
\[
    F_{0,t}=Q_{k,t}\Lambda_{k,t}^{1/2}, 
    \qquad
    D_{0,t} = \operatorname{diag}\left(C_t-F_{0,t}F_{0,t}^{\mathrm{T}}\right).
\]
The factor loading matrix is the rank-$k$ approximation of the correlation structure, while $D_{0,t}$ contains the residual diagonal variance. Starting from $(F_{0,t},D_{0,t})$, the EM updates in below produce the fitted correlation decomposition
\[
    C_t \approx F_t^*F_t^{*\mathrm{T}}+D_t^*.
\]
Finally, the decomposition is returned to the original volatility scale:
\[
    F_t=D_{\sigma,t}F_t^*,
    \qquad
    D_t=D_{\sigma,t}D_t^*D_{\sigma,t},
\]
which gives the covariance forecast
\[
     \hat{\Sigma}_t = F_tF_t^{\mathrm{T}}+D_t .
\]

\subsection*{Derivation of the EM updates for factor analysis}
\label{app:em_algo}

Given the correlation matrix $C_t$, we seek the maximum-likelihood estimates of $(F,D)$ under the factor model
\[
    r = Ff+\epsilon,
    \qquad
    f\sim\mathcal{N}(0,I_k),
    \qquad
    \epsilon\sim\mathcal{N}(0,D),
\]
where $D$ is diagonal. Since the factors $f$ are latent, we estimate $(F,D)$ using the EM algorithm~\cite{Dempster:1977,RubinThayer:1982}.

\paragraph{E-step.}
For a current estimate $(F^{(j)},D^{(j)})$, define
\[
    G^{(j)} = \left( I_k+ F^{(j)\mathrm{T}} D^{(j)-1} F^{(j)} \right)^{-1},
    \qquad 
    L^{(j)} = G^{(j)} F^{(j)\mathrm{T}} D^{(j)-1}.
\]
Under the joint Gaussian model, the conditional moments of the latent factors are
\[
    \mathbf{E}[f\mid r]=L^{(j)}r,
    \qquad
    \operatorname{Cov}(f\mid r)=G^{(j)}.
\]
Using the input correlation matrix $C_t$ from Appendix~\ref{app:factor_estimation}, the corresponding cross-moment and factor second moment are
\[
    M_{rf}^{(j)} = C_tL^{(j)\mathrm{T}},
    \qquad
    M_{ff}^{(j)} = L^{(j)}C_tL^{(j)\mathrm{T}}+G^{(j)}.
\]

\paragraph{M-step.}
Holding these conditional moments fixed, maximizing the expected complete-data log-likelihood gives
\[
    F^{(j+1)} = M_{rf}^{(j)}\left(M_{ff}^{(j)}\right)^{-1},
\]
followed by
\[
    D^{(j+1)} = \operatorname{diag}\!\left( C_t - 2M_{rf}^{(j)}F^{(j+1)\mathrm{T}} + F^{(j+1)}M_{ff}^{(j)}F^{(j+1)\mathrm{T}} \right).
\]
The diagonal restriction on $D$ allows each asset to have its own idiosyncratic variance while assuming that idiosyncratic returns are uncorrelated across assets.

\paragraph{Initialization and termination.}
The algorithm is initialized using the rank-$k$ eigendecomposition described in Appendix~\ref{app:factor_estimation}. The iterations continue until the relative changes in both $F$ and $D$ fall below a tolerance $\tau$,
\[
    \frac{\|F^{(j+1)}-F^{(j)}\|_F}{\|F^{(j)}\|_F}<\tau,
    \qquad
    \frac{\|D^{(j+1)}-D^{(j)}\|_F}{\|D^{(j)}\|_F}<\tau.
\]

\newpage

\section{Tables}

\begingroup
\setlength{\aboverulesep}{1pt}
\setlength{\belowrulesep}{1pt}
\setlength{\intextsep}{0pt}
\captionsetup{skip=4pt}

\begin{table}[H]
    \centering
    \scriptsize
    \setlength{\tabcolsep}{3pt}
    \caption{Individual ETFs.}
    \begin{tabular}{ll|SSS|SSSSS|lll}
        \toprule
            \textbf{Category} & 
            \textbf{ETF} & $\mathrm{\bm{\rho}}_{\textbf{SPY}}$ & $\mathrm{\bm{\rho}}_{\textbf{AGG}}$ & $\mathrm{\bm{\rho}}_{\textbf{GLD}}$ & 
            {$\boldsymbol{r}$} & {$\boldsymbol{\sigma}$} & {\textbf{MDD}} & \textbf{SR} & \textbf{LQ} & \textbf{Description} & \textbf{Issuer} & \textbf{Listed} \\
        \midrule
        Market & SPY   &  1.00 & -0.01 &  0.06 &  0.10 & 0.19 & 0.55 &  0.45 &  0.08 & S\&P 500             & SSGA      & 1993-01-22 \\
        Equity    & MDY   &  0.93 & -0.02 &  0.06 &  0.09 & 0.22 & 0.55 &  0.35 &  0.08 & S\&P MidCap 400      & SSGA      & 1995-05-04 \\
                  & QQQ   &  0.92 & -0.01 &  0.04 &  0.15 & 0.22 & 0.53 &  0.59 &  0.14 & NASDAQ 100           & Invesco   & 1999-03-10 \\
                  & IWM   &  0.89 & -0.04 &  0.06 &  0.08 & 0.24 & 0.59 &  0.25 & -0.18 & Russell 2000         & iShares   & 2000-05-29 \\
                  & VUG   &  0.96 &  0.01 &  0.07 &  0.12 & 0.20 & 0.51 &  0.51 &  0.26 & Growth index         & Vanguard  & 2004-02-01 \\
                  & VTV   &  0.96 & -0.06 &  0.05 &  0.09 & 0.19 & 0.59 &  0.36 &  0.01 & Value index          & Vanguard  & 2004-02-01 \\
                  & USMV  &  0.91 &  0.12 &  0.10 &  0.12 & 0.14 & 0.33 &  0.77 &  0.43 & MSCI min vol         & iShares   & 2011-10-20 \\
                  & SCHD  &  0.91 &  0.03 &  0.06 &  0.13 & 0.15 & 0.33 &  0.73 &  0.08 & Dividend equity      & Schwab    & 2011-10-20 \\
                  & QUAL  &  0.98 &  0.09 &  0.03 &  0.13 & 0.17 & 0.34 &  0.68 &  0.45 & MSCI quality         & iShares   & 2013-07-18 \\
        \midrule
        Sector & XLK   &  0.91 & -0.01 &  0.04 &  0.14 & 0.22 & 0.53 &  0.56 &  0.25 & Technology           & SSGA      & 1998-12-22 \\
        Equity     & XLP   &  0.77 &  0.01 &  0.03 &  0.09 & 0.14 & 0.32 &  0.52 &  0.25 & Consumer staples     & SSGA      & 1998-12-22 \\
                  & XLV   &  0.82 &  0.02 &  0.02 &  0.10 & 0.17 & 0.39 &  0.48 &  0.08 & Health care          & SSGA      & 1998-12-22 \\
                  & XLY   &  0.90 & -0.02 &  0.01 &  0.11 & 0.22 & 0.59 &  0.43 &  0.18 & Consumer discret.    & SSGA      & 1998-12-22 \\
                  & XLU   &  0.66 &  0.12 &  0.12 &  0.09 & 0.19 & 0.46 &  0.38 &  0.07 & Utilities            & SSGA      & 1998-12-22 \\
                  & XLI   &  0.91 & -0.04 &  0.05 &  0.10 & 0.21 & 0.62 &  0.38 &  0.01 & Industrial           & SSGA      & 1998-12-22 \\
                  & XLF   &  0.84 & -0.10 & -0.04 &  0.06 & 0.30 & 0.83 &  0.13 & -0.20 & Financial            & SSGA      & 1998-12-22 \\
                  & XLB   &  0.86 & -0.02 &  0.19 &  0.08 & 0.23 & 0.60 &  0.26 & -0.35 & Materials            & SSGA      & 1998-12-22 \\
                  & XLE   &  0.73 & -0.05 &  0.16 &  0.08 & 0.30 & 0.71 &  0.19 & -0.30 & Energy               & SSGA      & 1998-12-22 \\
                  & IYR   &  0.77 &  0.02 &  0.07 &  0.06 & 0.28 & 0.74 &  0.15 & -0.07 & Real estate          & iShares   & 2000-06-19 \\
                  & IBB   &  0.72 &  0.03 &  0.03 &  0.09 & 0.24 & 0.40 &  0.30 & -0.28 & Biotechnology        & iShares   & 2001-02-12 \\
        \midrule
        Country & EWJ   &  0.76 &  0.04 &  0.12 &  0.04 & 0.20 & 0.54 &  0.10 & -0.46 & MSCI Japan           & iShares   & 1996-03-12 \\
        Equity      & EWG   &  0.83 &  0.01 &  0.16 &  0.05 & 0.25 & 0.63 &  0.14 & -0.22 & MSCI Germany         & iShares   & 1996-03-12 \\
                  & EWU   &  0.84 &  0.00 &  0.16 &  0.04 & 0.23 & 0.64 &  0.08 & -0.28 & MSCI UK              & iShares   & 1996-03-12 \\
                  & EWA   &  0.81 &  0.04 &  0.20 &  0.07 & 0.27 & 0.67 &  0.17 & -0.31 & MSCI Australia       & iShares   & 1996-03-12 \\
                  & EWH   &  0.71 & -0.06 &  0.10 &  0.05 & 0.24 & 0.64 &  0.13 & -0.26 & MSCI Hong Kong       & iShares   & 1996-03-12 \\
                  & EWS   &  0.76 & -0.04 &  0.14 &  0.06 & 0.23 & 0.64 &  0.20 & -0.35 & MSCI Singapore       & iShares   & 1996-03-12 \\
                  & EWP   &  0.76 & -0.01 &  0.13 &  0.03 & 0.27 & 0.61 &  0.06 & -0.27 & MSCI Spain           & iShares   & 1996-03-12 \\
                  & EWW   &  0.74 &  0.01 &  0.15 &  0.05 & 0.28 & 0.65 &  0.13 & -0.46 & MSCI Mexico          & iShares   & 1996-03-12 \\
                  & EWI   &  0.77 & -0.01 &  0.13 &  0.01 & 0.28 & 0.70 & -0.01 & -0.42 & MSCI Italy           & iShares   & 1996-03-12 \\
                  & EWD   &  0.81 & -0.01 &  0.15 &  0.06 & 0.29 & 0.68 &  0.15 & -0.41 & MSCI Sweden          & iShares   & 1996-03-12 \\
                  & EWL   &  0.79 &  0.06 &  0.18 &  0.07 & 0.19 & 0.52 &  0.28 & -0.21 & MSCI Switzerland     & iShares   & 1996-03-12 \\
                  & EWC   &  0.80 &  0.04 &  0.28 &  0.06 & 0.22 & 0.61 &  0.21 & -0.06 & MSCI Canada          & iShares   & 1996-03-12 \\
                  & EWY   &  0.75 & -0.02 &  0.15 &  0.04 & 0.30 & 0.74 &  0.09 & -0.51 & MSCI South Korea     & iShares   & 2000-05-09 \\
                  & EWT   &  0.73 & -0.01 &  0.12 &  0.08 & 0.25 & 0.63 &  0.25 & -0.11 & MSCI Taiwan          & iShares   & 2000-06-23 \\
                  & EWZ   &  0.68 &  0.03 &  0.19 &  0.04 & 0.38 & 0.77 &  0.07 & -0.60 & MSCI Brazil          & iShares   & 2000-07-14 \\
        \midrule
        Government  & AGG   & -0.01 &  1.00 &  0.21 &  0.03 & 0.05 & 0.18 &  0.23 & -0.36 & US aggregate         & iShares   & 2003-09-29 \\
        FI          & TLT   & -0.32 &  0.70 &  0.17 &  0.03 & 0.15 & 0.48 &  0.10 & -0.41 & US treas. 20Y+       & iShares   & 2002-07-30 \\
                    & IEF   & -0.30 &  0.76 &  0.21 &  0.03 & 0.07 & 0.24 &  0.20 & -0.57 & US treas. 7-10Y      & iShares   & 2002-07-30 \\
                    & TIP   & -0.11 &  0.67 &  0.27 &  0.03 & 0.06 & 0.15 &  0.24 & -0.43 & US TIPS              & iShares   & 2003-12-07 \\
                    & MUB   &  0.13 &  0.49 &  0.11 &  0.03 & 0.05 & 0.14 &  0.34 & -0.15 & US municipal         & iShares   & 2007-09-10 \\
        \midrule
        Corporate   & LQD   &  0.19 &  0.70 &  0.15 &  0.04 & 0.09 & 0.25 &  0.25 & -0.31 & US iBoxx IG corp.    & iShares   & 2002-07-30 \\
        FI          & HYG   &  0.68 &  0.23 &  0.04 &  0.05 & 0.11 & 0.34 &  0.30 & -0.26 & US high yield corp.  & iShares   & 2007-04-11 \\
                    & LQDH  &  0.50 &  0.28 &  0.07 &  0.03 & 0.06 & 0.25 &  0.22 & -0.49 & Hedged IG corp.      & iShares   & 2014-06-17 \\
        \midrule
        Global      & EMB   &  0.38 &  0.40 &  0.18 &  0.04 & 0.11 & 0.35 &  0.26 & -0.18 & EM USD               & iShares   & 2007-12-19 \\
        FI          & VWOB  &  0.54 &  0.59 &  0.24 &  0.03 & 0.09 & 0.27 &  0.18 & -0.07 & EM govt.             & Vanguard  & 2013-06-04 \\
                    & BNDX  &  0.10 &  0.74 &  0.28 &  0.03 & 0.04 & 0.16 &  0.24 &  0.12 & Hedged intl.         & Vanguard  & 2013-06-04 \\
                    & IAGG  &  0.06 &  0.70 &  0.27 &  0.02 & 0.04 & 0.14 &  0.09 & -0.55 & Aggregate intl.      & iShares   & 2015-11-12 \\
        \midrule
        Physical & GLD  &  0.06 &  0.21 &  1.00 &  0.09 & 0.18 & 0.46 &  0.41 & -0.15 & Gold                 & SSGA      & 2004-11-19 \\
        Alt.        & SLV  &  0.23 &  0.17 &  0.80 &  0.03 & 0.31 & 0.76 &  0.06 & -0.47 & Silver               & iShares   & 2006-04-28 \\
                   & PPLT &  0.32 &  0.16 &  0.60 & -0.04 & 0.23 & 0.71 & -0.22 & -0.71 & Platinum             & abrdn     & 2010-01-08 \\
        \midrule
        Futures & USO  &  0.37 & -0.06 &  0.19 & -0.10 & 0.37 & 0.98 & -0.31 & -0.83 & Oil                  & USCF      & 2006-04-09 \\
        Alt.       & UGA  &  0.38 & -0.04 &  0.17 &  0.01 & 0.36 & 0.87 &  0.01 & -0.37 & Gasoline             & USCF      & 2008-02-28 \\
                   & CORN &  0.10 & -0.01 &  0.11 & -0.02 & 0.22 & 0.78 & -0.15 & -0.88 & Corn                 & Teucrium  & 2010-06-09 \\
                   & SOYB &  0.13 & -0.01 &  0.14 &  0.00 & 0.18 & 0.54 & -0.08 & -0.55 & Soybean              & Teucrium  & 2011-09-19 \\
                   & CANE &  0.12 &  0.00 &  0.10 & -0.06 & 0.24 & 0.81 & -0.28 & -0.79 & Sugar                & Teucrium  & 2011-09-19 \\
                   & WEAT &  0.03 &  0.00 &  0.10 & -0.11 & 0.27 & 0.82 & -0.46 & -1.03 & Wheat                & Teucrium  & 2011-09-19 \\
                   & CPER &  0.23 & -0.01 &  0.15 &  0.00 & 0.31 & 0.58 & -0.04 & -0.42 & Copper               & USCF      & 2011-11-15 \\
        \midrule
        Crypto & IBIT & 0.26 & 0.11 & 0.11 & 0.53 & 0.69 & 0.83 &  0.72 &  0.32 & Bitcoin   & iShares       & 2024-01-10 \\
        Alt.      & ETHA & 0.29 & 0.09 & 0.12 & 0.33 & 0.88 & 0.94 &  0.34 & -0.24 & Ethereum  & iShares       & 2024-07-23 \\
        \bottomrule
    \end{tabular} \\
    \label{tab:asset_universe}
\end{table}
\endgroup
\newpage

\begin{table}[H]
    \centering
    \scriptsize
    \setlength{\tabcolsep}{2pt}
    \caption{Equal weight portfolios ranked by Sharpe ratio.}
    \begin{tabular}{r|lSS|lSS|lSS} 
        \toprule
        \textbf{\#} & 
        \textbf{3 ETFs} & \textbf{SR} & \textbf{LQ} & 
        \textbf{4 ETFs} & \textbf{SR} & \textbf{LQ} & 
        \textbf{5 ETFs} & \textbf{SR} & \textbf{LQ} \\
        \midrule
        1  & QQQ IEF GLD & 0.81 &  0.44 & QQQ XLP TLT GLD & 0.84 & 0.62 & QQQ XLP IEF TIP GLD & 0.84 & 0.50 \\
        2  & XLK IEF GLD & 0.79 &  0.52 & QQQ XLP IEF GLD & 0.84 & 0.62 & QQQ XLP AGG IEF GLD & 0.84 & 0.55 \\
        3  & QQQ TIP GLD & 0.79 &  0.33 & XLK XLP TLT GLD & 0.82 & 0.59 & QQQ XLP TLT IEF GLD & 0.83 & 0.52 \\
        4  & QQQ XLP GLD & 0.78 &  0.70 & XLK XLP IEF GLD & 0.82 & 0.50 & XLK XLP IEF TIP GLD & 0.83 & 0.40 \\
        5  & QQQ AGG GLD & 0.78 &  0.46 & QQQ XLV TLT GLD & 0.81 & 0.51 & QQQ XLP TLT TIP GLD & 0.82 & 0.44 \\
        6  & QQQ TLT GLD & 0.78 &  0.36 & QQQ XLP TIP GLD & 0.81 & 0.60 & QQQ XLP AGG TLT GLD & 0.82 & 0.56 \\
        7  & XLK XLP GLD & 0.77 &  0.58 & QQQ XLV IEF GLD & 0.81 & 0.55 & XLK XLP AGG IEF GLD & 0.82 & 0.44 \\
        8  & XLK TIP GLD & 0.77 &  0.43 & QQQ XLP AGG GLD & 0.80 & 0.65 & QQQ XLP IEF LQD GLD & 0.82 & 0.48 \\
        9  & QQQ LQD GLD & 0.76 &  0.45 & XLK XLV TLT GLD & 0.80 & 0.47 & QQQ XLP AGG TIP GLD & 0.82 & 0.54 \\
        10 & XLK TLT GLD & 0.76 &  0.44 & QQQ AGG IEF GLD & 0.80 & 0.41 & XLK XLP TLT IEF GLD & 0.82 & 0.44 \\
        11 & XLK AGG GLD & 0.76 &  0.51 & QQQ IEF TIP GLD & 0.80 & 0.29 & QQQ XLV XLP TLT GLD & 0.81 & 0.65 \\
        12 & QQQ XLV GLD & 0.75 &  0.57 & QQQ XLK TLT GLD & 0.80 & 0.50 & QQQ XLV IEF TIP GLD & 0.81 & 0.53 \\
        13 & XLK LQD GLD & 0.75 &  0.40 & XLK XLV IEF GLD & 0.79 & 0.71 & QQQ XLV TLT IEF GLD & 0.81 & 0.41 \\
        14 & VUG IEF GLD & 0.74 &  0.44 & XLK XLP TIP GLD & 0.79 & 0.48 & QQQ XLK XLP TLT GLD & 0.81 & 0.68 \\
        15 & XLK XLV GLD & 0.74 &  0.73 & QQQ XLP LQD GLD & 0.79 & 0.62 & XLK XLP TLT TIP GLD & 0.81 & 0.47 \\
        16 & QQQ IEF TIP & 0.74 &  0.31 & XLK XLP AGG GLD & 0.79 & 0.53 & QQQ XLK TLT IEF GLD & 0.81 & 0.41 \\
        17 & QQQ XLP TLT & 0.74 &  0.40 & QQQ XLK IEF GLD & 0.79 & 0.49 & XLK XLP AGG TLT GLD & 0.81 & 0.47 \\
        18 & QQQ XLK GLD & 0.73 &  0.46 & XLK AGG IEF GLD & 0.78 & 0.46 & QQQ XLV AGG IEF GLD & 0.81 & 0.55 \\
        19 & QQQ XLU GLD & 0.73 &  0.72 & XLK IEF TIP GLD & 0.78 & 0.38 & QQQ XLV TLT TIP GLD & 0.80 & 0.35 \\
        20 & VUG XLP GLD & 0.73 &  0.65 & QQQ AGG TIP GLD & 0.78 & 0.31 & XLK XLV XLP TLT GLD & 0.80 & 0.70 \\
        21 & QQQ XLP IEF & 0.72 &  0.49 & QQQ IEF LQD GLD & 0.78 & 0.39 & XLK XLV IEF TIP GLD & 0.80 & 0.51 \\
        22 & QQQ AGG IEF & 0.72 &  0.29 & VUG XLP TLT GLD & 0.78 & 0.58 & XLK XLV TLT IEF GLD & 0.80 & 0.29 \\
        23 & XLP IEF GLD & 0.72 &  0.06 & VUG XLP IEF GLD & 0.78 & 0.69 & XLK XLP IEF LQD GLD & 0.80 & 0.37 \\
        24 & VUG TIP GLD & 0.72 &  0.33 & QQQ XLV TIP GLD & 0.78 & 0.58 & QQQ XLP TLT LQD GLD & 0.80 & 0.56 \\
        25 & XLV IEF GLD & 0.71 &  0.14 & QQQ VUG TLT GLD & 0.78 & 0.49 & QQQ XLK IEF TIP GLD & 0.80 & 0.39 \\
        26 & XLK XLP TLT & 0.71 &  0.38 & QQQ XLU IEF GLD & 0.77 & 0.74 & XLK XLP AGG TIP GLD & 0.80 & 0.42 \\
        27 & XLV XLP GLD & 0.71 &  0.36 & XLK XLP LQD GLD & 0.77 & 0.46 & QQQ XLV AGG TLT GLD & 0.80 & 0.47 \\
        28 & VUG TLT GLD & 0.71 &  0.36 & QQQ XLV AGG GLD & 0.77 & 0.59 & QQQ XLK TLT TIP GLD & 0.80 & 0.35 \\
        29 & VUG AGG GLD & 0.71 &  0.48 & QQQ XLU TLT GLD & 0.77 & 0.61 & QQQ XLP TIP LQD GLD & 0.80 & 0.47 \\
        30 & XLY IEF GLD & 0.71 &  0.27 & XLK XLV TIP GLD & 0.77 & 0.63 & XLK XLV AGG IEF GLD & 0.80 & 0.60 \\
        31 & XLK XLU GLD & 0.71 &  0.65 & XLK AGG TIP GLD & 0.77 & 0.40 & QQQ XLK AGG TLT GLD & 0.80 & 0.45 \\
        32 & QQQ VUG GLD & 0.71 &  0.56 & XLK IEF LQD GLD & 0.77 & 0.36 & QQQ XLV XLP IEF GLD & 0.80 & 0.61 \\
        33 & XLK IEF TIP & 0.71 &  0.19 & QQQ TIP LQD GLD & 0.77 & 0.30 & XLK XLV TLT TIP GLD & 0.80 & 0.32 \\
        34 & SPY IEF GLD & 0.70 &  0.57 & VUG XLK TLT GLD & 0.77 & 0.54 & QQQ XLK AGG IEF GLD & 0.80 & 0.43 \\
        35 & XLP AGG GLD & 0.70 &  0.05 & SPY QQQ TLT GLD & 0.77 & 0.63 & QQQ XLV IEF LQD GLD & 0.79 & 0.63 \\
        36 & QQQ XLV TLT & 0.70 &  0.18 & QQQ VUG IEF GLD & 0.76 & 0.63 & QQQ XLK XLP IEF GLD & 0.79 & 0.67 \\
        37 & XLK XLP IEF & 0.70 &  0.60 & QQQ XLK TIP GLD & 0.76 & 0.46 & XLK XLV AGG TLT GLD & 0.79 & 0.36 \\
        38 & XLP TIP GLD & 0.70 & -0.03 & XLV XLP IEF GLD & 0.76 & 0.37 & QQQ VUG XLP TLT GLD & 0.79 & 0.73 \\
        39 & VUG XLK GLD & 0.70 &  0.55 & XLK XLV AGG GLD & 0.76 & 0.77 & QQQ XLK XLV TLT GLD & 0.79 & 0.65 \\
        40 & VUG LQD GLD & 0.70 &  0.49 & QQQ XLI TLT GLD & 0.76 & 0.62 & QQQ XLP AGG LQD GLD & 0.79 & 0.54 \\
        41 & VUG XLV GLD & 0.70 &  0.53 & QQQ XLY TLT GLD & 0.76 & 0.53 & XLK XLP TLT LQD GLD & 0.79 & 0.43 \\
        42 & QQQ AGG TIP & 0.70 &  0.33 & VUG XLV TLT GLD & 0.76 & 0.51 & QQQ VUG TLT IEF GLD & 0.79 & 0.38 \\
        43 & XLY XLP GLD & 0.70 &  0.30 & QQQ XLV LQD GLD & 0.76 & 0.65 & QQQ XLV AGG TIP GLD & 0.79 & 0.61 \\
        44 & XLK AGG IEF & 0.69 &  0.23 & QQQ AGG LQD GLD & 0.76 & 0.42 & XLK XLV XLP IEF GLD & 0.79 & 0.71 \\
        45 & QQQ XLP TIP & 0.69 &  0.45 & QQQ XLK AGG GLD & 0.76 & 0.52 & VUG XLP IEF TIP GLD & 0.79 & 0.56 \\
        46 & XLV TIP GLD & 0.69 &  0.12 & XLV XLP TLT GLD & 0.76 & 0.25 & XLK XLP TIP LQD GLD & 0.78 & 0.35 \\
        47 & SPY QQQ GLD & 0.69 &  0.62 & XLK XLU IEF GLD & 0.76 & 0.73 & QQQ XLV TLT LQD GLD & 0.78 & 0.46 \\
        48 & QQQ XLY GLD & 0.69 &  0.36 & VUG XLP TIP GLD & 0.76 & 0.68 & XLK XLV IEF LQD GLD & 0.78 & 0.57 \\
        49 & XLP LQD GLD & 0.69 &  0.14 & QQQ AGG TLT GLD & 0.76 & 0.33 & QQQ XLK IEF LQD GLD & 0.78 & 0.39 \\
        50 & XLV AGG GLD & 0.69 &  0.20 & XLK XLU TLT GLD & 0.75 & 0.59 & QQQ XLK TLT LQD GLD & 0.78 & 0.45 \\
        51 & XLK XLV TLT & 0.69 &  0.31 & XLK XLY TLT GLD & 0.75 & 0.40 & VUG XLP AGG IEF GLD & 0.78 & 0.65 \\
        52 & QQQ IEF LQD & 0.69 &  0.26 & VUG XLK IEF GLD & 0.75 & 0.59 & VUG XLK XLP TLT GLD & 0.78 & 0.72 \\
        53 & QQQ XLK TLT & 0.69 &  0.37 & QQQ VTV TLT GLD & 0.75 & 0.75 & QQQ XLP XLU TLT GLD & 0.78 & 0.58 \\
        54 & XLK XLY GLD & 0.69 &  0.29 & XLY XLP TLT GLD & 0.75 & 0.47 & QQQ VUG IEF TIP GLD & 0.78 & 0.41 \\
        55 & XLY TIP GLD & 0.68 &  0.24 & VUG XLV IEF GLD & 0.75 & 0.61 & QQQ AGG IEF TIP GLD & 0.78 & 0.26 \\
        56 & XLY TLT GLD & 0.68 &  0.24 & QQQ TLT TIP GLD & 0.75 & 0.22 & XLK XLV AGG TIP GLD & 0.78 & 0.57 \\
        57 & SPY XLP GLD & 0.68 &  0.66 & XLY XLP IEF GLD & 0.75 & 0.44 & QQQ VUG TLT TIP GLD & 0.78 & 0.32 \\
        58 & XLV LQD GLD & 0.68 &  0.30 & SPY XLK TLT GLD & 0.75 & 0.66 & QQQ XLU IEF TIP GLD & 0.78 & 0.54 \\
        59 & QQQ XLI GLD & 0.68 &  0.54 & QQQ XLU TIP GLD & 0.75 & 0.65 & QQQ XLK XLU TLT GLD & 0.78 & 0.76 \\
        60 & QQQ XLV IEF & 0.68 &  0.18 & QQQ XLP IEF TIP & 0.75 & 0.47 & VUG XLK TLT IEF GLD & 0.78 & 0.44 \\
        \bottomrule
    \end{tabular} \\
    \label{tab:exhaustive_search_equal_weights}
\end{table}
\newpage

\begin{table}[H]
    \centering
    \scriptsize
    \setlength{\tabcolsep}{2pt}
    \caption{Fixed weight splits of SPY/AGG/GLD sorted by SR.}
    \begin{tabular}{lSSSSS|lSSSSS}
        \toprule
        \multicolumn{6}{c|}{\textbf{SPY/AGG/GLD}} & \multicolumn{6}{c}{\textbf{SPY/AGG/GLD vc8}} \\
        \midrule
        \textbf{Weights} & {$\boldsymbol{r}$} & {$\boldsymbol{\sigma}$} & {\textbf{MDD}} &  \textbf{SR} & \textbf{LQ} &
        \textbf{Weights} & {$\boldsymbol{r}$} & {$\boldsymbol{\sigma}$} & {\textbf{MDD}} &  \textbf{SR} & \textbf{LQ} \\
        \midrule
        30/40/30 & 0.07 & 0.08 & 0.22 & 0.67 &  0.56 & 30/30/40 & 0.07 & 0.08 & 0.15 & 0.67 &  0.40 \\
        30/50/20 & 0.07 & 0.07 & 0.20 & 0.67 &  0.62 & 40/20/40 & 0.07 & 0.08 & 0.17 & 0.67 &  0.26 \\
        40/30/30 & 0.08 & 0.10 & 0.26 & 0.67 &  0.51 & 40/30/30 & 0.07 & 0.08 & 0.18 & 0.66 &  0.33 \\
        40/20/40 & 0.09 & 0.11 & 0.28 & 0.66 &  0.39 & 50/00/50 & 0.07 & 0.08 & 0.18 & 0.66 &  0.18 \\
        30/30/40 & 0.08 & 0.10 & 0.24 & 0.66 &  0.42 & 50/10/40 & 0.07 & 0.08 & 0.19 & 0.66 &  0.20 \\
        40/40/20 & 0.08 & 0.09 & 0.25 & 0.65 &  0.66 & 40/10/50 & 0.07 & 0.08 & 0.17 & 0.66 &  0.23 \\
        50/10/40 & 0.10 & 0.12 & 0.32 & 0.65 &  0.37 & 30/20/50 & 0.07 & 0.08 & 0.15 & 0.66 &  0.18 \\
        20/50/30 & 0.07 & 0.08 & 0.18 & 0.65 &  0.32 & 30/40/30 & 0.07 & 0.07 & 0.17 & 0.66 &  0.45 \\
        20/60/20 & 0.06 & 0.06 & 0.17 & 0.65 &  0.42 & 60/00/40 & 0.07 & 0.09 & 0.20 & 0.65 &  0.25 \\
        40/10/50 & 0.10 & 0.12 & 0.30 & 0.65 &  0.30 & 40/40/20 & 0.07 & 0.07 & 0.19 & 0.65 &  0.54 \\
        50/20/30 & 0.09 & 0.11 & 0.30 & 0.65 &  0.48 & 40/00/60 & 0.07 & 0.08 & 0.17 & 0.65 &  0.07 \\
        50/00/50 & 0.10 & 0.13 & 0.33 & 0.64 &  0.28 & 50/20/30 & 0.07 & 0.08 & 0.19 & 0.65 &  0.37 \\
        30/20/50 & 0.09 & 0.11 & 0.26 & 0.64 &  0.27 & 60/10/30 & 0.07 & 0.08 & 0.20 & 0.65 &  0.39 \\
        60/00/40 & 0.10 & 0.14 & 0.36 & 0.63 &  0.36 & 30/50/20 & 0.06 & 0.07 & 0.18 & 0.64 &  0.56 \\
        20/40/40 & 0.07 & 0.09 & 0.20 & 0.63 &  0.24 & 20/40/40 & 0.07 & 0.08 & 0.15 & 0.64 &  0.22 \\
        40/00/60 & 0.10 & 0.13 & 0.31 & 0.63 &  0.19 & 30/10/60 & 0.07 & 0.08 & 0.15 & 0.64 &  0.01 \\
        50/30/20 & 0.08 & 0.10 & 0.29 & 0.63 &  0.54 & 20/30/50 & 0.07 & 0.08 & 0.15 & 0.64 &  0.14 \\
        30/60/10 & 0.06 & 0.07 & 0.20 & 0.63 &  0.58 & 70/00/30 & 0.07 & 0.09 & 0.20 & 0.63 &  0.32 \\
        60/10/30 & 0.10 & 0.13 & 0.35 & 0.62 &  0.54 & 60/20/20 & 0.07 & 0.08 & 0.20 & 0.63 &  0.30 \\
        30/10/60 & 0.09 & 0.12 & 0.28 & 0.61 &  0.16 & 50/30/20 & 0.07 & 0.08 & 0.19 & 0.63 &  0.42 \\
        40/50/10 & 0.07 & 0.08 & 0.24 & 0.61 &  0.47 & 20/50/30 & 0.06 & 0.07 & 0.16 & 0.63 &  0.28 \\
        20/70/10 & 0.05 & 0.06 & 0.17 & 0.61 &  0.36 & 30/00/70 & 0.07 & 0.08 & 0.17 & 0.62 & -0.10 \\
        20/30/50 & 0.08 & 0.10 & 0.22 & 0.61 &  0.18 & 70/10/20 & 0.07 & 0.09 & 0.20 & 0.62 &  0.23 \\
        60/20/20 & 0.09 & 0.12 & 0.34 & 0.60 &  0.44 & 40/50/10 & 0.06 & 0.07 & 0.18 & 0.62 &  0.41 \\
        70/00/30 & 0.10 & 0.14 & 0.39 & 0.60 &  0.45 & 20/20/60 & 0.07 & 0.08 & 0.18 & 0.61 & -0.01 \\
        30/00/70 & 0.10 & 0.14 & 0.30 & 0.59 &  0.07 & 20/60/20 & 0.05 & 0.06 & 0.17 & 0.60 &  0.42 \\
        50/40/10 & 0.08 & 0.10 & 0.29 & 0.59 &  0.41 & 80/00/20 & 0.07 & 0.09 & 0.21 & 0.60 &  0.18 \\
        20/20/60 & 0.09 & 0.12 & 0.25 & 0.58 &  0.14 & 30/60/10 & 0.05 & 0.06 & 0.18 & 0.60 &  0.58 \\
        10/60/30 & 0.06 & 0.07 & 0.16 & 0.58 &  0.06 & 60/30/10 & 0.07 & 0.08 & 0.19 & 0.60 &  0.21 \\
        70/10/20 & 0.10 & 0.14 & 0.40 & 0.57 &  0.36 & 20/10/70 & 0.07 & 0.08 & 0.20 & 0.59 & -0.11 \\
        10/70/20 & 0.05 & 0.06 & 0.16 & 0.57 &  0.10 & 50/40/10 & 0.06 & 0.08 & 0.19 & 0.59 &  0.29 \\
        10/50/40 & 0.06 & 0.08 & 0.17 & 0.57 &  0.02 & 70/20/10 & 0.07 & 0.08 & 0.20 & 0.59 &  0.14 \\
        60/30/10 & 0.08 & 0.12 & 0.35 & 0.56 &  0.35 & 20/00/80 & 0.07 & 0.08 & 0.21 & 0.58 & -0.16 \\
        20/10/70 & 0.09 & 0.13 & 0.27 & 0.56 &  0.06 & 10/40/50 & 0.06 & 0.08 & 0.17 & 0.58 &  0.02 \\
        80/00/20 & 0.10 & 0.16 & 0.45 & 0.55 &  0.30 & 80/10/10 & 0.07 & 0.09 & 0.20 & 0.58 &  0.09 \\
        10/40/50 & 0.07 & 0.10 & 0.20 & 0.55 &  0.00 & 10/50/40 & 0.06 & 0.07 & 0.16 & 0.57 & -0.01 \\
        70/20/10 & 0.09 & 0.13 & 0.40 & 0.54 &  0.26 & 90/00/10 & 0.07 & 0.09 & 0.20 & 0.57 &  0.06 \\
        30/70/00 & 0.05 & 0.07 & 0.20 & 0.54 &  0.24 & 20/70/10 & 0.05 & 0.06 & 0.17 & 0.57 &  0.36 \\
        40/60/00 & 0.06 & 0.08 & 0.24 & 0.54 &  0.18 & 10/30/60 & 0.06 & 0.08 & 0.20 & 0.56 & -0.01 \\
        20/00/80 & 0.10 & 0.15 & 0.30 & 0.53 &  0.00 & 10/60/30 & 0.05 & 0.07 & 0.16 & 0.55 & -0.02 \\
        10/30/60 & 0.08 & 0.11 & 0.23 & 0.53 & -0.02 & 40/60/00 & 0.05 & 0.07 & 0.18 & 0.55 &  0.18 \\
        50/50/00 & 0.07 & 0.10 & 0.30 & 0.52 &  0.13 & 60/40/00 & 0.06 & 0.08 & 0.19 & 0.55 &  0.07 \\
        10/80/10 & 0.04 & 0.05 & 0.17 & 0.52 &  0.12 & 70/30/00 & 0.06 & 0.08 & 0.20 & 0.55 &  0.03 \\
        80/10/10 & 0.10 & 0.15 & 0.45 & 0.52 &  0.20 & 10/20/70 & 0.06 & 0.08 & 0.22 & 0.54 & -0.05 \\
        10/20/70 & 0.08 & 0.13 & 0.28 & 0.51 & -0.04 & 80/20/00 & 0.06 & 0.09 & 0.20 & 0.54 &  0.00 \\
        60/40/00 & 0.08 & 0.11 & 0.36 & 0.51 &  0.14 & 50/50/00 & 0.06 & 0.08 & 0.19 & 0.54 &  0.11 \\
        20/80/00 & 0.05 & 0.06 & 0.18 & 0.50 &  0.16 & 10/10/80 & 0.06 & 0.08 & 0.23 & 0.53 & -0.06 \\
        90/00/10 & 0.10 & 0.17 & 0.50 & 0.50 &  0.16 & 90/10/00 & 0.06 & 0.09 & 0.20 & 0.53 & -0.02 \\
        70/30/00 & 0.08 & 0.13 & 0.41 & 0.49 &  0.15 & 30/70/00 & 0.05 & 0.06 & 0.18 & 0.53 &  0.24 \\
        10/10/80 & 0.09 & 0.14 & 0.33 & 0.49 & -0.05 & 10/70/20 & 0.05 & 0.06 & 0.16 & 0.53 &  0.00 \\
        80/20/00 & 0.09 & 0.15 & 0.46 & 0.48 &  0.15 & 10/00/90 & 0.06 & 0.08 & 0.24 & 0.52 & -0.07 \\
        10/00/90 & 0.09 & 0.16 & 0.38 & 0.47 & -0.07 & 00/50/50 & 0.06 & 0.08 & 0.20 & 0.49 &  0.00 \\
        00/60/40 & 0.06 & 0.08 & 0.19 & 0.47 &  0.16 & 00/40/60 & 0.06 & 0.08 & 0.23 & 0.49 & -0.13 \\
        00/50/50 & 0.06 & 0.10 & 0.21 & 0.47 &  0.09 & 00/30/70 & 0.06 & 0.08 & 0.25 & 0.48 & -0.19 \\
        00/70/30 & 0.05 & 0.07 & 0.19 & 0.46 &  0.15 & 10/80/10 & 0.04 & 0.05 & 0.17 & 0.48 &  0.06 \\
        90/10/00 & 0.10 & 0.17 & 0.51 & 0.46 &  0.11 & 00/20/80 & 0.06 & 0.08 & 0.26 & 0.47 & -0.18 \\
        00/40/60 & 0.07 & 0.11 & 0.27 & 0.46 &  0.00 & 00/60/40 & 0.05 & 0.07 & 0.18 & 0.47 &  0.08 \\
        00/30/70 & 0.07 & 0.13 & 0.32 & 0.45 & -0.06 & 20/80/00 & 0.04 & 0.05 & 0.18 & 0.47 &  0.16 \\
        00/20/80 & 0.08 & 0.14 & 0.37 & 0.44 & -0.12 & 00/10/90 & 0.06 & 0.08 & 0.27 & 0.47 & -0.15 \\
        00/80/20 & 0.04 & 0.06 & 0.19 & 0.43 &  0.02 & 00/70/30 & 0.05 & 0.07 & 0.19 & 0.43 &  0.00 \\
        00/10/90 & 0.09 & 0.16 & 0.41 & 0.42 & -0.14 & 00/80/20 & 0.04 & 0.06 & 0.19 & 0.38 & -0.06 \\
        10/90/00 & 0.04 & 0.05 & 0.18 & 0.40 &  0.05 & 10/90/00 & 0.03 & 0.05 & 0.17 & 0.36 & -0.03 \\
        00/90/10 & 0.04 & 0.05 & 0.19 & 0.36 & -0.20 & 00/90/10 & 0.03 & 0.05 & 0.19 & 0.30 & -0.20 \\
        \bottomrule
    \end{tabular} \\
    \label{tab:optimal_split_sbg}
\end{table}

\begin{table}[H]
    \centering
    \scriptsize
    \setlength{\tabcolsep}{2pt}
    \caption{Inverse volatility portfolios ranked by Sharpe ratio}
    \begin{tabular}{r|lSS|lSS|lSS}
        \toprule
        \textbf{\#} & 
        \textbf{3 ETFs} & \textbf{SR} & \textbf{LQ} & 
        \textbf{4 ETFs} & \textbf{SR} & \textbf{LQ} & 
        \textbf{5 ETFs} & \textbf{SR} & \textbf{LQ} \\
        \midrule
        1  & QQQ IEF GLD & 0.84 &  0.21 & QQQ XLP TLT GLD & 0.91 & 0.70 & QQQ XLK XLP TLT GLD & 0.90 & 0.68 \\
        2  & QQQ TLT GLD & 0.84 &  0.37 & QQQ XLP IEF GLD & 0.90 & 0.66 & QQQ XLK XLP IEF GLD & 0.88 & 0.60 \\
        3  & XLK IEF GLD & 0.82 &  0.29 & QQQ XLK TLT GLD & 0.90 & 0.50 & QQQ XLK XLV TLT GLD & 0.88 & 0.63 \\
        4  & XLK TLT GLD & 0.81 &  0.46 & QQQ XLV TLT GLD & 0.89 & 0.53 & QQQ XLV XLP TLT GLD & 0.88 & 0.62 \\
        5  & QQQ XLP TLT & 0.80 &  0.41 & QQQ XLK IEF GLD & 0.89 & 0.49 & QQQ VUG XLP TLT GLD & 0.87 & 0.77 \\
        6  & QQQ XLP GLD & 0.80 &  0.76 & XLK XLP TLT GLD & 0.88 & 0.73 & QQQ XLK XLU TLT GLD & 0.87 & 0.78 \\
        7  & VUG IEF GLD & 0.79 &  0.22 & XLK XLV TLT GLD & 0.88 & 0.51 & QQQ XLY XLP TLT GLD & 0.87 & 0.67 \\
        8  & QQQ XLP IEF & 0.78 &  0.45 & XLK XLP IEF GLD & 0.88 & 0.54 & QQQ XLK XLV IEF GLD & 0.86 & 0.52 \\
        9  & QQQ XLV TLT & 0.78 &  0.28 & QQQ XLV IEF GLD & 0.88 & 0.56 & SPY QQQ XLP TLT GLD & 0.86 & 0.83 \\
        10 & VUG TLT GLD & 0.78 &  0.40 & SPY QQQ TLT GLD & 0.88 & 0.72 & XLK XLV XLP TLT GLD & 0.86 & 0.61 \\
        11 & SPY IEF GLD & 0.78 &  0.41 & QQQ VUG TLT GLD & 0.87 & 0.63 & QQQ VUG XLP IEF GLD & 0.86 & 0.60 \\
        12 & QQQ AGG GLD & 0.78 &  0.24 & QQQ VTV TLT GLD & 0.87 & 0.85 & VUG XLK XLP TLT GLD & 0.86 & 0.76 \\
        13 & XLK XLP GLD & 0.78 &  0.73 & QQQ XLY TLT GLD & 0.87 & 0.64 & QQQ XLI XLP TLT GLD & 0.86 & 0.77 \\
        14 & QQQ TIP GLD & 0.77 & -0.06 & QQQ VUG IEF GLD & 0.87 & 0.60 & QQQ XLV XLP IEF GLD & 0.86 & 0.61 \\
        15 & SPY TLT GLD & 0.77 &  0.41 & QQQ XLI TLT GLD & 0.86 & 0.76 & XLK XLY XLP TLT GLD & 0.86 & 0.61 \\
        16 & QQQ XLK TLT & 0.77 &  0.35 & SPY QQQ IEF GLD & 0.86 & 0.68 & QQQ VUG XLV TLT GLD & 0.86 & 0.65 \\
        17 & XLK XLP TLT & 0.77 &  0.37 & XLK XLV IEF GLD & 0.86 & 0.52 & QQQ XLK XLU IEF GLD & 0.86 & 0.77 \\
        18 & QQQ XLV GLD & 0.77 &  0.61 & VUG XLK TLT GLD & 0.86 & 0.53 & QQQ XLV XLY TLT GLD & 0.86 & 0.58 \\
        19 & QQQ XLK IEF & 0.76 &  0.37 & QQQ VTV IEF GLD & 0.86 & 0.82 & QQQ XLY XLP IEF GLD & 0.85 & 0.71 \\
        20 & XLK XLV TLT & 0.76 &  0.30 & XLK XLY TLT GLD & 0.86 & 0.57 & QQQ VTV XLP TLT GLD & 0.85 & 0.86 \\
        21 & QQQ XLK GLD & 0.76 &  0.39 & QQQ XLY IEF GLD & 0.86 & 0.54 & QQQ VTV XLK TLT GLD & 0.85 & 0.83 \\
        22 & QQQ XLU GLD & 0.76 &  0.85 & SPY XLK TLT GLD & 0.86 & 0.74 & QQQ XLK XLY TLT GLD & 0.85 & 0.58 \\
        23 & XLK AGG GLD & 0.76 &  0.27 & VUG XLP TLT GLD & 0.85 & 0.69 & SPY QQQ XLP IEF GLD & 0.85 & 0.75 \\
        24 & XLK XLP IEF & 0.76 &  0.44 & QQQ XLI IEF GLD & 0.85 & 0.63 & QQQ XLK XLI TLT GLD & 0.85 & 0.63 \\
        25 & XLY IEF GLD & 0.76 &  0.36 & VUG XLK IEF GLD & 0.85 & 0.51 & QQQ VUG XLK TLT GLD & 0.85 & 0.64 \\
        26 & QQQ XLV IEF & 0.75 &  0.23 & VTV XLK TLT GLD & 0.85 & 0.76 & SPY QQQ XLK TLT GLD & 0.85 & 0.67 \\
        27 & SPY QQQ TLT & 0.75 &  0.36 & QQQ XLU TLT GLD & 0.85 & 0.58 & VUG XLK XLP IEF GLD & 0.85 & 0.68 \\
        28 & QQQ VTV TLT & 0.75 &  0.48 & SPY XLK IEF GLD & 0.84 & 0.59 & SPY XLK XLP TLT GLD & 0.85 & 0.92 \\
        29 & XLY TLT GLD & 0.75 &  0.45 & VUG XLP IEF GLD & 0.84 & 0.68 & XLK XLV XLY TLT GLD & 0.85 & 0.53 \\
        30 & XLK XLV GLD & 0.75 &  0.73 & XLK XLY IEF GLD & 0.84 & 0.54 & QQQ VUG XLU TLT GLD & 0.85 & 0.86 \\
        31 & XLK TIP GLD & 0.75 &  0.04 & XLK XLI TLT GLD & 0.84 & 0.70 & SPY QQQ XLV TLT GLD & 0.85 & 0.81 \\
        32 & QQQ VUG TLT & 0.75 &  0.40 & QQQ XLU IEF GLD & 0.84 & 0.55 & XLK XLV XLP IEF GLD & 0.85 & 0.61 \\
        33 & QQQ XLI TLT & 0.75 &  0.30 & VUG XLV TLT GLD & 0.84 & 0.61 & VUG XLK XLV TLT GLD & 0.85 & 0.74 \\
        34 & XLV TLT GLD & 0.74 &  0.20 & VTV XLK IEF GLD & 0.84 & 0.70 & QQQ XLV XLU TLT GLD & 0.84 & 0.65 \\
        35 & QQQ VUG IEF & 0.74 &  0.36 & XLY XLP TLT GLD & 0.83 & 0.59 & QQQ XLV XLI TLT GLD & 0.84 & 0.68 \\
        36 & VUG XLP GLD & 0.74 &  0.81 & QQQ XLP TIP GLD & 0.83 & 0.42 & QQQ XLI XLP IEF GLD & 0.84 & 0.69 \\
        37 & SPY QQQ IEF & 0.74 &  0.36 & XLV XLY TLT GLD & 0.83 & 0.53 & QQQ XLP XLU TLT GLD & 0.84 & 0.73 \\
        38 & QQQ VUG GLD & 0.74 &  0.51 & SPY XLP TLT GLD & 0.83 & 0.57 & XLK XLY XLP IEF GLD & 0.84 & 0.72 \\
        39 & VUG XLP TLT & 0.74 &  0.36 & XLK XLI IEF GLD & 0.83 & 0.60 & QQQ XLY XLU TLT GLD & 0.84 & 0.85 \\
        40 & XLK XLU GLD & 0.74 &  0.76 & VUG XLV IEF GLD & 0.82 & 0.56 & XLK XLI XLP TLT GLD & 0.84 & 0.86 \\
        41 & XLK XLV IEF & 0.73 &  0.23 & XLK XLU TLT GLD & 0.82 & 0.52 & QQQ VUG XLV IEF GLD & 0.84 & 0.61 \\
        42 & QQQ XLU TLT & 0.73 &  0.39 & XLY XLP IEF GLD & 0.82 & 0.49 & QQQ VTV XLP IEF GLD & 0.84 & 0.76 \\
        43 & QQQ VTV IEF & 0.73 &  0.49 & SPY VUG TLT GLD & 0.82 & 0.73 & QQQ VTV XLV TLT GLD & 0.84 & 0.78 \\
        44 & XLP TLT GLD & 0.73 &  0.15 & SPY XLP IEF GLD & 0.82 & 0.55 & QQQ VUG XLK IEF GLD & 0.84 & 0.46 \\
        45 & VUG XLK TLT & 0.73 &  0.30 & QQQ XLF TLT GLD & 0.82 & 0.60 & QQQ XLK XLY IEF GLD & 0.84 & 0.54 \\
        46 & QQQ XLY GLD & 0.73 &  0.37 & QQQ MDY TLT GLD & 0.82 & 0.55 & QQQ VTV XLK IEF GLD & 0.84 & 0.74 \\
        47 & SPY XLK TLT & 0.73 &  0.44 & VUG XLY TLT GLD & 0.82 & 0.65 & QQQ XLV XLY IEF GLD & 0.84 & 0.50 \\
        48 & VUG AGG GLD & 0.73 &  0.25 & SPY XLV TLT GLD & 0.82 & 0.64 & VTV XLK XLP TLT GLD & 0.84 & 0.81 \\
        49 & QQQ XLY TLT & 0.73 &  0.25 & QQQ XLK TIP GLD & 0.82 & 0.30 & SPY QQQ XLK IEF GLD & 0.84 & 0.57 \\
        50 & QQQ XLI IEF & 0.72 &  0.20 & XLK XLU IEF GLD & 0.82 & 0.51 & SPY XLK XLP IEF GLD & 0.83 & 0.83 \\
        51 & VUG XLK GLD & 0.72 &  0.55 & XLV XLP TLT GLD & 0.82 & 0.33 & QQQ XLP IBB TLT GLD & 0.83 & 0.47 \\
        52 & QQQ XLP TIP & 0.72 &  0.23 & VTV VUG TLT GLD & 0.81 & 0.70 & SPY QQQ XLU TLT GLD & 0.83 & 0.93 \\
        53 & VUG XLP IEF & 0.72 &  0.42 & XLK XLP TIP GLD & 0.81 & 0.47 & SPY XLK XLV TLT GLD & 0.83 & 0.77 \\
        54 & XLY XLP GLD & 0.72 &  0.39 & SPY VUG IEF GLD & 0.81 & 0.69 & QQQ XLK XLI IEF GLD & 0.83 & 0.52 \\
        55 & VTV XLK TLT & 0.72 &  0.40 & SPY XLY TLT GLD & 0.81 & 0.56 & QQQ VUG XLU IEF GLD & 0.83 & 0.84 \\
        56 & VUG TIP GLD & 0.72 & -0.03 & QQQ XLV TIP GLD & 0.81 & 0.55 & VUG XLK XLU TLT GLD & 0.83 & 0.82 \\
        57 & VUG XLK IEF & 0.72 &  0.29 & XLV XLY IEF GLD & 0.81 & 0.56 & QQQ XLP XLU IEF GLD & 0.83 & 0.71 \\
        58 & QQQ XLU IEF & 0.72 &  0.49 & VUG XLY IEF GLD & 0.81 & 0.63 & QQQ XLK TLT IEF GLD & 0.83 & 0.25 \\
        59 & XLK XLY GLD & 0.72 &  0.31 & VUG XLI TLT GLD & 0.81 & 0.68 & XLK XLV XLI TLT GLD & 0.83 & 0.65 \\
        60 & SPY QQQ GLD & 0.72 &  0.59 & QQQ XLF IEF GLD & 0.81 & 0.66 & VUG XLV XLP TLT GLD & 0.83 & 0.63 \\
        \bottomrule
    \end{tabular} \\
    \label{tab:exhaustive_search_inverse_volatility}
\end{table}

\newpage

\begin{table}[h]
    \footnotesize
    \centering
    \caption{Sensitivity check on final portfolios.}
    \begin{tabular}{ll|SSSSSSS}
        \toprule
        \multicolumn{8}{l}{\textit{2005--2025 (Full backtest window)}} \\
        \midrule
        \textbf{Portfolios} & \textbf{Strategy} & {$\boldsymbol{r}$} & {$\boldsymbol{\sigma}$} & \textbf{MDD} & \textbf{TO} & \textbf{SR} & \textbf{LQ} \\
        \midrule
        SPY/AGG (60/40)   & Fixed weights                  & 0.07 & 0.10 & 0.30 & 0.12 & 0.52 &  0.13 \\
        SPY               & Buy-and-hold vc8\%             & 0.06 & 0.09 & 0.20 & 0.29 & 0.53 & -0.03 \\
        SPY/AGG/GLD       & Equal weights                  & 0.08 & 0.09 & 0.24 & 0.17 & 0.68 &  0.49 \\
        SPY/TLT/GLD/IBIT  & Maximum diversification        & 0.10 & 0.09 & 0.28 & 0.42 & 0.90 &  0.54 \\
        Momentum Universe & Markowitz                      & 0.08 & 0.08 & 0.16 & 0.81 & 0.78 &  0.40 \\
        \midrule
        \multicolumn{8}{l}{\textit{2010--2025 (Excluding the financial crisis)}} \\
        \midrule
        \textbf{Portfolios} & \textbf{Strategy} & {$\boldsymbol{r}$} & {$\boldsymbol{\sigma}$} & \textbf{MDD} & \textbf{TO} & \textbf{SR} & \textbf{LQ} \\
        \midrule
        SPY/AGG (60/40)   & Fixed weights                  & 0.08 & 0.09 & 0.20 & 0.10 & 0.78 &  0.61 \\
        SPY               & Buy-and-hold vc8\%             & 0.07 & 0.09 & 0.20 & 0.28 & 0.70 & -0.04 \\
        SPY/AGG/GLD       & Equal weights                  & 0.08 & 0.08 & 0.17 & 0.13 & 0.75 &  0.26 \\
        SPY/TLT/GLD/IBIT  & Maximum diversification        & 0.11 & 0.09 & 0.28 & 0.38 & 1.02 &  0.84 \\
        Momentum Universe & Markowitz                      & 0.09 & 0.09 & 0.16 & 0.81 & 0.86 &  0.32 \\
        \midrule
        \multicolumn{8}{l}{\textit{2005--2020 (Excluding the COVID-19 pandemic and the start of the Russia-Ukraine war)}} \\
        \midrule
        \textbf{Portfolios} & \textbf{Strategy} & {$\boldsymbol{r}$} & {$\boldsymbol{\sigma}$} & \textbf{MDD} & \textbf{TO} & \textbf{SR} & \textbf{LQ} \\
        \midrule
        SPY/AGG (60/40)   & Fixed weights                  & 0.07 & 0.09 & 0.30 & 0.13 & 0.58 & 0.13 \\
        SPY               & Buy-and-hold vc8\%             & 0.06 & 0.09 & 0.18 & 0.32 & 0.57 & 0.00 \\
        SPY/AGG/GLD       & Equal weights                  & 0.08 & 0.09 & 0.24 & 0.19 & 0.71 & 0.26 \\
        SPY/TLT/GLD/IBIT  & Maximum diversification        & 0.10 & 0.08 & 0.15 & 0.43 & 1.09 & 0.44 \\
        Momentum Universe & Markowitz                      & 0.08 & 0.08 & 0.12 & 0.80 & 0.83 & 0.51 \\
        \bottomrule
    \end{tabular}
    \label{tab:final_portfolios_sensitivity_check}
\end{table}

\newpage

\section{Code}\label{app:code}

Throughout this project several million backtests have been run, most of which solve one convex problem per trading date. We first present code examples that illustrate how the backtest and our trading strategies are implemented in Python. We then collect some of the implementation choices that made this scale possible, which together reduced the running time of the exhaustive search from days to under an hour.

\subsection*{The backtest loop}
A backtest is a loop over trading days, and everything inside the loop should be plain vector and matrix operations. We load the adjusted closing prices once, convert them to a return matrix $R \in \mathbf{R}^{T \times n}$, and from that point on touch only NumPy arrays. Labeled data structures such as pandas dataframes are convenient but slow because of their per-operation overhead which dominates execution time. Restricting pandas to the boundary of the program and running the loop on NumPy arrays sped up each computed backtest by roughly 2 times. Listing~\ref{code:backtest_loop} shows the loop in full. 

\begin{lstlisting}[caption={The backtest loop}, label=code:backtest_loop]
for i in range(T):
    cash *= 1.0 + ffr[i]
    holdings *= 1.0 + returns[i]
    holdings *= universe[i]
    nav = cash + holdings.sum()
    weights = holdings / nav
    if i in trading_days:
        w_new = construct_portfolio(
            curr_weights=weights,
            alphas=alphas[i],
            F=Fs[i],
            D_half=D_halves[i],
            universe=universe[i],
        )
        turnover = 0.5 * np.abs(w_new - weights).sum()
        cash = (1.0 - w_new.sum() - spread * turnover) * nav
        holdings = w_new * nav
        weights = w_new
    navs[i] = nav
    composition[i] = weights
\end{lstlisting}

\subsection*{The strategies} Inside \texttt{construct\_portfolio} each strategy is a few lines of code, and all of the strategies of \S\ref{sec:portfolio_construction} are given in Listing~\ref{code:strategies}. The formula-based ones are plain NumPy, and the optimization-based ones are direct transcriptions of their formulations into CVXPY.

\newpage
\begin{lstlisting}[caption={The trading strategies}, label=code:strategies]
import cvxpy as cp
import numpy as np

def equal_weights(universe):
    return universe / universe.sum()

def inverse_volatility(sigma, universe):
    inv = universe / sigma
    return inv / inv.sum()

def volatility_control(w, F, D_half, sigma_tgt):
    sigma_hat = np.linalg.norm(np.r_[F.T @ w, D_half * w])
    return min(1.0, sigma_tgt / sigma_hat) * w

w = cp.Variable(n)
F = cp.Parameter((n, k))
D_half = cp.Parameter(n, nonneg=True)
risk_sqrt = cp.hstack([F.T @ w, cp.multiply(D_half, w)])
risk = cp.sum_squares(risk_sqrt)

min_vol = cp.Problem(
    cp.Minimize(risk), [cp.sum(w) == 1, w >= 0])

risk_parity = cp.Problem(
    cp.Minimize(0.5 * risk - rho * cp.sum(cp.log(w))),
    [cp.sum(w) <= 1, w >= 1e-8])

y = cp.Variable(n)
F_c = cp.Parameter((n, k))
d_c = cp.Parameter(n, nonneg=True)
mdp = cp.Problem(
    cp.Minimize(cp.sum_squares(F_c.T @ y)
                + cp.sum_squares(cp.multiply(d_c, y))),
    [cp.sum(y) == 1, y >= 0])

alpha = cp.Parameter(n)
w_pre = cp.Parameter(n)
cash = cp.Variable(nonneg=True)
trade = w - w_pre
markowitz = cp.Problem(
    cp.Maximize(alpha @ w - kappa * cp.norm1(trade)),
    [cp.sum(w) + cash == 1, w >= 0, w <= w_max,
     cp.norm2(risk_sqrt) <= sigma_tgt,
     cp.norm1(trade) <= T_tgt,
     cp.abs(trade) <= delta_w])
\end{lstlisting}

\subsection*{Two formulation habits} Both of the following are standard advice in the convex optimization literature, and both transfer directly to portfolio problems. First, risk is better expressed through a norm than through a quadratic form. The constraint $w^{\mathrm{T}}\Sigma w \le \sigma_{\mathrm{tgt}}^{2}$ and the constraint $\|L^{\mathrm{T}}w\|_2 \le \sigma_{\mathrm{tgt}}$, with $L$ the Cholesky factor of $\Sigma$, describe the same set, but the norm version hands the solver a second-order cone directly and avoids squaring the conditioning of the problem. When the covariance comes from the factor model of \S\ref{sec:factor_models}, the full matrix $F\Sigma^{f}F^{\mathrm{T}} + D$ should never be assembled. The portfolio variance splits into a factor part and an idiosyncratic part,
\[
w^{\mathrm{T}}\Sigma w
= \|\,\tilde{F}^{\mathrm{T}}w\,\|_2^{2} + \|\,D^{1/2}w\,\|_2^{2},
\]
where $\tilde{F} = F L_f$ and $L_f$ is the Cholesky factor of $\Sigma^{f}$, so the risk constraint involves one $k$-vector and one $n$-vector rather than an $n \times n$ matrix. This is the form used in Listing~\ref{code:strategies}. At $n = 59$ the saving is modest, but the formulation costs nothing, and it is the version that scales as the universe grows.

\subsection*{Parallelism}
A single backtest is serial, since the holdings on day $t$ depend on those on day $t-1$. The backtests themselves are independent of each other, so we run them in parallel. We use a pool of twelve worker processes, one per core. Processes rather than threads, since Python's global interpreter lock prevents threads from running Python code simultaneously. The price data is loaded and sliced once before the pool starts, so each worker only receives the arrays its own backtest needs.

\subsection*{Going faster} The setup above is fast enough for everything in this paper. If it were not, the same parametrized problems can be compiled. CVXPYgen~\cite{Schaller:2022} generates custom C code for a fixed problem structure, removing the remaining Python overhead from each solve, and slots into the loop with an unchanged interface. We did not need it for the results reported here, but for a larger universe, more frequent trading, or a wider hyperparameter sweep, it is the natural next step.

\newpage

\section{For the European investor}\label{app:EU_investor}
This paper only considers ETFs that are accessible to American investors. Since January 2018, EU regulation requires any fund sold to European retail investors to provide a key information document~\cite{priips2014}, which US-listed ETFs do not have. The ETFs in this report can therefore not be bought through an ordinary European brokerage account. Equivalents are instead available through European UCITS funds that track the same or closely related indices. Table~\ref{tab:eu_equivalents} lists the most relevant ETFs in this paper together with a European equivalent. 
 
\begin{table}[h]
    \footnotesize
    \setlength{\tabcolsep}{3pt}
    \centering
    \caption{European equivalents of selected ETFs presented in this paper.}
    \begin{tabular}{l|lll}
    \toprule
        \textbf{ETF} & \textbf{European equivalent} & \textbf{Ticker} & \textbf{ISIN}\\
        \midrule
        SPY  & iShares Core S\&P 500 UCITS ETF USD (Acc)                     & SXR8 & IE00B5BMR087 \\  
        QQQ  & iShares Nasdaq 100 UCITS ETF (Acc)                            & SXRV & IE00B53SZB19 \\  
        VUG  & Amundi Russell 1000 Growth UCITS ETF (Acc)                    & MWOW & IE0005E8B9S4 \\  
        VTV  & iShares Edge MSCI USA Value Factor UCITS ETF (Acc)            & QDVI & IE00BD1F4M44 \\  
        MDY  & SPDR S\&P 400 U.S. Mid Cap UCITS ETF (Acc)                    & SPY4 & IE00B4YBJ215 \\  
        IWM  & SPDR Russell 2000 U.S. Small Cap UCITS ETF (Acc)              & ZPRR & IE00BJ38QD84 \\  
        USMV & iShares Edge S\&P 500 Minimum Volatility UCITS ETF (Acc)      & IBCK & IE00B6SPMN59 \\  
        \midrule
        XLK  & SPDR S\&P U.S. Technology Select Sector UCITS ETF (Acc)       & ZPDT & IE00BWBXM948 \\  
        XLP  & SPDR S\&P U.S. Consumer Staples Select Sector UCITS ETF (Acc) & ZPDS & IE00BWBXM385 \\  
        XLV  & SPDR S\&P U.S. Health Care Select Sector UCITS ETF (Acc)      & ZPDH & IE00BWBXM617 \\  
        XLU  & SPDR S\&P U.S. Utilities Select Sector UCITS ETF (Acc)        & ZPDU & IE00BWBXMB69 \\  
        \midrule
        AGG  & iShares US Aggregate Bond UCITS ETF USD (Dist)                & EUNX & IE00B44CGS96 \\  
        IEF  & iShares USD Treasury Bond 7-10yr UCITS ETF (Dist)             & IUSM & IE00B1FZS798 \\  
        TLT  & iShares USD Treasury Bond 20+yr UCITS ETF (Dist)              & IS04 & IE00BSKRJZ44 \\  
        TIP  & iShares USD TIPS UCITS ETF (Acc)                              & IUST & IE00B1FZSC47 \\  
        \midrule
        GLD  & iShares Physical Gold ETC                                     & PPFB & IE00B4ND3602 \\  
        SLV  & WisdomTree Physical Silver                                    & VZLC & JE00B1VS3333 \\  
        \midrule
        IBIT & iShares Bitcoin ETP                                           & IB1T & XS2940466316 \\  
        ETHA & VanEck Ethereum ETN                                           & VETH & DE000A3GPSP7 \\  
        \bottomrule
    \end{tabular}
    \label{tab:eu_equivalents}
\end{table}

\noindent Several UCITS funds often track the same index. The funds listed here were selected because they are available through Avanza, a large Swedish online brokerage, so that a private investor can buy them directly. Other brokers may list other lines of the same fund, and the ISIN rather than the ticker is what identifies it.

The backtests of this paper assume that dividends are reinvested. This corresponds to an accumulating share class, marked (Acc), which reinvests the dividends inside the fund. Not every broker or fund offers this, and where we could not find an (Acc) equivalent we list the distributing class, marked (Dist), which pays the dividends out to the investor instead. The resulting return on your investment is mostly the same assuming you reinvest the dividends that is paid out. It does, however, affect what the price chart shows. The quoted price of a distributing fund usually excludes the dividends already paid out, so its growth may appear weaker than the return that is actually delivered.

\end{document}